\documentclass{article}
\pdfoutput=1
\usepackage{lscape}
\usepackage{enumerate}
\usepackage{color}
\usepackage{amsfonts}
\usepackage{amsmath}
\usepackage{comment}
\usepackage{textcomp}
\usepackage{amsthm, tikz}
\usepackage{amssymb}
\usepackage{color}
\usepackage{graphicx}
\usepackage{bbm}
\usepackage[matrix,arrow,curve]{xy}
\usepackage{array}
\usepackage{mathtools}

\def\be{\begin{eqnarray}}
\def\ee{\end{eqnarray}}
\def\nn{\nonumber}

\def\Tr{{\rm Tr}\,}

\def\l[{\phantom.[}

\def\GPK{starfish}
\def\cap{finger}

\def\bR{\bar R}
\def\bS{\bar S}
\def\bT{\bar T}

\hoffset= - 1.0in         

\begin{document}
\title{{\bf {Colored HOMFLY polynomials of knots presented as
{\it double fat} diagrams
}\vspace{.2cm}}
\author{{\bf A. Mironov$^{a,b,c,d}$}, \ {\bf A. Morozov$^{b,c,d}$}, \ {\bf An. Morozov$^{b,c,d,e}$}, \ {\bf P. Ramadevi$^{f}$}, \ {\bf Vivek  Kumar Singh$^{f}$}}
\date{ }
}

\maketitle

\vspace{-5.5cm}

\begin{center}
\hfill FIAN/TD-01/15\\
\hfill IITP/TH-02/15\\
\hfill ITEP/TH-04/15\\
\end{center}

\vspace{4.2cm}

\begin{center}

$^a$ {\small {\it Lebedev Physics Institute, Moscow 119991, Russia}}\\
$^b$ {\small {\it ITEP, Moscow 117218, Russia}}\\
$^c$ {\small {\it National Research Nuclear University MEPhI, Moscow 115409, Russia }}\\
$^d$ {\small {\it Institute for Information Transmission Problems, Moscow 127994, Russia}}\\
$^e$ {\small {\it Laboratory of Quantum Topology, Chelyabinsk State University, Chelyabinsk 454001, Russia }}\\
$^f$ {\small {\it Department of Physics, Indian Institute of Technology Bombay, Mumbai 400076, India}}
\end{center}

\vspace{1cm}

\begin{abstract}
Many knots and links in $S^3$ can be drawn as gluing of three manifolds with one or more four-punctured $S^2$ boundaries. We call these knot diagrams as double fat graphs whose invariants involve only the knowledge of the fusion and the braiding matrices of {\it four}-strand braids. Incorporating the properties of four-point conformal blocks in WZNW models, we conjecture colored HOMFLY polynomials for these double fat graphs where the color can be rectangular or non-rectangular representation. With the recent work of Gu-Jockers,
the fusion matrices for the non-rectangular $[21]$ representation, the first which involves multiplicity is known. We verify our conjecture by comparing with the $[21]$ colored
HOMFLY of many knots, obtained as closure of three braids. The conjectured form is computationally very effective leading to writing $[21]$-colored HOMFLY polynomials for many pretzel type knots  and non-pretzel type knots.
In particular, we find  class of pretzel mutants which are distinguished and another class of mutants which cannot be distinguished by $[21]$ representation. The difference between the [21]-colored HOMFLY of two mutants seems to have a general form, with $A$-dependence completely defined by the old conjecture due to Morton and Cromwell. In particular, we check it for an entire multi-parametric family of mutant knots evaluated using evolution method.
\end{abstract}

\vspace{1cm}

\section{Introduction}
$R$-colored HOMFLY polynomial \cite{knotpols}for  any knot  $\mathcal K$ is defined as the Wilson loop average in
Chern-Simons theory on three-manifold $\mathcal M$ \cite{CS}
\be
H_R^{{\cal K}\subset {\cal M}} \sim \int
\left( {\rm Tr}_R \, P\!\exp\oint_{\cal K} {\bf A}\right)
 \exp\left\{\frac{\kappa}{4\pi}\int_{{\cal M}}
\Tr_{adj}\left({\bf A}d{\bf A} +\frac{2}{3}{\bf A}^3\right)\right\}D{\bf A}~.
\ee
In fact, these polynomials are  Laurent polynomials (that is, all the coefficients are integers)
in two variables $q= \exp\left(\frac{2\pi i}{\kappa+N}\right)$ and $A=q^N$ for knots in $\mathcal M=S^3$.
Here $N$ denotes the rank of the $SU(N)$ gauge group and $k$ is the Chern-Simons coupling.

These polynomials can be lifted to a positive integer superpolynomial
\cite{DGR}-\cite{KhR} by introducing an additional
parameter ${\bf t}$. The deviation from ${\bf t}= -1$ is known as $\beta$-deformation \cite{betadefo}.
There is no combinatoric approach of obtaining superpolynomials for any knot.
It is still a mystery to obtain superpolynomials for non-rectangular representations $R$ (i.e. when the corresponding Young diagram $R$ is not rectangular)\cite{Ano21,GGS,ArthMM}.
Even the evaluation of  colored HOMFLY polynomials for such representations is a difficult task.

Two recent achievements open a way to perform explicit calculation of the colored HOMFLY
polynomials beyond rectangular representations.
\begin{itemize}
\item The first of them is the evaluation of the 3-strand knot polynomials for representation $R=[21]$
in \cite{Ano21,AnoMcabling} and the data of  Racah matrices $S$ and $\bar S$ given
in Ref.\cite{GuJ}.
\item The second is the conjectured expression \cite{Kaul,NRZ2,GMMMS} for knots in $S^3$ obtained
from gluing three-manifolds with one or more four-punctured $S^2$ boundaries. We
call such diagrams as double fat tree diagrams (see definitions below) .
The unreduced colored HOMFLY polynomial involving the Racah matrices
for such knot diagrams will be
\be
\boxed{
H_R^{\cal K} =  \sum_{\{X_\mu,a_i^{(\mu)}\}} \!\!\!\!
\sigma_{_{\!X|a}}
\prod_{{\rm vertices\ \mu}}^{} d_{_{X_\mu}}
\prod_{{\rm propagators}\ \mu\nu}^{} \frac{d_R}{\sqrt{d_{_{X_\mu}}d_{_{X_\nu}}}}\
{\cal A}_{X_\mu X_\nu}^{a_i^{(\mu)}a_j^{(\nu)}}
}
\label{basic}
\ee
We indicate  each of the three-manifolds $\mu$'s  with $k$  four-punctured $S^2$ boundaries
as a $k_{\mu}$-valent vertex. Further, we place representation $X_{\mu}\in  R\otimes \bar R ~{\rm or}~
R\otimes R$, depending on the configuration and orientations of  the knot diagram (see example).
We will have additional indices $a_i^{(\mu)} = 1,\ldots,m_{X_\mu}$ in the summation besides $X_{\mu}$'s,
whenever $X$ has a non-trivial multiplicity $m_X$.
The weight factor is the product of quantum dimensions $d_X$ of $X$.
Additional factors are just signs, $\sigma=\pm 1$,
which  appear only for non-trivial multiplicities. We will  see in sec.\ref{signs} that  these $\sigma$'s
are crucially important for non-rectangular $R$ including $R=[21]$.
The form of ${\cal A}_{X_\mu X_\nu}^{a_i^{(\mu)}a_j^{(\nu)}}$
will depend on braiding and fusion matrices as explained in  eqn.(\ref{Amatr}).
\end{itemize}
{\bf Definitions}\\
{\it Tree diagram}:
We denote  the knots drawn as double fat  diagrams as ordinary trees with vertices $\mu$'s of arbitrary valencies $k_{\mu}$'s:
\be
\begin{picture}(300,60)(-150,-15)
\put(0,0){\circle{20}}
\put(50,0){\circle{20}}
\put(80,30){\circle{20}}
\put(80,-30){\circle{20}}
\put(110,0){\circle{20}}
\put(140,30){\circle{20}}
\put(10,0){\line(1,0){30}}
\put(57,7){\line(1,1){16}}
\put(57,-7){\line(1,-1){16}}
\put(87,23){\line(1,-1){16}}
\put(90,30){\line(1,0){40}}
\put(20,-30){\circle{20}}
\put(-30,-30){\circle{20}}
\put(43,-7){\line(-1,-1){16}}
\put(-20,-30){\line(1,0){30}}
\put(45,-2){\mbox{$X_\mu$}}
\put(75,28){\mbox{$X_\nu$}}
\end{picture}
\label{tree}
\ee
\vspace{1cm}

\noindent
{\it An example}\\
Suppose a knot  can be presented in the following form
(which we call double fat diagram)
\bigskip

\begin{figure}[h!]
\centering\leavevmode
\includegraphics[width=6 cm]{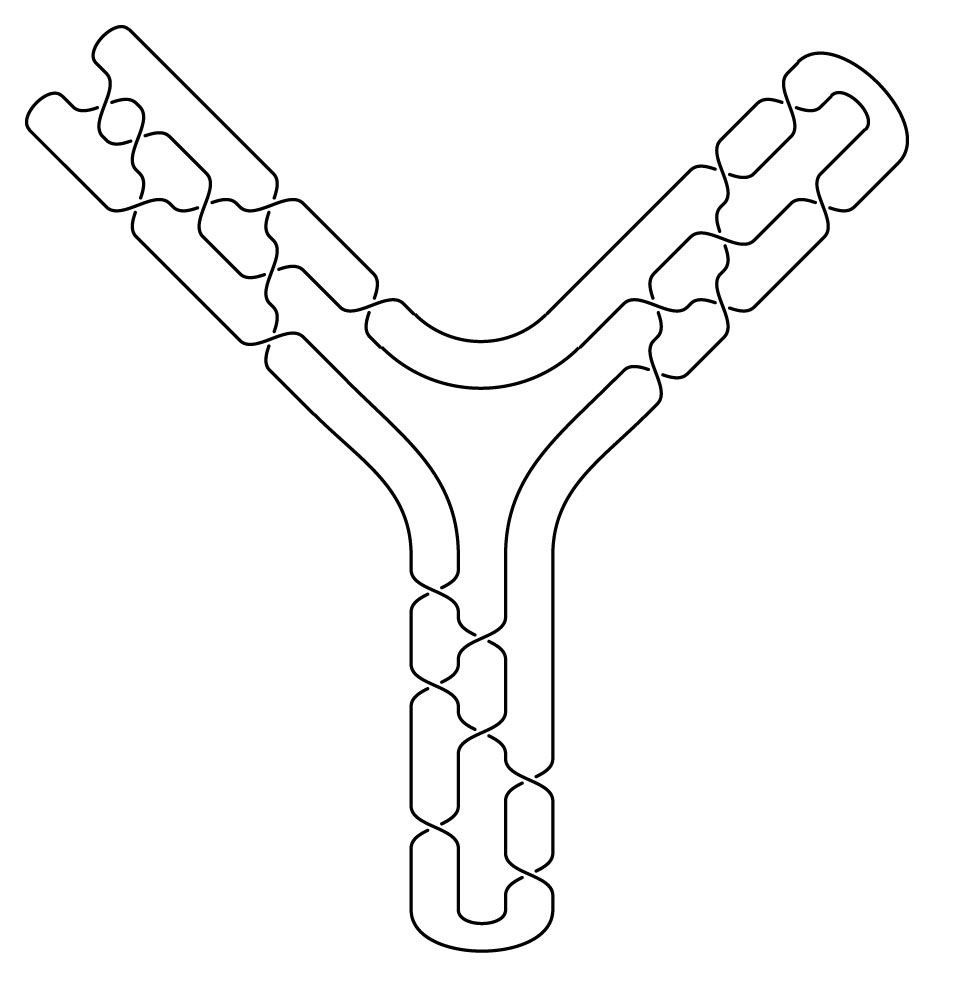}
\caption{ Knot drawn as double fat diagram}
\label{example}
\end{figure}


\noindent
The above double fat knot diagram  is presentable as a single trivalent vertex connected by edges  to three
monovalent vertices. In this example,  it is three four-strand braids (propagators, or edges) connected by double lines into the vertex $X$, we call it the double fat diagram.
In the field theory context, this particular diagram has a peculiar shape of a {\it starfish} with three fingers. The colored HOMFLY polynomial of such knots will be

\bigskip

\be
\begin{picture}(300,40)(50,-15)
\put(0,0){\circle{20}}
\put(50,0){\circle{20}}
\put(80,30){\circle{20}}
\put(80,-30){\circle{20}}
\put(10,0){\line(1,0){30}}
\put(57,7){\line(1,1){16}}
\put(57,-7){\line(1,-1){16}}
\put(45,-2){\mbox{$X$}}
\put(130,-1){\mbox{$=$}}
\put(140,-2){\mbox{~~$\sum_X \sum_{a_1,a_2,a_3=1}^{\mu_X} \sigma_{X|a_1,a_2,a_3}\cdot \frac{{d_R}^3}{d_{X}^{1/2}} {\cal A}_{X1}^{a_1a_2}(1)  {\cal A}_{X1}^{a_2a_3}(2)  {\cal A}_{X1}^{a_3a_1}(3)$}}
\end{picture}
\label{three}
\ee

\bigskip
\vskip.2cm

\noindent
The subscript $1$ in ${\cal A}_{X1}^{a_ia_j}$ indicates the singlet representation (\ref{parpre}). In this paper, we will call  graphs involving a $k$-valent vertex connected to $k$ monovalent vertices as {\it starfish} diagram with $k$ fingers.
\bigskip

\noindent
{\it Edge in the tree diagram}:
We always consider each edge in (\ref{tree}) being a 4-strand braid:
\be
\begin{picture}(300,220)(-60,-45)
\qbezier(-10,0)(20,15)(50,0)\qbezier(-10,0)(20,-15)(50,0)
\qbezier(-10,140)(20,155)(50,140)\qbezier[40](-10,140)(20,125)(50,140)
\put(-10,0){\line(0,1){140}}  \put(50,0){\line(0,1){140}}
\put(5,50){\line(-1,-5){14}}
\put(3,-20){\line(1,4){10}}
\put(28,20){\line(1,-4){10}}
\put(49,-20){\line(-1,5){14}}
\put(-50,17){\line(1,0){150}}  \put(-55,20){\mbox{$S$}}
\put(10,20){\line(1,0){20}}
\put(10,40){\line(1,0){20}}
\put(10,20){\line(0,1){20}}
\put(30,20){\line(0,1){20}}
\put(-50,43){\line(1,0){150}}  \put(-55,46){\mbox{$S^\dagger$}} \put(90,28){\mbox{$T_0^{n_1}$}}

\put(3,50){\line(1,0){14}}
\put(3,65){\line(1,0){14}}
\put(3,50){\line(0,1){15}}
\put(17,50){\line(0,1){15}}
\put(23,50){\line(1,0){14}}
\put(23,65){\line(1,0){14}}
\put(23,50){\line(0,1){15}}
\put(37,50){\line(0,1){15}}
\put(-50,72){\line(1,0){150}}  \put(-55,75){\mbox{$S$}} \put(83,55){\mbox{$T_-^{m_1}T_+^{l_1}$}}
\put(10,75){\line(1,0){20}}
\put(10,95){\line(1,0){20}}
\put(10,75){\line(0,1){20}}
\put(30,75){\line(0,1){20}}
\put(-50,98){\line(1,0){150}}  \put(-55,101){\mbox{$S^\dagger$}}\put(90,84){\mbox{$T_0^{n_2}$}}
\put(3,105){\line(1,0){14}}
\put(3,120){\line(1,0){14}}
\put(3,105){\line(0,1){15}}
\put(17,105){\line(0,1){15}}
\put(23,105){\line(1,0){14}}
\put(23,120){\line(1,0){14}}
\put(23,105){\line(0,1){15}}
\put(37,105){\line(0,1){15}}
\put(-50,123){\line(1,0){150}}  \put(-55,126){\mbox{$S$}} \put(83,110){\mbox{$T_-^{m_2}T_+^{l_2}$}}
\put(5,65){\line(0,1){40}}
\put(35,65){\line(0,1){40}}
\put(14,40){\line(0,1){10}}
\put(26,40){\line(0,1){10}}
\put(14,65){\line(0,1){10}}
\put(26,65){\line(0,1){10}}
\put(14,95){\line(0,1){10}}
\put(26,95){\line(0,1){10}}
\put(5,120){\line(-1,5){8}}
\put(14,120){\line(-1,5){8}}
\put(35,120){\line(1,5){8}}
\put(26,120){\line(1,5){8}}
\put(17,27){\mbox{$n_1$}}
\put(17,83){\mbox{$n_2$}}
\put(4,56){\mbox{$m_1$}}
\put(27,56){\mbox{$l_1$}}
\put(4,110){\mbox{$m_2$}}
\put(27,110){\mbox{$l_2$}}
\put(-12,-35){\mbox{$X,a$}}
\put(34,-35){\mbox{$\bar X,b$}}
\put(-12,170){\mbox{$Y,c$}}
\put(34,170){\mbox{$\bar Y,d$}}
\put(200,50){\line(1,0){20}}
\put(200,50){\line(0,1){20}}
\put(200,70){\line(1,0){20}}
\put(220,50){\line(0,1){20}}
\put(205,50){\line(0,-1){20}}
\put(215,50){\line(0,-1){20}}
\put(205,70){\line(0,1){20}}
\put(215,70){\line(0,1){20}}
\put(208,57){\mbox{$2$}}
\put(234,58){\mbox{$=$}}
\qbezier(265,50)(280,60)(265,70)
\qbezier(265,50)(250,60)(265,70)
\qbezier(265,50)(255,45)(255,20)
\qbezier(265,50)(275,45)(275,20)
\qbezier(265,70)(255,75)(255,100)
\qbezier(265,70)(275,75)(275,100)
\put(250,40){\line(1,0){30}}
\put(250,40){\line(0,1){40}}
\put(250,80){\line(1,0){30}}
\put(280,40){\line(0,1){40}}
\put(265,50){\circle*{4}}
\put(265,70){\circle*{4}}
\end{picture}
\label{edge}
\ee

\noindent
Boxes in this picture denote vertical 2-strand braids of given lengths,
which can be parallel or antiparallel, depending on the directions of arrows
(which in their turn depend on parity of the lengths).
The number of boxes in each propagator can be arbitrary. For a given braid word in the above picture,
we will have to insert the appropriate $S,S^{\dagger}$ fusion matrices and suitable
powers of the braiding matrices $T_-,T_+,T_0$ as highlighted. These four-strand braids are called propagators which connects two different vertices.  We put represention $X$ and $Y$ on the two vertices as explained below.

\noindent
{\it Vertex in tree diagram}:
Each vertex $\mu$ in (\ref{tree}) is the cyclic junctions of $k_\mu$
4-strand braids:

\begin{picture}(300,230)(-50,-140)
\qbezier(0,0)(20,20)(40,0)\qbezier(0,0)(20,-20)(40,0)
\qbezier[30](0,60)(20,40)(40,60)\qbezier(0,60)(20,80)(40,60)
\put(0,0){\line(0,1){60}} \put(40,0){\line(0,1){60}}
\qbezier(100,0)(120,20)(140,0)\qbezier(100,0)(120,-20)(140,0)
\qbezier[30](100,60)(120,40)(140,60)\qbezier(100,60)(120,80)(140,60)
\put(100,0){\line(0,1){60}} \put(140,0){\line(0,1){60}}
\qbezier(200,0)(220,20)(240,0)\qbezier(200,0)(220,-20)(240,0)
\qbezier[30](200,60)(220,40)(240,60)\qbezier(200,60)(220,80)(240,60)
\put(200,0){\line(0,1){60}} \put(240,0){\line(0,1){60}}
\put(290,20){\mbox{$\ldots$}}
\qbezier(350,0)(370,20)(390,0)\qbezier(350,0)(370,-20)(390,0)
\qbezier[30](350,60)(370,40)(390,60)\qbezier(350,60)(370,80)(390,60)
\put(350,0){\line(0,1){60}} \put(390,0){\line(0,1){60}}
\qbezier(26,0)(70,-54)(114,0)\qbezier(35,0)(70,-40)(105,0)
\qbezier(126,0)(170,-54)(214,0)\qbezier(135,0)(170,-40)(205,0)
\qbezier(226,0)(260,-27)(280,-27)\qbezier(235,0)(260,-20)(280,-20)
\qbezier(310,-27)(330,-27)(364,0)\qbezier(310,-20)(330,-20)(355,0)
\qbezier(5,0)(190,-220)(385,0)\qbezier(14,0)(190,-200)(376,0)
\qbezier(5,0)(190,-220)(385,0)\qbezier(14,0)(190,-200)(376,0)
\qbezier(87,-7)(103,-10)(104,-22)\qbezier[20](87,-9)(90,-21)(104,-22)
\put(106,-26){\mbox{$X,a_1$}}
\qbezier(187,-7)(203,-10)(204,-22)\qbezier[20](187,-9)(190,-21)(204,-22)
\put(206,-26){\mbox{$X,a_2$}}
\qbezier(336,-4)(350,-10)(350,-22)\qbezier[20](336,-4)(339,-20)(350,-22)
\put(305,4){\mbox{$X,a_{k-1}$}}
\qbezier(312,-47)(325.5,-50)(330,-65)\qbezier[20](312,-47)(315,-60)(330,-65)
\put(332,-70){\mbox{$X,a_k$}}
\end{picture}

\noindent
With each such vertex $\mu$ one associates a representation $X_\mu$.
The crucial feature of this construction is the selection rule for propagators:
the two representations $X,\bar X$ at one end of the four-strand braid (\ref{edge}) differ by conjugation,
while the additional indices $a$ and $b$(multiplicity indices) can be different.

Now we are in a position to write the explicit form for matrix element ${\cal A}_{Ycd,Xab,}$
for every propagator connecting two vertices with index $Xab$ and $Ycd$.
The form for the matrix, in terms of the fusion or Racah matrix $S$ and
the braiding matrices $T_0,T_-,T_+$, will be
\be
{\cal A}\  =    S\cdot \prod_\alpha^{\leftarrow} \
\Big(\ T_-^{\,m_\alpha}T_+^{\,l_\alpha}\, S^\dagger \, T_0^{\,n_\alpha}\, S\ \Big)
\label{Amatr}
\ee
Interchanging $m_\alpha$ and $l_\alpha$ at each level $\alpha$ inside every
\cap is a mutation transform (provided one considers only the tree diagrams), and in fact one can perform a mutation to change
$m_\alpha,l_\alpha \rightarrow m_\alpha+k_\alpha,l_\alpha-k_\alpha$ with any $k_\alpha$.
 For symmetric representations $R$, the HOMFLY polynomial depends
only on the sums $m_\alpha+l_\alpha$ as there is no distinction between $T_-$ and $T_+$.
However,  for  non-rectangular $R$,
 the  invariant depends on $m_\alpha$ and $l_\alpha$ separately leading to a possibility
 of distinguishing mutants.

\bigskip

In the case of the fundamental representation $R=[1]$ of $SU_q(2)$,
i.e. for $N=2$, expressions (\ref{Amatr}) involve
\be
S = S^\dagger = \left(\begin{array}{cc} \frac{1}{[2]} & \frac{\sqrt{[3]}}{[2]} \\
\frac{\sqrt{[3]}}{[2]} & -\frac{1}{[2]} \end{array}\right), \ \ \ \
T = \left(\begin{array}{cc} 1&0\\ 0 & -q^2 \end{array}\right)
\ee
in vertical framing. We need to add frame corrections to $T$-matrix resulting in different form for parallel and antiparallel braiding.
For higher $N$,  there are two different Racah matrices,  $S$ for parallel and
$\bar S$ for antiparallel braids respectively and the respective frame corrected braiding matrices $T,\bar T$  as given below:
\begin{eqnarray}
S= \frac{1}{[N]}\left(\begin{array}{cc} \sqrt{\frac{[N][N-1]}{[2]}} & \sqrt{\frac{[N][N+1]}{[2]}} \\ \\
\sqrt{\frac{[N][N+1]}{[2]}} & -\sqrt{\frac{[N][N-1]}{[2]}} \end{array}\right), \ \ \ \
T = -\frac{1}{qA}\left(\begin{array}{cc} 1&0\\ 0 & -q^2 \end{array}\right)
\label{Sfund}\\
\bar S = \frac{1}{[N]}\left(\begin{array}{cc} 1 &  \sqrt{[N-1][N+1]}  \\ \\
\sqrt{[N-1][N+1]}  & -1 \end{array}\right), \ \ \ \
\bar T = \left(\begin{array}{cc} 1&0\\ 0 & -A \end{array}\right)
\label{barSfund}
\end{eqnarray}
where $A=q^N$.
Two types of matrices $S$ remain sufficient for symmetric, rectangular and non-rectangular representations.
However, for  non-rectangular $R$ we will have additional indices $a$ and $b$ added to $X$ to incorporate the
multiplicity of some representations.

For any representation $R$, the first row in the Racah matrices $S$ and $\bar S$ are always constructed from the quantum dimensions:
\begin{eqnarray}
S_{1X}^{ab} &=& \frac{\sqrt{d_X}}{d_R}\,\delta^{ab}, \ \ \ X\in R\otimes R, \ \ \ 1\in R\otimes\bar R \nn \\
\bar S_{1X}^{ab} &=& \frac{\sqrt{d_X}}{d_R}\,\delta^{ab}, \ \ \ X\in R\otimes \bar R \ \ \ \ \ \ \ \ \ \ \
\label{firstline}
\end{eqnarray}
This property is crucial for self-consistency of (\ref{basic}) when an edge is a tadpole.
In fact, the role of the end-vertex in a tree diagram (with the sum over the corresponding $X$)
is to imitate gluing of the caps like
\be
\begin{picture}(300,95)(-70,-27)
\qbezier(0,-20)(10,40)(30,40)\qbezier(60,-20)(50,40)(30,40)
\qbezier(8,-20)(15,32)(30,32)\qbezier(52,-20)(45,32)(30,32)
\qbezier(-10,0)(30,-15)(70,0)\qbezier(-10,0)(30,15)(70,0)
\qbezier(-10,0)(-10,55)(30,55)\qbezier(30,55)(70,55)(70,0)
\qbezier(-5,-12)(2.5,-8)(12,-16)\qbezier(-5,-12)(2.5,-18)(12,-16)
\put(13,-25){\mbox{$X,a$}}
\qbezier(65,-12)(57.5,-8)(48,-16)\qbezier(65,-12)(57.5,-18)(48,-16)
\put(65,-25){\mbox{$\bar X,b$}}
\put(100,20){\mbox{$= \ \ \  \left\{\begin{array}{ccc} \bar S_{1X}^{\,ab} & & X\in R\otimes \bar R \\
&{\rm if} & \\ S_{1X}^{\,ab}& & X \in R\otimes R\end{array}\right.$}}
\end{picture}
\label{capXX}
\ee
\noindent
with a singlet representation $1\in R\otimes \bar R$,
or
\be
\begin{picture}(300,94)(-100,-31)
\qbezier(0,-20)(10,40)(15,40)\qbezier(10,-20)(20,37)(15,40)
\qbezier(60,-20)(50,40)(45,40)\qbezier(50,-20)(40,40)(45,40)
\qbezier(-10,0)(30,-15)(70,0)\qbezier(-10,0)(30,15)(70,0)
\qbezier(-10,0)(-10,55)(30,55)\qbezier(30,55)(70,55)(70,0)
\qbezier(-5,-12)(2.5,-8)(15,-16)\qbezier(-5,-12)(2.5,-18)(15,-16)
\put(17,-25){\mbox{$X$}}
\qbezier(65,-12)(57.5,-8)(45,-16)\qbezier(65,-12)(57.5,-18)(45,-16)
\put(65,-25){\mbox{$\bar X$}}
\put(100,20){\mbox{$= \ \delta_{1X}$}}
\end{picture}
\label{capX0}
\ee
\noindent
Relations (\ref{firstline}) ensure that each tadpole (end-edge of a branch in the tree,
which we call \cap) is nothing but the matrix element ${\cal A}_{1X}^{ab}$.
Hence,  we omit the cycles at the end-edges of the tree branches.
Additional universally applicable simplifications emerge from the elementary $S,T$-matrix
identities, like the celebrated $(ST)^3=1$ in the case of $R=[1]$ for $SU_q(2)$,
and their generalization to arbitrary representations of $SU_q(N)$ (\ref{SbTbS}).

\bigskip

The set of fat-tree diagrams is quite ample.
According to \cite{NRZ2},
which  contains fat diagram presentations of  knots upto 10 crossings in
Rolfsen table \cite{katlas} (with the possible exception of $10_{161}$),
we believe that most of the  knots must have an equivalent  fat diagram presentation.
In fact,  many complicated higher crossing  knot invariants  looks calculable
when presented as double fat graphs. Particularly, the  knots in Ref.\cite{NRZ2}
belongs to family of  starfish diagram with  three-fingers.
We will see that there are plenty of  {\it mutants}  within the starfish family,
which belong to the familiar class called  {\it pretzel} knots \cite{pret}.
Interestingly, we can calculate $R=[21]$ colored HOMFLY polynomials for such mutant pairs and explicitly check whether the mutants are distinguishable or not.

\bigskip

It is appropriate to mention that the eq.(\ref{basic}) describing
the colored HOMFLY polynomials for the fat tree diagrams is by no means a trivial formula.  Particularly, consistency of  knot equivalences led to
formal theorems for $k$-valent vertex states\cite{KaulRama}.  However,
its origins are not immediately clear  neither from the conformal blocks \cite{Wit,KaulRama} nor  from the Reshetikhin-Turaev \cite{RT,MMMknots12,RTmod} approaches to knot/link polynomials.
There is lot of evidence for the eq.(\ref{basic}) from
the conformal block method \cite{Kaul,NRZ2,GMMMS}
and its advanced versions like
the evolution \cite{evo}, cabling \cite{AnoMcabling} and differential expansion \cite{ArthMM} methods for symmetric representations. In this paper,
we will do similar verification for $R=[21]$.
Moreover, eq.(\ref{basic}) looks typical for topological field theories
(like the Hurwitz model in \cite{Hurw,MMN}) and it can serve as a basis of a new intuition and calculus
in knot theory, involving a kind of pant decomposition of link diagrams.


The paper is organized as follows. In the first part we describe the elementary building blocks for double far graphs and explain
how to construct knots out of them. In the second part we first describe some peculiarities of the representation theory especially
related to the first non-trivial mixed representation $R=[21]$, and then consider the colored HOMFLY polynomials in representation $R=[21]$ for various knots. In particular, we check that these HOMFLY polynomials do differ between the notorious Conway-Kinoshita-Terasaka (KTC)
mutant pair, in accordance with the expectations \cite{Mort}. Moreover, the manifest difference between their HOMFLY polynomials is in complete agreement with \cite{Mura}. We also discuss other mutant pairs.  In particular, we  find a difference between the HOMFLY polynomials for a whole 2-parametric family of mutants. However, representation $R=[21]$  is not sufficient to distinguish
between another set of mutants.  We list explicit $[21$ colored HOMFLY polynomials for various knots  in the Appendix. We  have  a table that describes the features of all knots upto 10 crossings.

Note that all answers for the HOMFLY polynomials in representation $R=[21]$ in this paper for the knots that have a three braid representation we compared with the results obtained by the cabling method of \cite{Ano21,AnoMcabling}. Besides, we made a few self-consistency checks, see s.\ref{tests}.

Throughout the paper we use the notation $[n]$ both for the quantum number and for the representation (e.g., $[21]$). Hopefully this would not lead to a misunderstanding. Other notations are
\be
D_i = \frac{Aq^i-A^{-1}q^{-i}}{q-q^{-1}}\ \ \ \ \ \ \ \ \ \ \ \ \ \ \ \ \ \{x\}=x-{1\over x}
\ee
and $d_R$ is the quantum dimension of the representation $R$.

\newpage

\part{Diagrammar}

\section{Chern-Simons evolution}

Expression (\ref{Amatr}) for each propagator (\ref{edge}) in (\ref{basic})
is usually interpreted   as the ordinary time evolution
in Chern-Simons theory, which is provided by monodromies of conformal block
made from braiding matrices $T_{\pm},T_0$ which are of three types
(depending on the pair of adjacent strands that it acts on) and Racah matrices $S$,
see \cite{Wit} for the original idea and \cite{KaulRama,NRZ2,GMMMS} for recent
applications. In the case of arbitrary representation $R$,  the
form of the $S$ matrix and the eq.(\ref{Amatr}) needs to be formulated
accurately. This section describes our notation, which will be used in the rest of
the text.

$S$ is a matrix, with two peculiar multi-indices $Xab$ and $Ycd$, which are collections of square matrices of different sizes, depending on irreducible representations $X$ and $Y$.
Two more potentially convenient notation are
\be
\begin{picture}(400,110)(-300,-30)
\put(-285,25){\mbox{$
S^{1234}_{Ycd,Xab} = S\Big[\underbrace{R_1,R_2}_{X,a}|\underbrace{R_3,R_4}_{\bar X,b}\Big]
= S^{Ycd}_{Xab}\left[\begin{array}{cc}R_1 & R_2\\ R_3& R_4 \end{array}\right]\ :
$}}
\qbezier(-200,35)(-188,50)(-176,35)
\put(-195,45){\mbox{$Y,c$}}
\qbezier(-218,35)(-189,75)(-160,35)
\put(-172,55){\mbox{$\bar Y,d$}}
\put(0,0){\line(0,1){60}}
\put(10,0){\line(0,1){20}}
\qbezier(10,20)(10,26)(20,30)\qbezier(20,30)(30,34)(30,40)
\put(30,40){\line(0,1){20}}
\qbezier(60,20)(60,26)(50,30)\qbezier(50,30)(40,34)(40,40)
\put(40,40){\line(0,1){20}}
\put(60,0){\line(0,1){20}}
\put(70,0){\line(0,1){60}}
\qbezier[30](0,20)(0,39)(40,40)\qbezier[30](40,40)(80,41)(80,60)
\put(-14,3){\mbox{$R_1$}}\put(12,3){\mbox{$R_2$}} \put(46,3){\mbox{$R_3$}}
\put(72,3){\mbox{$R_4$}}
\put(-5,-15){\mbox{$X,a$}} \put(75,-15){\mbox{$\bar X,b$}}
\put(25,68){\mbox{$Y,c$}} \put(65,68){\mbox{$\bar Y,d$}}
\end{picture}
\label{Spic}
\ee
\noindent
The punctured line helps to illustrate the meaning of indices, it indicates that, after applying the Racah matrix $S$,
the representation $R_1$ becomes close to $R_4$.
Note that the initial state $Xab$ stands to the right/bottom of the final $Ycd$,
but representations are ordered oppositely: from the left to the right and from the
top to the bottom.
When we write explicit matrices in sec.\ref{Racah},
the left/upper indices (like $Ycd$) label rows,
while the right/bottom ($Xab$) -- columns.

\bigskip

The composition of two evolutions gives the identity:

\begin{picture}(300,160)(-250,-10)
\put(-200,50){\mbox{$
\sum\limits_{Y,c,d} S^{2341}_{Zef,Ycd}\cdot S^{1234}_{Ycd,Xab} = \delta_{X\bar Z} \delta_{af}\delta_{be}
$}}
\put(0,0){\line(0,1){20}}
\put(10,0){\line(0,1){20}}
\qbezier(10,20)(10,26)(20,30)\qbezier(20,30)(30,34)(30,40)
\put(30,40){\line(0,1){20}}
\qbezier(80,20)(80,26)(60,30)\qbezier(60,30)(40,34)(40,40)
\put(40,40){\line(0,1){20}}
\put(80,0){\line(0,1){20}}
\put(90,0){\line(0,1){80}}
\qbezier[30](0,20)(0,39)(50,40)\qbezier[30](50,40)(100,41)(100,60)
\put(-14,3){\mbox{$R_1$}}\put(12,3){\mbox{$R_2$}} \put(66,3){\mbox{$R_3$}}
\put(92,3){\mbox{$R_4$}}
\put(45,109){\mbox{$R_3$}}\put(72,109){\mbox{$R_4$}}
\put(85,109){\mbox{$R_1$}}\put(112,109){\mbox{$R_2$}}
\put(-5,-15){\mbox{$X,a$}} \put(75,-15){\mbox{$\bar X,b$}}
\put(10,68){\mbox{$Y,c$}} \put(105,68){\mbox{$\bar Y,d$}}
\put(55,125){\mbox{$Z,e$}} \put(95,125){\mbox{$\bar Z,f$}}
\qbezier[50](-10,60)(70,60)(130,60)
\put(30,60){\line(0,1){20}}
\put(40,60){\line(0,1){20}}
\qbezier(40,80)(40,86)(50,90)\qbezier(50,90)(60,94)(60,100)
\put(60,100){\line(0,1){20}}
\put(70,100){\line(0,1){20}}
\qbezier(70,100)(70,94)(80,90)\qbezier(80,90)(90,86)(90,80)
\put(90,60){\line(0,1){20}}
\put(100,60){\line(0,1){60}}
\qbezier[30](30,80)(30,99)(75,100)\qbezier[30](75,100)(110,101)(110,120)
\end{picture}

\bigskip

\noindent
If combined with unitarity of $S$ (orthogonality when $S$ is real),
\be
\sum_{Y,c,d} S^{1234}_{Ycd,Zef}\cdot \Big(S^{1234}\Big)^{\ast}_{Ycd,Xab} = \delta_{XZ} \delta_{ae}\delta_{bf}
\ee
this implies:
\be
S^{2341}_{Zef,Ycd} = S^{1234}_{Ycd,\bar Zfe} 
\ee

\bigskip

To study knots, we require  two of the four representations $R_1,R_2,R_3,R_4$ to be  $R$ and the other
two to be conjugate representation  $\bR$.  This implies that there are two independent matrices, which we denote $S$ and $\bar S$:
\be
S^{R\bR R\bR}_{Ycd,Xab} = \bar S_{Ycd,Xab}, \ \ \ \ \ \ \ Xab, Ycd \in R\otimes \bar R, \nn \\
S^{\bR R\bR R}_{Ycd,Xab} = S^{R\bR R\bR}_{Xab,\bar Ydc} = \bar S_{Xab,\bar Ydc}, \nn \\
\nn \\
S^{RR\bR\bR}_{Ycd,Xab} = S_{Ycd,Xab}, \ \ \ \ Xab\in R\otimes R, \ Ycd\in R\otimes \bR, \nn \\
S^{\bR\bR RR}_{Ycd,Xab} = S^{R\bR\bR R}_{Xab,\bar Ydc} = S_{\bar Y dc,\bar X ba}, \nn \\
S^{R\bR\bR R}_{Ycd,Xab} = S^{RR\bR\bR}_{Xab,\bar Y dc} = S_{Xab,\bar Ydc}, \nn \\
S^{\bar R RR\bR}_{Ycd,Xab} = S^{\bR\bR RR}_{Xab,\bar Ydc} = S_{\bar Xba,Ycd}
\label{Stypes}
\ee
Note that $S$ converts two parallel braids into two antiparallel,
while $\bar S$ converts antiparallel into antiparallel:
\be
\begin{array}{|ccc|}
\hline && \\
 {\rm antiparallel} & \stackrel{\overline{S}=\overline{S}^\dagger}{\longleftarrow}
 & {\rm antiparallel} \\
 && \\
   {\rm antiparallel} & \stackrel{S}{\longleftarrow} & {\rm parallel} \\
 && \\
 {\rm parallel} & \stackrel{S^\dagger }{\longleftarrow} & {\rm antiparallel} \\
 && \\ \hline
\end{array}
\label{Saction}
\ee

\bigskip
Operators $T$ describe crossing of {\it two} adjacent strands and are much simpler than $S$.
The only delicate point is that there are three different pairs of adjacent strands
in the 4-strand braid and therefore there are three different $T$-matrices, which we denote
$T^+,T^0$ and $T^-$. Also, the two intersecting strands can be either parallel or antiparallel,
what we denote by $T$ and $\bar T$ respectively.
This brings the number of different $T$-insertions in (\ref{Amatr}) to six.

\begin{picture}(300,100)(-150,-25)
\put(0,20){\line(0,1){40}}
\put(0,0){\line(1,2){10}}
\put(10,0){\line(-1,2){10}}
\qbezier(10,20)(10,26)(20,30)\qbezier(20,30)(30,34)(30,40)
\put(30,40){\line(1,2){10}}
\qbezier(60,20)(60,26)(50,30)\qbezier(50,30)(40,34)(40,40)
\put(40,40){\line(-1,2){10}}
\put(60,0){\line(1,2){10}}
\put(70,0){\line(-1,2){10}}
\put(70,20){\line(0,1){40}}
\put(5,10){\circle*{4}}
\put(35,50){\circle*{4}}
\put(65,10){\circle*{4}}
\put(-15,8){\mbox{$T^-$}}
\put(75,8){\mbox{$T^+$}}
\put(45,48){\mbox{$T^0$}}
\put(-5,-12){\mbox{$X,a$}}
\put(55,-12){\mbox{$\bar X,b$}}
\put(25,65){\mbox{$Y,c$}}
\put(150,48){\mbox{$T^0_{Yc} = t_Y\epsilon_c  $}}
\put(-130,-2){\mbox{$T^-_{Xa} = t_X\epsilon_a  $}}
\put(150,-2){\mbox{$T^+_{\bar Xb} = t_{\bar X}\epsilon_b  $}}
\end{picture}

\noindent
One can promote these operators to matrices of the same type as $S$:
\be
T^-_{Xab,X'a'b'} = t_X\epsilon_a \cdot\delta_{XX'}\delta_{aa'}\delta_{bb'}\nn\\
T^0_{Xab,X'a'b'} = t_X\epsilon_a \cdot\delta_{XX'}\delta_{aa'}\delta_{bb'}\nn\\
T^+_{Xab,X'a'b'} = t_X\epsilon_b \cdot\delta_{XX'}\delta_{aa'}\delta_{bb'}
\ee
The eigenvalues are the standard ones in the theory of cut-and-join $\hat W$-operators
\cite{MMN} and the same which appear in the Rosso-Jones formula \cite{RJ,DMMSS}:
\be
\left\{\begin{array}{c}
t_{X\in R\otimes R} \cr
\bar t_{X \in R\otimes \bR}
\end{array}\right\}
= \frac{\epsilon_X}{A^{|R|}q^{2\varkappa_R}} \cdot q^{\varkappa_X}
\\
\varkappa_X = \sum_{(i,j)\in X}(i-j)
\ee
All $\epsilon_X$'s take values $\pm 1$.

We usually omit the indices $\pm$, 0 of $T$ later on, since these $T$'s are all the same as matrices.

\section{Pretzel fingers and propagators
\label{Prefi}}
Our next goal is to classify different types of expressions (\ref{Amatr})
for the propagator (\ref{edge}) -- according to the possible number and parities of
parameters $n_\alpha$, $m_\alpha$ and $l_\alpha$ and to directions of strands (arrows).
A special role is played by the terminal branches
in the tree (\ref{tree}), which we call {\it fingers}.
They are represented by matrix elements
${\cal A}_{1X}^{ab} = {\cal A}_{1,Xab}$,
and all the $T$-matrices ($X=1$) standing to the very left
of all $S$ can be omitted.
According to (\ref{Saction}), the last (most left) $S$ in such matrix element can be either
$\bar S$ or $S$ but not $S^\dagger$.

In this section we begin from the simplest possible types of propagators and fingers,
belonging to the  {\it pretzel} type.
For pretzels all $m_\alpha=l_\alpha=0$ and there is just one parameter $n$
in each of the $k$ of the fingers.
Still, there are propagators and thus fingers of three different kinds, depending on directions of arrows
and the parity of $n$.
In the case of pretzels, the notation with bars is sufficient to
distinguish between all these cases,
still we add explicit indices $par,ea,oa$ (depending on parallel oriented braids, antiparallel
oriented even braids or antiparallel oriented odd braids) to avoid any confusion:
\be
\begin{picture}(350,165)(-300,-55)
\put(-270,0){\line(1,0){30}}
\put(-270,0){\line(0,1){30}}
\put(-240,30){\line(-1,0){30}}
\put(-240,30){\line(0,-1){30}}
\put(-257,13){\mbox{$n$}}
\put(-273,-40){\line(1,5){8}}
\put(-237,-40){\line(-1,5){8}}
\put(-290,-50){\mbox{$X,a$}}
\put(-240,-50){\mbox{$\bar X,b$}}
\put(-272,-35){\vector(0,1){2}}
\put(-238,-36){\vector(0,1){2}}
\put(-290,75){\mbox{$Z,e$}}
\put(-240,75){\mbox{$\bar Z,f$}}
\put(-237,70){\line(-1,-5){8}}
\put(-273,70){\line(1,-5){8}}
\put(-272,66){\vector(0,1){2}}
\put(-238,65){\vector(0,1){2}}
\put(-285,70){\vector(0,-1){110}}
\put(-225,70){\vector(0,-1){110}}
\put(-285,66){\vector(0,-1){2}}
\put(-225,66){\vector(0,-1){2}}
\qbezier(-295,-20)(-255,0)(-215,-20)
\qbezier(-295,-20)(-255,-40)(-215,-20)
\qbezier(-295,50)(-255,70)(-215,50)
\qbezier[50](-295,50)(-255,30)(-215,50)
\put(-295,-20){\line(0,1){70}}
\put(-215,-20){\line(0,1){70}}
\put(-360,70){\mbox{$Z\ \in\ R\otimes \bar R$}}
\put(-360,13){\mbox{$Y\ \in\ R\otimes R$}}
\put(-360,-40){\mbox{$X\ \in\ R\otimes \bar R$}}
\put(-200,43){\mbox{$=\sum_{Y,c,d} S^{RR\bR\bR}_{\bar Zfe,Ycd}\, t_{Yc}^n\,S^{\bar RR R\bar R}_{Ycd,Xab}$}}
\put(-180,7){\mbox{$  \stackrel{(\ref{Stypes})}{=}
\sum_{Y,c,d} S_{Zef,Ycd}\, t_{Yc}^n\, S_{\bar Xba,Ycd}$}}
\put(-170,-40){\mbox{$\boxed{{\cal A}^{\rm par}(n) = S  T^n S^\dagger}$}}
\put(-160,95){\mbox{{\bf parallel braid:}}}
\put(0,95){\mbox{$\downarrow\uparrow \ \ \uparrow\downarrow$}}
\put(0,0){\line(1,0){30}}
\put(0,0){\line(0,1){30}}
\put(30,30){\line(-1,0){30}}
\put(30,30){\line(0,-1){30}}
\put(13,13){\mbox{$n$}}
\put(-3,-40){\line(1,5){8}}
\put(33,-40){\line(-1,5){8}}
\put(-20,-50){\mbox{$X,a$}}
\put(30,-50){\mbox{$\bar X,b$}}
\put(-2,-35){\vector(0,1){2}}
\put(32,-36){\vector(0,1){2}}
\qbezier(5,30)(5,50)(-5,50)
\qbezier(-15,30)(-15,50)(-5,50)
\qbezier(25,30)(25,50)(35,50)
\qbezier(45,30)(45,50)(35,50)
\put(-15,30){\vector(0,-1){70}}
\put(45,30){\vector(0,-1){70}}
\qbezier(-25,-20)(15,0)(55,-20)
\qbezier(-25,-20)(15,-40)(55,-20)
\qbezier(-25,-20)(-35,70)(15,70)
\qbezier(55,-20)(65,70)(15,70)
\put(70,13){\mbox{$=\ (S T^nS^\dagger)_{1X}^{ab} $}}
\end{picture}
\label{parpre}
\ee

\be
\begin{picture}(350,165)(-300,-55)
\put(-270,0){\line(1,0){30}}
\put(-270,0){\line(0,1){30}}
\put(-240,30){\line(-1,0){30}}
\put(-240,30){\line(0,-1){30}}
\put(-269,13){\mbox{{\rm even}\ $n$}}
\put(-273,-40){\line(1,5){8}}
\put(-237,-40){\line(-1,5){8}}
\put(-290,-50){\mbox{$X,a$}}
\put(-240,-50){\mbox{$\bar X,b$}}
\put(-272,-35){\vector(0,1){2}}
\put(-238,-36){\vector(0,-1){2}}
\put(-290,75){\mbox{$Z,e$}}
\put(-240,75){\mbox{$\bar Z,f$}}
\put(-237,70){\line(-1,-5){8}}
\put(-273,70){\line(1,-5){8}}
\put(-272,66){\vector(0,1){2}}
\put(-238,65){\vector(0,-1){2}}
\put(-285,70){\vector(0,-1){110}}
\put(-225,-40){\vector(0,1){110}}
\put(-285,66){\vector(0,-1){2}}
\put(-225,-36){\vector(0,1){2}}
\qbezier(-295,-20)(-255,0)(-215,-20)
\qbezier(-295,-20)(-255,-40)(-215,-20)
\qbezier(-295,50)(-255,70)(-215,50)
\qbezier[50](-295,50)(-255,30)(-215,50)
\put(-295,-20){\line(0,1){70}}
\put(-215,-20){\line(0,1){70}}
\put(-360,70){\mbox{$Z\ \in\ R\otimes \bar R$}}
\put(-360,13){\mbox{$Y\ \in\ R\otimes \bR$}}
\put(-360,-40){\mbox{$X\ \in\ R\otimes \bar R$}}
\put(-200,43){\mbox{$=\sum_{Y,c,d} S^{R\bR R\bR}_{\bar Zfe,Ycd}\, \bar t_{Yc}^n\,S^{\bar RR \bR R}_{Ycd,Xab}$}}
\put(-180,7){\mbox{$  \stackrel{(\ref{Stypes})}{=}
\sum_{Y,c,d} \bar S_{Zef,Ycd}\, \bar t_{Yc}^n \, \bar S_{\bar X_{ab},\bar Ydc}$}}
\put(-170,-40){\mbox{$\boxed{{\cal A}^{\rm ea}(\bar n) =  \bar S  \bar T^n \bar S}$}}
\put(-160,95){\mbox{{\bf even antiparallel braid:}}}
\put(0,95){\mbox{$\downarrow\uparrow \ \ \downarrow\uparrow$}}
\put(0,0){\line(1,0){30}}
\put(0,0){\line(0,1){30}}
\put(30,30){\line(-1,0){30}}
\put(30,30){\line(0,-1){30}}
\put(1,13){\mbox{{\rm even} $n$}}
\put(-3,-40){\line(1,5){8}}
\put(33,-40){\line(-1,5){8}}
\put(-20,-50){\mbox{$X,a$}}
\put(30,-50){\mbox{$\bar X,b$}}
\put(-2,-35){\vector(0,1){2}}
\put(32,-36){\vector(0,-1){2}}
\qbezier(5,30)(5,50)(-5,50)
\qbezier(-15,30)(-15,50)(-5,50)
\qbezier(25,30)(25,50)(35,50)
\qbezier(45,30)(45,50)(35,50)
\put(-15,30){\vector(0,-1){70}}
\put(45,30){\line(0,-1){70}}
\put(45,-36){\vector(0,1){2}}
\qbezier(-25,-20)(15,0)(55,-20)
\qbezier(-25,-20)(15,-40)(55,-20)
\qbezier(-25,-20)(-35,70)(15,70)
\qbezier(55,-20)(65,70)(15,70)
\put(70,13){\mbox{$=\ (\bar S \bar T^n\bar S)_{1X}^{ab} $}}
\end{picture}
\label{{antiparevenpre}}
\ee

\be
\begin{picture}(350,165)(-300,-55)
\put(-270,0){\line(1,0){30}}
\put(-270,0){\line(0,1){30}}
\put(-240,30){\line(-1,0){30}}
\put(-240,30){\line(0,-1){30}}
\put(-268,13){\mbox{{\rm odd} $n$}}
\put(-273,-40){\line(1,5){8}}
\put(-237,-40){\line(-1,5){8}}
\put(-290,-50){\mbox{$X,a$}}
\put(-240,-50){\mbox{$\bar X,b$}}
\put(-272,-35){\vector(0,-1){2}}
\put(-238,-36){\vector(0,1){2}}
\put(-290,75){\mbox{$Z,e$}}
\put(-240,75){\mbox{$\bar Z,f$}}
\put(-237,70){\line(-1,-5){8}}
\put(-273,70){\line(1,-5){8}}
\put(-272,66){\vector(0,1){2}}
\put(-238,65){\vector(0,-1){2}}
\put(-285,70){\vector(0,-1){110}}
\put(-225,-40){\vector(0,1){110}}
\put(-285,66){\vector(0,-1){2}}
\put(-225,-36){\vector(0,1){2}}
\qbezier(-295,-20)(-255,0)(-215,-20)
\qbezier(-295,-20)(-255,-40)(-215,-20)
\qbezier(-295,50)(-255,70)(-215,50)
\qbezier[50](-295,50)(-255,30)(-215,50)
\put(-295,-20){\line(0,1){70}}
\put(-215,-20){\line(0,1){70}}
\put(-360,70){\mbox{$Z\ \in\ R\otimes \bar R$}}
\put(-360,13){\mbox{$Y\ \in\ R\otimes \bR$}}
\put(-360,-40){\mbox{$X\ \in\ R\otimes R$}}
\put(-200,43){\mbox{$=\sum_{Y,c,d} S^{R\bR R\bR}_{\bar Zfe,Ycd}\, \bar t_{Yc}^n\,S^{\bR\bar RRR}_{Ycd,Xab}$}}
\put(-180,7){\mbox{$  \stackrel{(\ref{Stypes})}{=}
\bar S_{Zef,Ycd}\, \bar t_{Yc}^n\, S_{Ycd,Xab}$}}
\put(-170,-40){\mbox{$\boxed{{\cal A}^{\rm oa}(n) =  \bar S  \bar T^n S}$}}
\put(-160,95){\mbox{{\bf odd antiparallel braid:}}}
\put(0,95){\mbox{$\downarrow\downarrow \ \ \uparrow\uparrow$}}
\put(0,0){\line(1,0){30}}
\put(0,0){\line(0,1){30}}
\put(30,30){\line(-1,0){30}}
\put(30,30){\line(0,-1){30}}
\put(2,13){\mbox{{\rm odd} $n$}}
\put(-3,-40){\line(1,5){8}}
\put(33,-40){\line(-1,5){8}}
\put(-20,-50){\mbox{$X,a$}}
\put(30,-50){\mbox{$\bar X,b$}}
\put(-2,-35){\vector(0,-1){2}}
\put(32,-36){\vector(0,1){2}}
\qbezier(5,30)(5,50)(-5,50)
\qbezier(-15,30)(-15,50)(-5,50)
\qbezier(25,30)(25,50)(35,50)
\qbezier(45,30)(45,50)(35,50)
\put(-15,30){\vector(0,-1){70}}
\put(45,30){\line(0,-1){70}}
\put(45,-36){\vector(0,1){2}}
\qbezier(-25,-20)(15,0)(55,-20)
\qbezier(-25,-20)(15,-40)(55,-20)
\qbezier(-25,-20)(-35,70)(15,70)
\qbezier(55,-20)(65,70)(15,70)
\put(70,13){\mbox{$=\ (\bar S \bar T_0^n S)_{1X}^{ab} $}}
\end{picture}
\label{{antiparoddpre}}
\ee

\bigskip

\noindent
The choice of $S$-matrices in above examples
is easy to understand from (\ref{Saction}).  Recall,  $S$
converts parallel states (from $R\otimes R$) into antiparallel (from $R\otimes \bar R$),
while $S^\dagger$ does the reverse conversion, and $\bar S$ relates
antiparallel to antiparallel states.

In the final expressions (boxed) we suppressed most indices, thus implying that we deal with matrices
in extended space, where basis is labeled by the multi-index ${Xab}$.
We explain the details of this formalism in sec.\ref{Racah},
while {\it in sec.\ref{exapols} we use the symbolical notation: just $X$ for the multi-index and
$\overline{\sum_X}$ for appropriate contractions, including non-trivial sign-factors $\sigma$} (which is symbolized by the bar over the sum).

\bigskip
These are the propogators which are relevant for the knots belonging to pretzel family and the knots in the present paper.

\section{Other building blocks}

\subsection{Non-pretzel fingers}
Generalization to arbitrary non-pretzel fingers and propagators (\ref{edge})
with arbitrary parameters $m_\alpha,l_\alpha,n_\alpha$ is exhaustively described
by (\ref{Amatr}), one just needs to put appropriate $S$ and $T$ matrices for
the given choice of arrows, like we did for the pretzel fingers in sec.\ref{Prefi}.
Since this is straightforward we do not provide this additional list.

What is more important, some diagrams, which do not {\it a priori} look like (\ref{edge}),
actually belong to this family, and this is the reason for the double fat graph
description to present  so many different knots and links.
In this section, we mention just two examples of this kind,
exploited in our further considerations.

\subsection{Horizontal loop}
The first example is simple:
\be
\begin{picture}(300,120)(20,-10)
\put(0,0){\line(0,1){40}}  \put(0,45){\line(0,1){55}}
\qbezier(-15,0)(-15,65)(-2,63)
\qbezier(2,62.5)(15,58)(12,50)
\qbezier(12,50)(-2,36)(-8,50)
\qbezier(-12,58)(-15,75)(-15,100)
\put(-17,-12){\mbox{$X,a$}}
\put(-17,105){\mbox{$X,b$}}
\put(130,50){\circle*{4}}
\put(109,70){\circle*{4}}
\put(151,70){\circle*{4}}
\put(60,47){\mbox{$=$}}
\qbezier(100,10)(100,85)(130,85)\qbezier(160,10)(160,85)(130,85)
\qbezier(110,10)(110,30)(130,50) \qbezier(150,10)(150,30)(130,50)
\qbezier(130,50)(75,95)(130,95)\qbezier(130,50)(185,95)(130,95)
\put(95,-2){\mbox{$X,a$}}
\put(145,-2){\mbox{$\bar X,b$}}
\put(250,47){\mbox{${\cal A}^{\rm hl} = \Big(ST_-T_+S^\dagger T_0S\Big)_{1X}^{ab}
  $}}
\end{picture}
\label{horloop}
\ee
\noindent
It is nearly obvious that this
is just the non-pretzel finger with parameters $n_1= m_2=l_2=\pm 1$
(signs and the types of $S$ and $T$ matrices depend on the types of intersection
and directions of arrows).
No $S$-matrix identities are needed to get an expression for it.

One could easily insert a horizontal braid of arbitrary length and consider a sequence
of such horizontal loops:
\be
\begin{picture}(300,150)(-10,-30)
\put(0,-13){\line(0,1){58}}
\put(-35,20){\line(1,0){20}}
\put(-35,36){\line(1,0){20}}
\put(-35,20){\line(0,1){16}}
\put(-15,20){\line(0,1){16}}
\put(-32,25){\mbox{$m_1$}}
\qbezier(-15,23)(10,23)(10,28) \qbezier(-15,33)(10,33)(10,28)
\put(-30,48){\mbox{$\ldots$}}
\put(0,55){\line(0,1){58}}
\put(-35,80){\line(1,0){20}}
\put(-35,64){\line(1,0){20}}
\put(-35,80){\line(0,-1){16}}
\put(-15,80){\line(0,-1){16}}
\put(-32,70){\mbox{$m_k$}}
\qbezier(-15,77)(10,77)(10,72) \qbezier(-15,67)(10,67)(10,72)
\qbezier(-50,11)(-50,23)(-35,23)
\qbezier(-50,45)(-50,33)(-35,33)
\qbezier(-50,89)(-50,77)(-35,77)
\qbezier(-50,55)(-50,67)(-35,67)
\put(-15,-25){\mbox{$X,a_1$}}
\put(-17,118){\mbox{$X,a_{k+1}$}}
\qbezier(-50,11)(-50,-1)(-30,-1)
\qbezier(-10,-13)(-10,-1)(-30,-1)
\qbezier(-50,89)(-50,101)(-30,101)
\qbezier(-10,113)(-10,101)(-30,101)
\put(0,32.5){\circle*{4}}
\put(0,23.5){\circle*{4}}
\put(0,76.5){\circle*{4}}
\put(0,67.5){\circle*{4}}
\put(80,47){\mbox{${\cal A}^{\rm hl}(m_1,\ldots,m_k)_X^{a_1a_{k+1}}
= \sum\limits_{a_2,\ldots,a_k}\ \prod\limits_{i=1}^k \Big(ST_-T_+S^\dagger T_0^{m_i}S\Big)_{1X}^{a_ia_{i+1}}
  $}}
\end{picture}
\label{hormultloop}
\ee
\subsection{Horizontal braid}
The isolated horizontal braid

\begin{figure}[h!]
\centering\leavevmode
\includegraphics[width=15 cm]{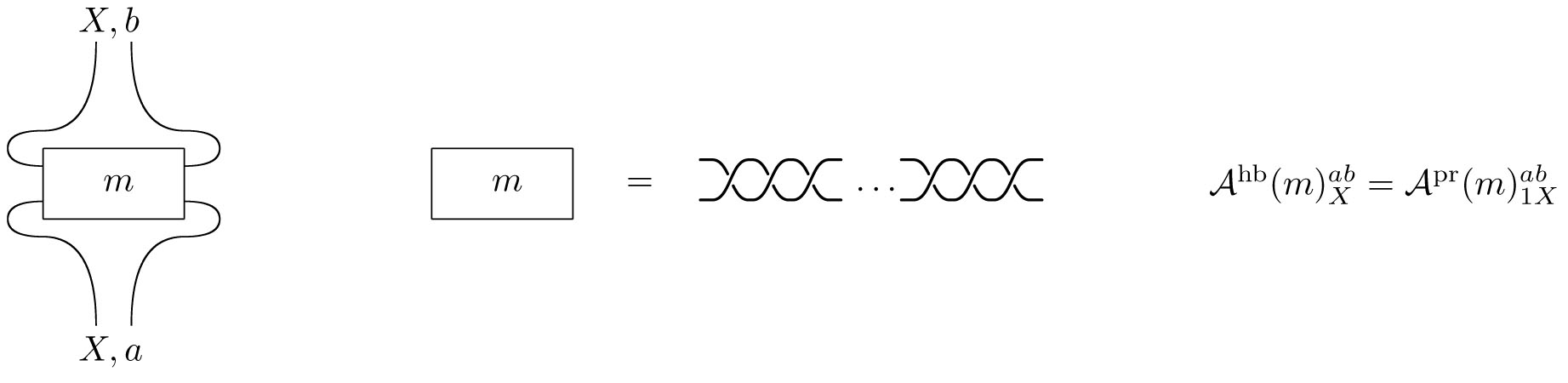}
\label{elehor}
\end{figure}

\noindent
is nothing but the pretzel finger.
The type of the finger depends on orientation of lines and the parity of $m$.

\bigskip

More interesting is the horizontal braid located in between the two
lines in the double fat propagator as shown:
\be
\begin{picture}(300,120)(-50,-60)
\put(-50,-10){\line(1,0){40}}
\put(-50,-10){\line(0,1){20}}
\put(-50,10){\line(1,0){40}}
\put(-10,-10){\line(0,1){20}}
\qbezier(-70,-40)(-70,-5)(-50,-5)
\qbezier(-70,40)(-70,5)(-50,5)
\qbezier(10,-40)(10,-5)(-10,-5)
\qbezier(10,40)(10,5)(-10,5)
\put(-80,-40){\line(0,1){80}}
\put(20,-40){\line(0,1){80}}
\put(-85,-50){\mbox{$X,a$}}
\put(5,-50){\mbox{$\bar X,b$}}
\put(-85,43){\mbox{$Y,c$}}
\put(5,43){\mbox{$\bar Y,d$}}
\put(-33,-2){\mbox{$m$}}
\put(150,-10){\line(1,0){40}}
\put(150,-10){\line(0,1){20}}
\put(150,10){\line(1,0){40}}
\put(190,-10){\line(0,1){20}}
\qbezier(130,-50)(130,-5)(150,-5)
\qbezier(130,50)(130,5)(150,5)
\qbezier(210,-50)(210,-5)(190,-5)
\qbezier(210,50)(210,5)(190,5)
%
\put(220,-50){\line(0,1){100}}
\put(115,-60){\mbox{$X,a$}}
\put(205,-60){\mbox{$\bar X,b$}}
\put(115,53){\mbox{$Y,c$}}
\put(205,53){\mbox{$\bar Y,d$}}
\put(167,-2){\mbox{$m$}}
\put(130,-45.5){\circle*{4}}
\put(130,45.5){\circle*{4}}
\put(207,-21){\circle*{4}}
\put(207,21){\circle*{4}}
\put(65,-2){\mbox{$=$}}
\qbezier(120,-50)(120,-45)(165,-40)\qbezier(165,-40)(212,-40)(212,0)
\qbezier(120,50)(120,45)(165,40)\qbezier(165,40)(212,40)(212,0)
\qbezier[70](100,15)(150,15)(245,15)
\qbezier[70](100,-15)(150,-15)(245,-15)
\put(240,-25){\mbox{$Z,e,f$}}
\put(240,18){\mbox{$Z,g,h$}}
\end{picture}
\label{horbraid1}
\ee

This is a contraction of three blocks, where the middle one is exactly the horizontal braid above.
Schematically, it is
\be
\sum_Z  {1\over\sqrt{d_Z}}\Big(T^{-1}S^\dagger T^{-1}S\Big)_{YZ} {\cal A}^{\rm hb}(m)_{1Z}
\Big(TS^\dagger TS\Big)_{XZ}
\ \stackrel{(\ref{SbTbS})}{=}\
\sum_Z  {1\over\sqrt{d_Z}}S_{YZ}S_{XZ}  {\cal A}_{1Z}^{\rm pr}(m)\,
\ee

In fact there are seven versions of this relation,
for different arrow directions and different parities of $m$. Four of them describe even $m$:

\vspace{2cm}
\setlength{\unitlength}{0.5pt}
\begin{picture}(50,50)(-100,-60)
\put(-50,-10){\line(1,0){40}}
\put(-50,-10){\line(0,1){20}}
\put(-50,10){\line(1,0){40}}
\put(-10,-10){\line(0,1){20}}
\qbezier(-70,-40)(-70,-5)(-50,-5)
\qbezier(-70,40)(-70,5)(-50,5)
\qbezier(10,-40)(10,-5)(-10,-5)
\qbezier(10,40)(10,5)(-10,5)
\put(-80,-40){\vector(0,1){80}}
\put(20,40){\vector(0,-1){80}}
\put(-70,38){\vector(0,1){2}}
\put(-52,-5){\vector(1,0){2}}
\put(-8,5){\vector(-1,0){2}}
\put(10,-38){\vector(0,-1){2}}
\put(-85,-60){\mbox{$X,a$}}
\put(5,-60){\mbox{$\bar X,b$}}
\put(-85,43){\mbox{$Y,c$}}
\put(5,43){\mbox{$\bar Y,d$}}
\put(-43,-5){\mbox{\small{$2m$}}}
\label{hb1}
\end{picture}
\begin{picture}(50,50)(-300,-70)
\put(-50,-10){\line(1,0){40}}
\put(-50,-10){\line(0,1){20}}
\put(-50,10){\line(1,0){40}}
\put(-10,-10){\line(0,1){20}}
\qbezier(-70,-40)(-70,-5)(-50,-5)
\qbezier(-70,40)(-70,5)(-50,5)
\qbezier(10,-40)(10,-5)(-10,-5)
\qbezier(10,40)(10,5)(-10,5)
\put(-80,-40){\vector(0,1){80}}
\put(20,40){\vector(0,-1){80}}
\put(-70,38){\vector(0,1){2}}
\put(-8,-5){\vector(-1,0){2}}
\put(-8,5){\vector(-1,0){2}}
\put(-70,-38){\vector(0,-1){2}}
\put(-85,-60){\mbox{$X,a$}}
\put(5,-60){\mbox{$\bar X,b$}}
\put(-85,43){\mbox{$Y,c$}}
\put(5,43){\mbox{$\bar Y,d$}}
\put(-43,-5){\mbox{\small{$2m$}}}
\label{hb2}
\end{picture}
\begin{picture}(50,50)(-500,-80)
\put(-50,-10){\line(1,0){40}}
\put(-50,-10){\line(0,1){20}}
\put(-50,10){\line(1,0){40}}
\put(-10,-10){\line(0,1){20}}
\qbezier(-70,-40)(-70,-5)(-50,-5)
\qbezier(-70,40)(-70,5)(-50,5)
\qbezier(10,-40)(10,-5)(-10,-5)
\qbezier(10,40)(10,5)(-10,5)
\put(-80,-40){\vector(0,1){80}}
\put(20,40){\vector(0,-1){80}}
\put(10,38){\vector(0,1){2}}
\put(-52,-5){\vector(1,0){2}}
\put(-52,5){\vector(1,0){2}}
\put(10,-38){\vector(0,-1){2}}
\put(-85,-60){\mbox{$X,a$}}
\put(5,-60){\mbox{$\bar X,b$}}
\put(-85,43){\mbox{$Y,c$}}
\put(5,43){\mbox{$\bar Y,d$}}
\put(-43,-5){\mbox{\small{$2m$}}}
\label{hb3}
\end{picture}
\begin{picture}(50,50)(-700,-90)
\put(-50,-10){\line(1,0){40}}
\put(-50,-10){\line(0,1){20}}
\put(-50,10){\line(1,0){40}}
\put(-10,-10){\line(0,1){20}}
\qbezier(-70,-40)(-70,-5)(-50,-5)
\qbezier(-70,40)(-70,5)(-50,5)
\qbezier(10,-40)(10,-5)(-10,-5)
\qbezier(10,40)(10,5)(-10,5)
\put(-80,-40){\vector(0,1){80}}
\put(20,40){\vector(0,-1){80}}
\put(-70,-38){\vector(0,-1){2}}
\put(-52,5){\vector(1,0){2}}
\put(-8,-5){\vector(-1,0){2}}
\put(10,38){\vector(0,1){2}}
\put(-85,-60){\mbox{$X,a$}}
\put(5,-60){\mbox{$\bar X,b$}}
\put(-85,43){\mbox{$Y,c$}}
\put(5,43){\mbox{$\bar Y,d$}}
\put(-43,-5){\mbox{\small{$2m$}}}
\label{hb4}
\end{picture}

\vspace{2cm}

These graphs are described by
\be
\sum_{Z\in R\otimes\bar R}   {1\over\sqrt{d_Z}}\bar S_{YZ}\bar S_{XZ}  {\cal A}_{1Z}(2m)
\ee
with ${\cal A}_{1Z}(2m)$ being ${\cal A}^{ea}$ in the first and the fourth pictures and ${\cal A}^{par}$ in the two remaining pictures.


Similarly, odd $m$ are described by the following four pictures:

\vspace{2cm}

\begin{picture}(50,50)(-50,-60)
\put(-50,-10){\line(1,0){70}}
\put(-50,-10){\line(0,1){20}}
\put(-50,10){\line(1,0){70}}
\put(20,-10){\line(0,1){20}}
\qbezier(-70,-40)(-70,-5)(-50,-5)
\qbezier(-70,40)(-70,5)(-50,5)
\qbezier(40,-40)(40,-5)(20,-5)
\qbezier(40,40)(40,5)(20,5)
\put(-80,-40){\vector(0,1){80}}
\put(50,40){\line(0,-1){80}}
\put(-70,38){\vector(0,1){2}}
\put(-52,-5){\vector(1,0){2}}
\put(22,-5){\vector(-1,0){2}}
\put(40,38){\vector(0,1){2}}
\put(-85,-60){\mbox{$X,a$}}
\put(35,-60){\mbox{$\bar X,b$}}
\put(-85,43){\mbox{$Y,c$}}
\put(35,43){\mbox{$\bar Y,d$}}
\put(-43,-5){\mbox{{\small{$2m+1$}}}}

\label{hb5}
\end{picture}

\begin{picture}(50,50)(-250,-110)
\put(-50,-10){\line(1,0){70}}
\put(-50,-10){\line(0,1){20}}
\put(-50,10){\line(1,0){70}}
\put(20,-10){\line(0,1){20}}
\qbezier(-70,-40)(-70,-5)(-50,-5)
\qbezier(-70,40)(-70,5)(-50,5)
\qbezier(40,-40)(40,-5)(20,-5)
\qbezier(40,40)(40,5)(20,5)
\put(-80,-40){\vector(0,1){80}}
\put(50,40){\vector(0,-1){80}}
\put(-70,38){\vector(0,1){2}}
\put(22,-5){\vector(-1,0){2}}
\put(22,5){\vector(-1,0){2}}
\put(-70,-38){\vector(0,-1){2}}
\put(-85,-60){\mbox{$X,a$}}
\put(35,-60){\mbox{$\bar X,b$}}
\put(-85,43){\mbox{$Y,c$}}
\put(35,43){\mbox{$\bar Y,d$}}
\put(-43,-5){\mbox{\small{$2m+1$}}}

\label{hb6}
\end{picture}

\begin{picture}(50,50)(-450,-160)
\put(-50,-10){\line(1,0){70}}
\put(-50,-10){\line(0,1){20}}
\put(-50,10){\line(1,0){70}}
\put(20,-10){\line(0,1){20}}
\qbezier(-70,-40)(-70,-5)(-50,-5)
\qbezier(-70,40)(-70,5)(-50,5)
\qbezier(40,-40)(40,-5)(20,-5)
\qbezier(40,40)(40,5)(20,5)
\put(-80,-40){\vector(0,1){80}}
\put(50,40){\vector(0,-1){80}}
\put(40,38){\vector(0,1){2}}
\put(-52,-5){\vector(1,0){2}}
\put(-52,5){\vector(1,0){2}}
\put(40,-38){\vector(0,-1){2}}
\put(-85,-60){\mbox{$X,a$}}
\put(35,-60){\mbox{$\bar X,b$}}
\put(-85,43){\mbox{$Y,c$}}
\put(35,43){\mbox{$\bar Y,d$}}
\put(-43,-5){\mbox{\small{$2m+1$}}}
\label{hb7}
\end{picture}

\begin{picture}(50,50)(-650,-210)
\put(-50,-10){\line(1,0){70}}
\put(-50,-10){\line(0,1){20}}
\put(-50,10){\line(1,0){70}}
\put(20,-10){\line(0,1){20}}
\qbezier(-70,-40)(-70,-5)(-50,-5)
\qbezier(-70,40)(-70,5)(-50,5)
\qbezier(40,-40)(40,-5)(20,-5)
\qbezier(40,40)(40,5)(20,5)
\put(-80,-40){\vector(0,1){80}}
\put(50,-40){\vector(0,1){80}}
\put(-70,-38){\vector(0,-1){2}}
\put(-52,5){\vector(1,0){2}}
\put(22,5){\vector(-1,0){2}}
\put(40,-38){\vector(0,-1){2}}
\put(-85,-60){\mbox{$X,a$}}
\put(35,-60){\mbox{$\bar X,b$}}
\put(-85,43){\mbox{$Y,c$}}
\put(35,43){\mbox{$\bar Y,d$}}
\put(-43,-5){\mbox{\small{$2m+1$}}}
\label{hb8}
\end{picture}
\setlength{\unitlength}{1pt}

\vspace{-2cm}

The first graph here does not emerge as a subgraph of the tree double fat diagrams, and the three remaining are described by
\be
\sum_{Z\in R\otimes\bar R}   {1\over\sqrt{d_Z}}\bar S_{YZ}\bar S_{XZ}  {\cal A}_{1Z}(2m+1)
\ee
with ${\cal A}_{1Z}(2m+1)$ being ${\cal A}^{par}$ in the second and the third pictures and ${\cal A}^{oa}$ in the fourth one.

One can perform the simple self-consistency checks
by attaching (\ref{capXX}) and (\ref{capX0}) to the top and bottom of these configurations.

\section{Some examples of double fat diagrams and knot polynomials
\label{exapols}}
Contractions of the pretzel fingers produce a wide variety of the pretzel link and knots.
 We also look at knots obtained by inclusion of non-pretzel fingers with non-vanishing
$m_\alpha$ and/or $l_\alpha$, non-trivial propagators with horizontal braids and the appropriate contractions of the fingers.

It is {\it not} our goal in the present paper to provide a systematic description
of the variety of double fat diagrams. Instead we demonstrate how knot  polynomials
are built, when the diagram is given in this form.
We present the polynomials for many knots which belongs to the knot families presented
in the following subsections.

\subsection{Pretzel knots and links}
The simplest subfamily inside the tree double fat set is that of the pretzel links/knots,
for which eq.(\ref{basic}) was analyzed in great detail in \cite{GMMMS}.
If considered as made from the two-strand braids, the pretzel link/knot looks like a $(g=k-1)$-loop
diagram naturally lying on the surface of genus $g$:

\begin{figure}[h!]
\centering\leavevmode
\includegraphics[width=18 cm]{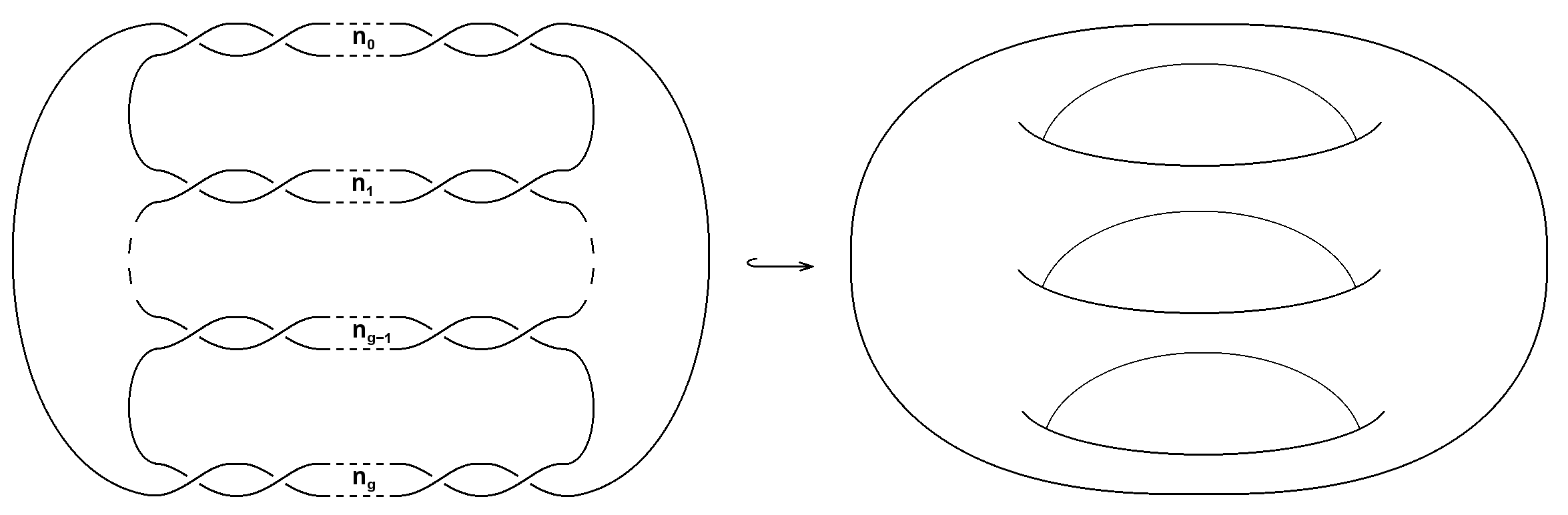}
\label{braidg}
\end{figure}

However, if composed from the four-strand braids, pretzel links/knots
become trees of type (\ref{tree}), with just a
single vertex, i.e. are of a peculiar
{\it starfish/fingerfish} type (it is not a {\it hedgehog},
because all the fingers are cyclically ordered):
\be
\begin{picture}(300,90)(-10,-40)
\put(0,0){\circle{20}} \put(-3,-3){\mbox{$X$}}
\put(-10,0){\line(-1,0){30}}
\put(-7,7){\line(-1,1){20}}
\put(0,10){\line(0,1){30}}
\put(7,7){\line(1,1){20}}
\put(10,0){\line(1,0){30}}
\put (25,-15){\mbox{$\ldots$}}
\put(-7,-7){\line(-1,-1){20}}
\put(-90,3){\mbox{${\cal A}_{1X}^{a_1a_2}(n_1)$}}
\put(-75,27){\mbox{${\cal A}_{1X}^{a_2a_3}(n_2)$}}
\put(-75,-33){\mbox{${\cal A}_{1X}^{a_ka_1}(n_k)$}}
\put(80,-2){\mbox{$H_R^{red} = d_R^{k-1}\cdot
{\sum\limits_{X,\{a\}}}\!\!\!^\sigma \
d_X^{1-k/2} \prod\limits_{i=1}^k {\cal A}_{1X}^{a_ia_{i+1}}(n_i)
\equiv d_R^{k-1} \overline{\sum_X} d_X^{1-k/2} \prod\limits_{i=1}^k {\cal A}_{1X} (n_i) $}}
\end{picture}
\label{pretzel}
\ee
It is clear that parallel and even antiparallel
fingers can be combined in arbitrary \GPK \ combinations connected by the states
$X\in R\otimes\bar R$ while if one finger
is odd antiparallel, all connections are via $X\in R\otimes R$,
thus all the other fingers should also be odd antiparallel.

Pretzel family contains quite a lot of interesting knots and links.
Mutations permute different parameters $n_i$, but colored HOMFLY in
all (anti)symmetric and rectangular representations,
when there are no indices $a_i$
are invariant under all permutations of $n_i$.
In non-rectangular representations dependence of propagators and fingers
on the indices $a_i$  forces the invariant to be unchanged only for
the cyclic permutation which we detail in sec.\ref{signs}.

However, these details do not affect the 3-finger pretzel knots,
of which the simplest subfamily
is that of the twist knots $Tw_k$,
made from three antiparallel fingers, two of them being of unit length:
$Tw_k = Pr_{\bar 1,\bar 1,\overline{2k-1}}$
(clearly, there are no non-trivial mutations in this case).
The simplest among the twist knots is the trefoil, which is also a
torus $2$-strand knot.
\subsubsection{Trefoil}
Trefoil is the pretzel knot $(\bar 1,\bar 1,\bar 1)$, thus it has a three-finger
representation
\be
H_R^{3_1} =
d_R^{2}\sum_{X\in R\otimes R} d_X^{-1/2} (\bar S  \bar T S )_{1X}^3  \stackrel{(\ref{SbTbS})}{=} \
d_R^{2}\sum_X d_X^{-1/2} (\bar T^{-1}S  T^{-1})_{1X}^3 = d_R^{2}\sum_X \frac{S_{1X}}{d_X^{1/2}}\cdot
S_{1X}^2T_X^{-3} \
= \nn \\
 \stackrel{(\ref{firstline})}{=} \ d_R\!\! \sum_{X\in R\otimes R} S_{1X}^2T_X^{-3}
= d_R\cdot \Tr S^\dagger T^{-3} S
\label{terftrans}
\ee
where at the r.h.s. we have  the familiar  formula for the two-strand knot.
\subsubsection{Twist knots: thin, triple-boundary (pretzel)
$(\overline{1}|\overline{1}|\overline{2k-1})$}
The twist knots are the closures of braids where the number of strands grows linearly
with the number of intersections, and thus braid representation is inconvenient
for looking for generic formulas.

Instead, they
possess a two-finger (i.e. two-bridge) realization so that
\be
H^{Tw(k)} = \Big(\bar S \bar T^{2k} \bar S^\dagger  \bar T^2 \bar S\Big)_{00}
\label{twisttwof}
\ee
and the evolution in $k$ can be straightforwardly studied in this formula
in various representations \cite{evo,MMM21}.

Remarkably, the same family possesses  a three-finger (pretzel) realization,
with all three fingers odd antiparallel, which is
more interesting for the purposes of the present text:
\be
H^{(\overline{1,1,2k-1})}
=d_R^{2} \sum_{X\in R\otimes R} d_X^{-1/2} (\bar S  \bar TS  )_{1X}(\bar S  \bar TS  )_{1X}
(\bar S  \bar T^{2k-1}S   )_{1X} \ \stackrel{(\ref{SbTbS})}{=} d_R^{2}\
\sum_{X\in R\otimes R} \frac{S_{1X}}{d_X^{1/2}}\cdot
S_{1X}T_X^{-2} (\bar S  \bar T^{2k-1}S   )_{1X}
=
\ee
$$
 \stackrel{(\ref{firstline})}{=} \ d_R\!\! \sum_{X\in R\otimes R} S_{1X}T_X^{-2}
 (S^\dagger\bar T^{2k-1}\bar S)_{X0}
 =d_R\cdot(ST^{-2}S^\dagger \bar T^{2k-1}\bar S)_{00}
\ \boxed{=} \ d_R\cdot(\bar T\bar S \bar T^2  \bar S  \bar T^{2k}\bar S)_{00}
\ \stackrel{\bar T_{00}=1}{=} \
d_R\cdot( \bar S \bar T^2  \bar S \bar T^{2k}\bar S)_{00}
$$
This  is exactly the same as (\ref{twisttwof}).
The boxed equality in the second line is the double application of (\ref{SbTbS}) in the form
\vspace{-0.6cm}
\be
T^{-1}S^\dagger = S^\dagger  \bar T\bar S \bar T
\ \ \ \Longrightarrow \ \ \
S T^{-2} S^\dagger = \overbrace{S S^\dagger}^{I}
\bar T\bar S \bar T^2\bar S \bar T
= \bar T\bar S \bar T^2\bar S \bar T
\ee

\subsubsection{Odd antiparallel pretzel knots}

More generally, if there is any odd number of odd antiparallel fingers,
\be
H_{R}^{Pr(\bar n_1, \ldots \bar n_{2k+1})} =
d_R^{2k} \overline{\sum_{X\in R\otimes R}} {d_X^{\frac{1}{2}-k}}\,
{\cal A}^{\rm oa}_{1X}(\bar n_1)
\ldots   {\cal A}^{\rm oa}_{1X}(\bar n_{2k+1})
\ \ \ \ \ \ \ \ \ {\rm with} \ \ \ \ \ \ \ \
{\cal A}^{\rm oa}_{1X}({\bar n}) = \Big(\bar S^\dagger \bar  T^n   S\Big)_{1X}
\ee
If the number of fingers is even, we get links rather than knots.

\subsubsection{Pure parallel pretzel knots}
If the pretzel knot is made from the parallel fingers only, their number should be even
and exactly one length should be even, otherwise we get a link rather than a knot.
The answer for the HOMFLY polynomial is
\be
H_{R}^{Pr(n_1, \ldots n_{2k})} =
d_R^{2k-1} \overline{\sum_{X\in R\otimes \bar R}} {d}_X^{\,1-k}\,
{\cal A}^{\rm par}_{1X}(n_1)
\ldots   {\cal A}^{\rm par}_{1X}(  n_{2k})
\ \ \ \ \ \ \ \ \ {\rm with} \ \ \ \ \ \ \ \
{\cal A}^{\rm par}_{1X}({n}) = \Big(S \bar {T}^n S^\dagger \Big)_{1X}
\ee
\subsubsection{Mixed parallel-antiparallel pretzel knots}
The remaining family of the pretzel links/knots contains even number of parallel fingers
and odd number of antiparallel, which have even lengths.
The corresponding HOMFLY polynomial is
\be
H_{R}^{Pr(n_1, \ldots, n_{2k},\bar{m}_1,\ldots,\bar{m}_{2l+1} )} =
d_R^{2k+2l} \overline{\sum_{X\in R\otimes \bar R}} {d}_X^{\frac{1}{2}-k-l}\,
{\cal A}^{\rm par}_{1X}(n_1)
\ldots   {\cal A}^{\rm par}_{1X}(n_{2k})\cdot
{\cal A}^{\rm ea}_{1X}(\bar{m}_1)
\ldots   {\cal A}^{\rm ea}_{1X}(\bar m_{2l+1})
\ee
with
\be
{\cal A}^{\rm par}_{1X}({n}) = \Big(S \bar {T}^n S^\dagger \Big)_{1X}
\ \ \ \ \ \ {\rm and} \ \ \ \ \
{\cal A}^{\rm ea}_{1X}({\bar m}) = \Big(\bar S^\dagger \bar  T^n \bar S\Big)_{1X}
\ee
\subsection{Non-pretzel fingers}
Ten intersection  chiral knot $10_{71}$ has a neat presentation as a three-finger
with each finger being non-pretzel type \cite{Kaul}. Using the states
for such non-pretzel fingers, the HOMFLY polynomial for this knot is:
\be
\label{ten71}
H_{R}^{10_{71}} =
\overline{\sum_X} \frac{d_{R}^2}{\sqrt{d_X}}
\Big(\bar S\bar T^2\bar S\bar T^{-2} \bar S\Big)_{1X}
\Big(ST^2S^\dagger \bar T^{-1}\bar S\Big)_{1X}
\Big(ST^{-3}S^\dagger \bar T^{-1}\bar S\Big)_{1X}
\ee
With the help of (\ref{ramaid})
the last matrix element can be changed for
\be
\Big(ST^{-3}S^\dagger \bar T^{-1}\bar S\Big)_{1X} \ \longrightarrow \ \
\Big( ST^{-2}S^\dagger\bar T\Big)_{1X}
\ee
Note that beyond $SU_q(2)$ this is {\it not} a matrix identity but both LHS or RHS of
the above equation inserted in eq.(\ref{ten71})
give the same polynomial invariant.
\subsection{Double fat graph presentation of knots from three-strand braids }
The next example of non-pretzel knots is a large 4-parametric family:
consider a closure of the  four boxes in the $3$-strand braid as shown:
\be
\begin{picture}(300,100)(-80,-50)
\put(-50,-2){\mbox{${\Large {\rm qTr}}$}}
\qbezier(-20,-25)(-30,0)(-20,25)
\qbezier(280,-25)(290,0)(280,25)
\put(0,0){\vector(1,0){40}}
\put(0,20){\vector(1,0){40}}
\put(40,-5){\line(1,0){30}}
\put(40,25){\line(1,0){30}}
\put(40,-5){\line(0,1){30}}
\put(70,-5){\line(0,1){30}}
\put(70,0){\vector(1,0){20}}
\put(70,20){\vector(1,0){70}}
\put(0,-20){\vector(1,0){90}}
\put(90,5){\line(1,0){30}}
\put(90,-25){\line(1,0){30}}
\put(90,-25){\line(0,1){30}}
\put(120,-25){\line(0,1){30}}
\put(120,0){\vector(1,0){20}}
\put(120,-20){\vector(1,0){70}}
\put(140,-5){\line(1,0){30}}
\put(140,25){\line(1,0){30}}
\put(140,-5){\line(0,1){30}}
\put(170,-5){\line(0,1){30}}
\put(170,20){\vector(1,0){90}}
\put(170,0){\vector(1,0){20}}
\put(190,5){\line(1,0){30}}
\put(190,-25){\line(1,0){30}}
\put(190,-25){\line(0,1){30}}
\put(220,-25){\line(0,1){30}}
\put(220,0){\vector(1,0){40}}
\put(220,-20){\vector(1,0){40}}
\put(52,8){\mbox{$p$}}
\put(102,-12){\mbox{$q$}}
\put(152,8){\mbox{$r$}}
\put(202,-12){\mbox{$s$}}
\end{picture}
\label{4s3braid}
\ee
In the fundamental representation the
unreduced HOMFLY polynomial is
\be
H_{[1]}^{(pqrs)} = \Big({q\over A}\Big)^{p+q+r+s} d_{[3]} + \Big(-{1\over qA}\Big)^{p+q+r+s}d_{[111]} +
d_{[21]}\cdot \Tr_{2\times 2} \Big(S^\dagger T^sST^rS^\dagger T^qST^p\Big)
\label{Hpqrsbraid}
\ee
In the following subsections, we redraw these four-parametric $3$-strand braid knots as double fat graphs.
Particularly, these two equivalent diagrams enables evaluation of knot invariants from both evolution method and
from the double fat graph method.

\subsubsection{$s=1$: the starfish case}
The choice of  one of the four parameters, say $s=1$ gives
knots which can be redrawn as  a three-finger starfish realization:
\be
\begin{picture}(300,200)(-70,-55)
\put(-50,-5){\line(1,0){40}}
\put(-50,25){\line(1,0){40}}
\put(-50,-5){\line(0,1){30}}
\put(-10,-5){\line(0,1){30}}
\put(30,65){\line(1,0){32}}
\put(30,115){\line(1,0){32}}
\put(30,65){\line(0,1){50}}
\put(62,65){\line(0,1){50}}
\put(140,-5){\line(1,0){40}}
\put(140,25){\line(1,0){40}}
\put(140,-5){\line(0,1){30}}
\put(180,-5){\line(0,1){30}}
\put(-38,8){\mbox{$p$}}
\put(153,8){\mbox{$q$}}
\put(43,91){\mbox{$r$}}
\qbezier(40,115)(40,130)(25,130)
\qbezier(25,130)(0,130)(0,60)
\qbezier(52,115)(52,130)(67,130)
\qbezier(67,130)(90,130)(90,60)
\qbezier(-50,5)(-68,5)(-65,-10)
\qbezier(-65,-10)(-65,-30)(110,-30)
\qbezier(-50,15)(-65,15)(-65,30)
\qbezier(-65,30)(-65,50)(-10,50)
\qbezier(-10,50)(0,50)(0,60)
\qbezier(-10,15)(40,15)(40,65)
\put(-10,5){\vector(1,0){150}}
\qbezier(90,60)(90,15)(140,15)
\qbezier(52,65)(52,20)(145,-15)
\qbezier(145,-15)(200,-25)(200,-10)
\qbezier(180,5)(200,5)(200,-10)
\qbezier(180,15)(220,15)(215,-10)
\qbezier(215,-10)(215,-30)(110,-30)
\put(138,15){\vector(1,0){2}}
\put(40,63){\vector(0,1){2}}
\put(52,63){\vector(0,1){2}}
\put(-52,15){\vector(1,0){2}}
\put(-52,5){\vector(1,0){2}}
%
\qbezier[100](-15,70)(10,10)(-15,-50)
\qbezier[100](-10,80)(46,30)(102,80)
\qbezier[100](105,70)(75,10)(105,-50)
\qbezier[50](-15,70)(-40,10)(-15,-50)
\qbezier[50](-10,80)(46,130)(102,80)
\qbezier[50](105,70)(135,10)(105,-50)
\qbezier[100](-15,70)(-100,70)(-100,10)\qbezier[100](-15,-50)(-100,-50)(-100,10)
\qbezier[100](-10,80)(-10,150)(46,150)\qbezier[100](102,80)(102,150)(46,150)
\qbezier[100](105,70)(250,70)(250,10)\qbezier[100](105,-50)(250,-50)(250,10)
\put(101,5){\circle*{5}}
\end{picture}
\label{pevqoddrev}
\ee

\noindent
The tree is still (\ref{pretzel}) with just three fingers,
but now we allow a non-vanishing parameter $m$ in one of them.
In this case
in the two \cap s with $T^p$ and $T^q$ we have the antiparallel
strands at the entrance converted to the parallel ones and then back,
this provides the factors $(S \, T^p\, S^\dagger)_{1X}$.
In the third \cap \ the antiparallel strands are first converted into antiparallel, then
into parallel and then back into the antiparallel ones, this provides the factor
$\Big(S\, T^r \, S^\dagger\, \bar T^{-1}\, \bar S\Big)_{1X}$
(read from the right to the left):
\be
H_{R}^{(pqr1)} =d_R \overline{\sum_{X\in R\otimes \bar R}} d_X^{-1/2}
(S  T^qS^\dagger\bar T^{-1}\bar S)_{1X}
(S T^pS^\dagger)_{1X} (S T^rS^\dagger)_{1X}  = \nn\\
= d_R \sum_{X\in R\otimes \bar R} d_X^{-1/2} \overline{ \sum_{a,b,c=1}^{m_X}} {\cal A}(q,-1)_{1X}^{ab}
\cdot{\cal A}^{\rm par}_{1X}(p)^{bc}\cdot {\cal A}^{\rm par}_{1X}(r)^{ca}
\label{Hpqr13}
\ee
The last two factors in the sum are already familiar ${\cal A}^{\rm par}(p)$
and ${\cal A}^{\rm par}(q)$ from (\ref{parpre}),
while the new one is
\be
{\cal A}(q,-1)_{0,Xab} =
\sum_{Zef,Ycd}
S^{RR\bar R\bar R}_{0,Zef}\, t_{Ze}^{q}\, S^{R\bR\bR R}_{\bar Z fe,Ycd}\,
\bar t_{Yc}^{-1}\, S^{R\bR R\bR}_{Ycd,Xab}
\stackrel{(\ref{Stypes})}{=}
\sum_{Zef,Ycd}
S_{0,Zef}\, t_{Ze}^{q}\, S_{Ycd,Zef}\, \bar t_{Yc}^{-1}\, \bar S_{Ycd,Xab}
\ee
(like all the sums {\it inside} particular fingers, this is an ordinary matrix multiplication
without any sign $\sigma$-factors).

The third finger is
\be
{\cal A}^{{\rm par}_{\bar 1}}(n) = {\cal A}^{{\rm par}}(n)\cdot \bar  T^{-1}\bar S
=  S T^nS^\dagger \bar  T^{-1}\bar S
\ee
and
\be
\overline{\sum_{X,Y}} \frac{d_{[21]}^3}{\sqrt{d_Xd_Y}}
{\cal A}^{\rm par}_{1X}(n_1){\cal A}^{\rm par}_{1X}(n_3)
\bar S_{XY}{\cal A}^{\rm par}_{0Y}(n_2){\cal A}^{\rm par}_{0Y}(n_4)
\ee

Configuration (\ref{pevqoddrev}) is not always a knot: it can also be a link
with two or even three components:
\be
\begin{array}{ccc|c}
p& q & r & \# \ {\rm of\ link} \\
&&& {\rm components} \\
\hline
{\rm even} &{\rm even} & {\rm odd} & 1 \\
{\rm odd} & {\rm even} &{\rm even} & 1 \\
{\rm odd} &{\rm odd} &{\rm odd} & 1 \\
\hline
{\rm even} &{\rm even} &{\rm even} & 2 \\
{\rm even} &{\rm odd} & {\rm odd} & 2  \\
{\rm odd} &{\rm even} & {\rm odd} & 2  \\
{\rm odd} &{\rm odd} & {\rm even} & 2  \\
\hline
{\rm even} &{\rm odd} & {\rm even} & 3  \\
\end{array}
\ee
Among knots, the notable members of the three-finger family are the thick knots $10_{124}$ and $10_{139}$, which also have pretzel
realizations: $10_{124} = (5,1,3,1)$ and $10_{139} = (4,2,3,1)$.

\subsubsection{Generic four-box 3-strand braids beyond starfish family}
For $s\neq 1$ the braid (\ref{4s3braid}) cannot be converted by the Reidemeister moves
to a starfish configuration.
Instead, it is equivalent to
\be
\begin{picture}(300,150)(-20,-90)
\put(0,0){\line(1,0){20}}
\put(0,0){\line(0,1){20}}
\put(0,20){\line(1,0){20}}
\put(20,0){\line(0,1){20}}
\put(8,8){\mbox{$p$}}
\put(40,0){\line(1,0){20}}
\put(40,0){\line(0,1){20}}
\put(40,20){\line(1,0){20}}
\put(60,0){\line(0,1){20}}
\put(48,8){\mbox{$r$}}
\qbezier(15,20)(22,50)(25,10)
\qbezier(45,20)(38,50)(35,10)
\qbezier(25,10)(30,-20)(35,10)
\qbezier(15,0)(30,-30)(45,0)
\qbezier(5,20)(-2,60)(-10,-20)
\qbezier(55,20)(62,60)(70,-20)
\put(5,0){\vector(-1,-4){5}}
\put(60,-20){\vector(-1,4){5}}
\put(17,26){\vector(-1,-3){2}}
\put(30,-15){\vector(1,0){2}}
\put(3,26){\vector(1,-3){2}}
\put(68,-8){\vector(1,-4){2}}
\put(-15,-30){\mbox{$X,a$}}
\put(58,-30){\mbox{$\bar X,c$}}
\put(200,0){\line(1,0){20}}
\put(200,0){\line(0,1){20}}
\put(200,20){\line(1,0){20}}
\put(220,0){\line(0,1){20}}
\put(208,8){\mbox{$q$}}
\put(240,0){\line(1,0){20}}
\put(240,0){\line(0,1){20}}
\put(240,20){\line(1,0){20}}
\put(260,0){\line(0,1){20}}
\put(248,8){\mbox{$s$}}
\qbezier(215,20)(222,50)(225,10)
\qbezier(245,20)(238,50)(235,10)
\qbezier(225,10)(230,-20)(235,10)
\qbezier(215,0)(230,-30)(245,0)
\qbezier(205,20)(198,60)(190,-20)
\qbezier(255,20)(262,60)(270,-20)
\put(200,-20){\vector(1,4){5}}
\put(255,0){\vector(1,-4){5}}
\put(257,26){\vector(-1,-3){2}}
\put(230,-15){\vector(-1,0){2}}
\put(243,26){\vector(1,-3){2}}
\put(192,-8){\vector(-1,-4){2}}
\put(185,-30){\mbox{$Y,d$}}
\put(256,-30){\mbox{$\bar Y,f$}}
\qbezier(65,-35)(67,-50)(125,-45)\qbezier(125,-45)(185,-45)(187,-35)
\qbezier(-3,-35)(-10,-55)(120,-55)\qbezier(120,-55)(195,-55)(198,-35)
\qbezier(-10,-35)(-15,-65)(120,-65)\qbezier(120,-65)(270,-65)(265,-35)
\put(100,-46){\vector(-1,0){2}}
\put(120,-55){\vector(1,0){2}}
\put(140,-65){\vector(-1,0){2}}
\put(80,-37){\vector(1,1){2}}
\qbezier(71,-35)(73,-43)(80,-37)
\qbezier(80,-37)(200,60)(220,60)
\qbezier(220,60)(260,75)(280,40)
\qbezier(277,-35)(295,0)(280,40)
\qbezier(271,-35)(273,-45)(277,-35)
\end{picture}
\ee
This is actually a double fat tree as shown below:
\be
\begin{picture}(300,50)(-130,-30)
\put(0,0){\circle{20}}
\put(-4,-4){\mbox{$X$}}
\put(-7,7){\line(-1,1){20}}
\put(-7,-7){\line(-1,-1){20}}
\put(-32,20){\mbox{$r$}}
\put(-32,-25){\mbox{$p$}}
\put(10,0){\line(1,0){50}}
\put(12,3){\mbox{$c$}}
\put(-17,-2){\mbox{$b$}}
\put(12,-9){\mbox{$a$}}
\put(51,3){\mbox{$d$}}
\put(83,-2){\mbox{$e$}}
\put(51,-9){\mbox{$f$}}
\put(31,4){\mbox{$\bar S$}}
\put(70,0){\circle{20}}
\put(66,-4){\mbox{$Y$}}
\put(77,7){\line(1,1){20}}
\put(77,-7){\line(1,-1){20}}
\put(100,20){\mbox{$q$}}
\put(100,-25){\mbox{$s$}}
\end{picture}
\label{2vert4}
\ee
where the propagator (edge), connecting two trivalent vertices,  is actually performing an operation like (\ref{Spic})
on antiparallel strands, i.e. is represented by the matrix $\bar S$.
According to (\ref{basic}), the corresponding HOMFLY polynomial is
\be
H^{(pqrs)} =   {\sum_{X,Y\in R\otimes \bar R}} \frac{d_R^2}{\sqrt{d_Xd_Y}}
\overline{\sum_{a,b,c=1}^{m_X}}
\overline{ \sum_{d,e,f=1}^{m_Y}}
{\cal A}^{\rm par}(p)_{1X}^{ab} {\cal A}^{\rm par}(r)_{1X}^{bc }
\cdot \bar S_{XY}^{ca,fd}
{\cal A}^{\rm par}(q)_{1Y}^{de} {\cal A}^{\rm par}(s)_{1Y}^{ef }
\label{4b3str}
\ee
This provides a new type of expression already for $s=1$.
The simplest examples of knots with $s\neq 1$ gave $10$ intersections:
$10_{79}= (3,-2,2,-3)$ and $10_{152}= (3,2,2,3)$.
\subsection{Diagrams with non-tadpole propagators}
This class of knots (\ref{2vert4}) and (\ref{4b3str}) clearly indicates that not all the pretzel fingers are
contracted directly. We require an additional propagator connecting two trivalent vertices leading to
 our basic formula (\ref{basic}) with  two independent summations. We will call left trivalent vertex as
 $ppS$ to highlight two fingers with parallel braids and connected by propagator $S$.
\subsubsection{The form for ppS block}
A new building block appearing in (\ref{4b3str}) is
\be
{\cal A}^{ppS}_Y(n_1,n_2) =
d_{R}\overline{\sum_{X\in R\otimes \bar R}} \frac{1}{\sqrt{d_X}}\,
{\cal A}^{\rm par}_{1X}(n_1){\cal A}^{\rm par}_{1X}(n_2)
\bar S_{XY}
\label{defAppS}
\ee
This expression remains the same if $\bar S$ is transposed,
$\bar S_{XY} \longrightarrow \bar S_{YX}$.

The number of parallel fingers in this building block can exceed two (but needs to be even),
however, for our purposes below, the two will be enough.

For the 4-box 3-strand braid one can rewrite (\ref{4b3str}) as
\be
H_{R}^{(pqrs)} = d_{R}^3 \overline{\sum_{X,Y}} \frac{1}{\sqrt{d_Xd_Y}}\,
{\cal A}^{\rm par}_{1X}(p){\cal A}^{\rm par}_{1X}(r) \bar S_{XY}
{\cal A}^{\rm par}_{1Y}(q){\cal A}^{\rm par}_{1Y}(s)
= d_{R}^2 \overline{\sum_Y} \frac{1}{\sqrt{d_Y}}\,{\cal A}^{ppS}_Y(p,r)
{\cal A}^{\rm par}_{1Y}(q){\cal A}^{\rm par}_{1Y}(s)
\ee
We could have similar tree diagrams with other type of propagators which we discuss now.
\subsubsection{Double braids}
In fact, propagator  in (\ref{4b3str}) can be further generalized. One of the
important generalization is provided by the horizontal braid (\ref{horbraid1}):
\be
\begin{picture}(300,100)(-50,-50)
\put(-50,-10){\line(1,0){40}}
\put(-50,-10){\line(0,1){20}}
\put(-50,10){\line(1,0){40}}
\put(-10,-10){\line(0,1){20}}
\qbezier(-70,-40)(-70,-5)(-50,-5)
\qbezier(-70,40)(-70,5)(-50,5)
\qbezier(10,-40)(10,-5)(-10,-5)
\qbezier(10,40)(10,5)(-10,5)
\put(-80,-40){\line(0,1){80}}
\put(20,-40){\line(0,1){80}}
\put(-85,-50){\mbox{$X,a$}}
\put(10,-50){\mbox{$\bar X,b$}}
\put(-85,43){\mbox{$Y,c$}}
\put(10,43){\mbox{$\bar Y,d$}}
\put(-33,-2){\mbox{$m$}}
\put(100,-2){\mbox{$= \ \sum\limits_Z\sum\limits_{ef,gh}^{m_Z}
\frac{1}{\sqrt{d_Z}} \bar S_{XZ}^{ab,ef}\bar S_{YZ}^{cd,gh}\, \cdot$}}
\put(250,-10){\line(1,0){40}}
\put(250,-10){\line(0,1){20}}
\put(250,10){\line(1,0){40}}
\put(290,-10){\line(0,1){20}}
\qbezier(240,-10)(240,-5)(250,-5)
\qbezier(240,10)(240,5)(250,5)
\qbezier(300,-10)(300,-5)(290,-5)
\qbezier(300,10)(300,5)(290,5)
\qbezier(240,-10)(240,-15)(250,-15)
\qbezier(240,10)(240,15)(250,15)
\qbezier(300,-10)(300,-15)(290,-15)
\qbezier(300,10)(300,15)(290,15)
\qbezier(265,-40)(265,-15)(250,-15)
\qbezier(265,40)(265,15)(250,15)
\qbezier(275,-40)(275,-15)(290,-15)
\qbezier(275,40)(275,15)(290,15)
\put(330,-40){\line(0,1){80}}
\put(320,-40){\line(0,1){80}}
\put(265,-50){\mbox{$Z,e$}}
\put(315,-50){\mbox{$\bar Z,f$}}
\put(265,43){\mbox{$Z,g$}}
\put(315,43){\mbox{$\bar Z,h$}}
\put(267,-2){\mbox{$m$}}
\end{picture}
\label{horbraid}
\ee
The right two lines in the above diagram do not change the representation,
(that is $\delta_{fh}$),
the left part of this diagram is just the pretzel finger ${\cal A}_{1Z}^{eg}$.
The type of $S$-matrices and the pretzel finger depends on the directions of arrows and on the parity
of the braid length $m$. The simplest application of (\ref{horbraid}) is to the double braids of \cite{evo},
which has also the pretzel representation $(-1,\overline{2k},m)$. We present more knots using horizontal braids as propagators.
\subsubsection{Pretzel fingers connected via horizontal braid
\label{KTCfam}}
We consider examples where pretzel fingers appear from both sides of the
horizontal braid:
\be
\begin{picture}(300,170)(-130,-135)
\put(0,0){\line(1,0){20}}
\put(0,0){\line(0,1){20}}
\put(0,20){\line(1,0){20}}
\put(20,0){\line(0,1){20}}
\put(8,8){\mbox{$p$}}
\put(40,0){\line(1,0){20}}
\put(40,0){\line(0,1){20}}
\put(40,20){\line(1,0){20}}
\put(60,0){\line(0,1){20}}
\put(48,8){\mbox{$q$}}
\qbezier(15,20)(22,50)(25,10)
\qbezier(45,20)(38,50)(35,10)
\qbezier(25,10)(30,-20)(35,10)
\qbezier(15,0)(30,-30)(45,0)
\qbezier(5,20)(-2,60)(-10,-20)
\qbezier(55,20)(62,60)(70,-20)
\put(5,0){\vector(-1,-4){5}}
\put(60,-20){\vector(-1,4){5}}
%
\put(17,26){\vector(-1,-3){2}}
\put(30,-15){\vector(1,0){2}}
\put(3,26){\vector(1,-3){2}}
\put(68,-8){\vector(1,-4){2}}
\put(-15,-30){\mbox{$X,a$}}
\put(58,-30){\mbox{$\bar X,c$}}
\put(0,-100){\line(1,0){20}}
\put(0,-100){\line(0,-1){20}}
\put(0,-120){\line(1,0){20}}
\put(20,-100){\line(0,-1){20}}
\put(8,-112){\mbox{$s$}}
\put(40,-100){\line(1,0){20}}
\put(40,-100){\line(0,-1){20}}
\put(40,-120){\line(1,0){20}}
\put(60,-100){\line(0,-1){20}}
\put(48,-112){\mbox{$r$}}
\qbezier(15,-120)(22,-150)(25,-110)
\qbezier(45,-120)(38,-150)(35,-110)
\qbezier(25,-110)(30,-80)(35,-110)
\qbezier(15,-100)(30,-70)(45,-100)
\qbezier(5,-120)(-2,-160)(-10,-80)
\qbezier(55,-120)(62,-160)(70,-80)
\put(55,-100){\vector(1,4){5}}
\put(0,-80){\vector(1,-4){5}}
%
\put(43,-126){\vector(1,3){2}}
\put(30,-85){\vector(-1,0){2}}
%
\put(57,-126){\vector(-1,3){2}}
\put(-8,-92){\vector(-1,4){2}}
\put(-15,-76){\mbox{$Y,d$}}
\put(58,-76){\mbox{$\bar Y\!,f$}}
\qbezier(-12,-32)(-13,-45)(-12,-65)
\qbezier(71.5,-32)(73,-45)(71.5,-65)
\qbezier(-3,-32)(-6,-37)(15,-37)
\qbezier(-3,-65)(-6,-60)(15,-60)
\qbezier(62,-32)(65,-37)(45,-37)
\qbezier(62,-65)(65,-60)(45,-60)
\put(15,-34){\line(0,-1){29}}  \put(45,-34){\line(0,-1){29}}
\put(15,-34){\line(1,0){30}}  \put(15,-63){\line(1,0){30}}
\put(26,-50){\mbox{$m$}}
\end{picture}
\label{mutfam}
\ee
The central block is just the horizontal braid
described by the corresponding version of (\ref{horbraid}).
Therefore, the reduced HOMFLY polynomial for the diagram (\ref{mutfam}) with even $m$ is
\be
\sum_{X,Y\in R\otimes \bar R} \frac{d_R^2}{\sqrt{d_Xd_Y}} \overline{\sum_{a,b,c=1}^{m_X}}
 \overline{\sum_{d,e,f=1}^{m_Y}}
{\cal A}^{\rm par}(p)_{1X}^{ab} {\cal A}^{\rm par}(q)_{1X}^{bc }
 \left(\sum_{Z \in R\otimes \bR}\overline{\sum_{g,h,i}^{m_Z}}
 \frac{1}{\sqrt{d_Z}}\bar S_{XZ}^{ca,gh}
{\cal A}^{\rm ea}(m)_{1Z}^{hi }\bar S_{YZ}^{ig,fd}\right)
{\cal A}^{\rm par}(r)_{1Y}^{de} {\cal A}^{\rm par}(s)_{1Y}^{ef } \nn \\
=\overline{\sum_Z} \frac{1}{\sqrt{d_Z}}\,
{\cal A}^{ppS}_Z(p,q)\, {\cal A}^{ppS}_Z(r,s)
{\cal A}^{\rm ea}_{1Z}(m)~~~~~~~~~~~~~~~~~~~~~~~~~~~~~~~~~~~~~~~~~~
\label{Mpqmrs}
\ee
The four external fingers involve parallel braids, while that of the even length $m$ in the propagator
is antiparallel.
The fat tree is the same (\ref{2vert4}), only the internal propagator is more sophisticated.
Mutation is a permutation of $p$ and $q$ or of $r$ and $s$.
\bigskip

\noindent{\it Familiar special limits}\\
$\bullet$ At $m=0$ we get a composite knot, thus the reduced HOMFLY polynomial should factorize.
In (\ref{Mpqmrs}) this follows from (\ref{capX0}), which implies that
${\cal A}^{\rm ea}_{1Z}(0) = \delta_{1,Z}$.
\bigskip

\noindent$\bullet$ At $s=0$  the parallel finger ${\cal A}^{\rm par}_{1Y}(0) = \delta_{1,Y}$, therefore
${\cal A}^{ppS}_Y(0,n) = d_{R}\bar S_{1Y}{\cal A}_{11}(n) =\sqrt{d_Y}
{\cal A}_{11}(n)$ and
\be
H_{R}^{(pq|m|r0)} = {\cal A}^{\rm par}_{11}(r)
\overline{\sum_Z} {\cal A}^{ppS}_{1Z}(p,q) {\cal A}^{\rm ea}_{1Z}(m)
\ee
$\bullet$ If  we also take $q=0$, we obtain a composite knot, made out of two
2-strand constituents:
\be
H_{R}^{(p0|m|r0)} = H_{R}^{(0p|m|r0)} =
{\cal A}^{\rm par}_{11}(p){\cal A}^{\rm par}_{11}(r)
\underbrace{\sum_Z \sqrt{d_Z}\cdot {\cal A}^{\rm ea}_{1Z}(m)}_{=1} =
{\cal A}^{\rm par}_{11}(p){\cal A}^{\rm par}_{11}(r)
\ee
The polynomial form for these double fat graph knots presented  in this part I can be worked out for  knots  carrying  fundamental  or symmetric representation.
This is plausible due to the explicit form of  the  Racah matrices \cite{NRZ1}, \cite[2nd paper]{GMMMS}
known for all symmetric representations. In all these cases all the $\sigma$-factors are unities, and there is no need to deal specifically with
overlined sums.
The same formula continue to work for representation $[21]$, but then some $\sigma$-factors
are $-1$. Expressions for $R=[21]$ are rather involved using Racah matrices explicitly calculated in \cite{GuJ}.  Our main
focus in Part II will be to compute $R=[21]$ colored HOMFLY polynomials.

\newpage

\part{Examples}
Before we proceed with the explicit knot polynomial computations for $R=[21]$, we require expressions for the Racah matrices. We will briefly recapitulate them in the following section.
\section{Racah matrices \label{Racah}}
\subsection{Generalities}
The Racah matrices relate the two expansions
\be
(R_1\otimes R_2)\otimes R_3 = \oplus_{P } W^P_{R_1R_2}\otimes (P \otimes R)
= \sum_{Q}  \left(\oplus_P W^P_{R_1R_2}\otimes W^Q_{PR_3}\right)\otimes Q
\ee
and
\be
R_1\otimes (R_2\otimes R_3) = \oplus_{S }  W^S_{R_2R_3}\otimes (R_1\otimes S)
= \sum_{Q}  \left(\oplus_S W^Q_{R_1S}\otimes W^S_{R_2R_3}\right)\otimes Q
\ee
that is, it is a linear operator
\be
\left(\oplus_S W^Q_{R_1S}\otimes W^S_{R_2R_3}\right) \ = \
\hat S\left[\begin{array}{cc} R_3&Q\\R_1&R_2\end{array}\right]\
\left(\oplus_P W^P_{R_1R_2}\otimes W^Q_{PR_3}\right)
\ee
When the vector spaces $W$ are one-dimensional ("no multiplicities" case),
one can represent this linear operator as a matrix with indices $P\in R_1\otimes R_2$ and
$S\in R_2\otimes R_3$.
When $W$ are multidimensional, there is no any distinguished basis and the concrete
form of the Racah matrix depends on conventions, and on four additional indices
labeling the bases in the four $W$-spaces for each given pair $P$ and $S$.

For the purposes of knot (rather than link) theory, all the four representations
$R_1,R_2,R_3,Q$ are either $R$ or its conjugate $\bar R$.
In result, there are two essentially different Racah matrices:

$\bullet$ $S_{S,c,d|P,a,b}$  with  $P\in R\otimes R$, but $S\in R\otimes \bar R$,
and

$\bullet$ $\bar S_{S,c,d|P,a,b}$  with both $P,S \in R\otimes \bar R$.

\noindent
These are the only two kinds of the Racah matrices that show up in our discussion of the double fat tree diagrams.
Note that $S$ is essentially asymmetric, while $\bar S$ can be symmetric. In fact, $\bar S$ is symmetric
for $R$'s taken to be symmetric representation which belongs to the  multiplicity-free case.

Another important fact is that the singlet representation $1\in R\otimes R$, but
$1\notin R\otimes R$ (except for the case of $N=2$). Therefore, the matrix elements $1X$ in (\ref{basic})
cannot have $S^\dagger$ at the very left, but only $S$
or $\bar S$. In general, as we already saw earlier, (\ref{Saction})
$S$ converts the parallel strands into antiparallel, while $\bar S$
antiparallel to antiparallel.

Our \cap s never contain three parallel strands, thus these two types of relations are sufficient for our consideration.
However, beyond this paper the third Racah matrix connecting the parallel strands to parallel  (it was called "mixing matrix" in \cite{MMMknots12}) plays a big role. Actually, it is much simpler: it does not depend on $N$ (i.e. on $A$). For symmetric representations, it coincides with restrictions of both $S$ and $\bar S$ to $N=2$, but for more complicated representations the story is a little more involved.

The Racah matrices $S$ are always unitary
and satisfy (\ref{firstline}).
There are also additional non-trivial relations like \cite{ramaid,GuJ}
\be
\bar S^\dagger  \bar T S = \bar T^{-1}S T^{-1} \ \
\stackrel{{\rm conjugation}}{\Longleftrightarrow} \ \
S^\dagger  \bar T \bar S = T^{-1}S^\dagger \bar T^{-1}
\label{SbTbS}
\ee

\begin{figure}[ht!]
\centering\leavevmode
\includegraphics[width=8cm]{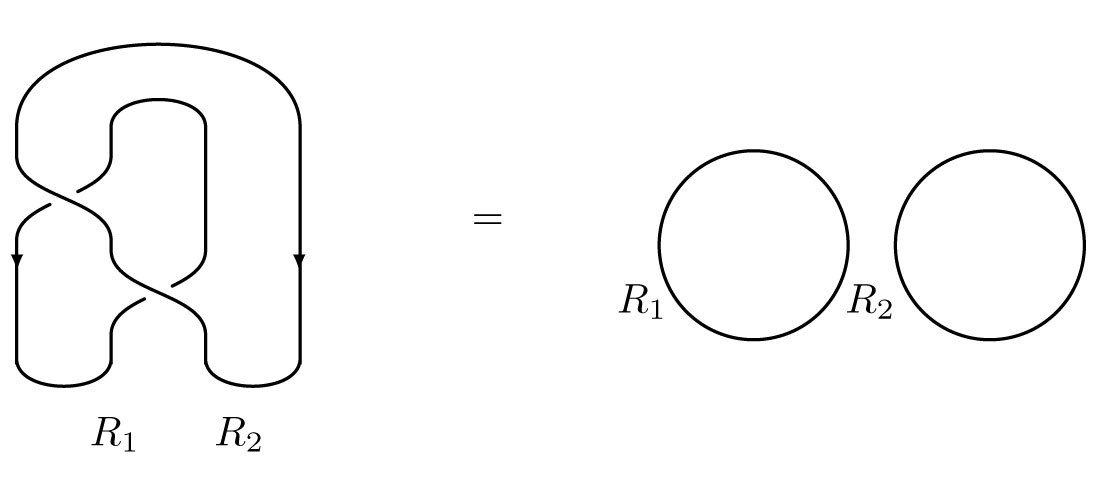}
\caption{knot equivalence}
\label{equvi}
\end{figure}

\noindent
This identity can be illustrated by Fig.\ref{equvi}.
The knot equivalence gives the following relation:
\be
d_{R_1}d_{R_2}\cdot (STS^\dagger \bar T\bar S)_{_{11}}  = d_{R_1}d_{R_2}
\ee
and one can insert $\bar T$ to the very left, since it trivially acts on the singlet state. This immediately gives the matrix element $(...)_{_{11}}$ of the second equality of (\ref{SbTbS}).

Further generalization  with indices restored (\ref{SbTbS}), we get
\be
\sum_{X,a,b,c,d}\Big(\bar S^\dagger\Big)_{YX}^{ef,ab}  \bar T_X^{ab,dc} \bar S_{XZ}^{cd,gh} =
\sum_{k,l,m,n}
(T^{-1})_Y^{ef,lk}S_{YZ}^{lk,mn}(\bar T^{-1})_{Z}^{mn,gh}
\label{ramaid}
\ee
An important application of this identity is the possibility of shifting
any pretzel finger of length $1$ to any position \cite{pret} at any representation $R$,
though permutations of arbitrary fingers are {\it not} a symmetry of the knot.
\subsection{Racah matrices for few symmetric representations}
For symmetric representations $R$, there are no multiplicities, no indices $a,b,c,d$
and the Racah matrices are canonically defined.

\bigskip

$\bullet$ When $R$ is the fundamental representation $R=[1]$, they are given by (\ref{Sfund})
and (\ref{barSfund}), or in a generalizable notation,
\be
S = \frac{1}{\sqrt{[2]D_0}}\left(\begin{array}{cc|c}
 [11] & [2]  & \\
&&\\
\hline
&&\\
 \sqrt{D_{-1}}& \sqrt{D_1} & 0  \\
&&\\
 \sqrt{D_1} & -\sqrt{D_{-1}}&{\rm adj}
\end{array}\right),
\ \ \ \ \ \ \ \
\bar S = \frac{1}{\chi_{[1]}}\left(\begin{array}{cc|c }
0 & {\rm adj} &  \\
&&\\
\hline
&&\\
1 & \sqrt{\bar d_1} & 0 \\
&&\\
 \sqrt{\bar d_1} & -1 & {\rm adj}\\
\end{array}\right)
\ee
where $D_i = \frac{Aq^i-A^{-1}q^{-i}}{q-q^{-1}}$, and $\bar d_1 \equiv d_{[\bar 11]}= D_1D_{-1}$,
$\bar d_2\equiv d_{[\bar 22]} = \frac{1}{[2]^2}D_3D_0^2D_{-1}$, $\ldots$,
$\bar d_m = D_{2m-1}\left(\prod_{j=0}^{m-2}\frac{D_j}{[j+2]}\right)^2\! D_{-1}$ are the quantum dimensions of the representations from the decomposition $[r]\otimes [r]=1+[\bar 11]+[\bar 22]+\ldots$, where $[\bar 11]$ denotes $[21^{N-2}]$, $[\bar 22]$ denotes $[42^{N-2}]$ etc.

\bigskip

$\bullet$ When $R$ is the first symmetric representation $R=[2]$,
$[2]\otimes [2] = [4]+[31]+[22]$, $[2]\otimes\overline{[2]} = 1+[\bar 11]+[\bar 22]$,
\be
S = \frac{[2]}{D_1D_0}\left(\begin{array}{ccc|c} \l[22] & [31] & [4] & \\ &&&\\
\hline &&&\\
\frac{\chi_{[22]}}{\sqrt{ d_0}}=\sqrt{\chi_{22}}
& \frac{\chi_{31}}{\sqrt{ d_1}} =\sqrt{\chi_{31}}&
\frac{\chi_4}{\sqrt{ d_2}}=\sqrt{\chi_4}
& 0\\ &&&\\
\sqrt{\frac{\overline{ d}_1}{ d_0}}\cdot\frac{D_1D_0^2}{[2]^2[3]}
&\sqrt{\frac{\overline{ d}_1}{ d_1}}\cdot\frac{(D_2-D_0)D_1D_0}{[2][4]}
& -\sqrt{\frac{\overline{ d}_1}{ d_2}}\cdot\frac{D_3D_1D_0}{[3][4]} & {\rm adj}\\ &&&\\
\sqrt{\frac{\overline{ d}_2}{ d_0}}\cdot\frac{D_1D_0}{[2][3]}
& -\sqrt{\frac{\overline{ d}_2}{ d_2}}\cdot\frac{D_1D_0}{[4]}
& \sqrt{\frac{\overline{ d}_2}{ d_2}}\cdot \frac{D_1D_0}{[3][4]} & [\bar 22]
\end{array}\right)
\ee
with
\be
 d_m\equiv d_{[r+m,r-m]}=\chi_{[r+m,r-m]} ={[2m+1]\over
 [r+m+1]![r-m]!}\prod_{i=0}^{2r-1}D_j\prod_{j=0}^{r-m-1}{D_{j-1}\over D_{r+m+j}}\ ,
, \ \ \ \ \ r=2
\ee
and
\be
\bar S = \frac{1}{\chi_{[2]}}\left(\begin{array}{ccc|c}
 0 & {\rm adj} & [\bar 22] &  \\ &&&\\
\hline &&& \\
 1 & \sqrt{\bar d_1} & \sqrt{\bar d_2} & 0 \\
&&&\\
 \sqrt{\bar d_1} & \frac{D_1}{[2]D_2}\Big(D_3D_{-1}-1\Big) & -\frac{D_0}{D_2}\sqrt{D_3D_1}
 & {\rm adj} \\ &&&\\
\sqrt{\bar d_2} &  -\frac{D_0}{D_2}\sqrt{D_3D_1} & \frac{D_0}{D_2} & [\bar 22]
\end{array}\right)
\ee

\bigskip

$\bullet$ Similarly, for representation $R=[3]$ and so on.

\subsection{Representation $[21]$}

In the case of $R=[21]$ there are seven different items $X$ in the decomposition
\be
\l[21]\otimes \underbrace{\overline{[21]}}_{[2^{N-2}1]} = [432^{N-4}1]\oplus[42^{N-2}]
\oplus[42^{N-3}11]\oplus[332^{N-3}]\oplus[332^{N-4}11]\oplus \underline{2\cdot [32^{N-2}1]}\oplus \underbrace{[2^N]}_{singlet}
\label{21bar21}
\ee
The corresponding dimensions and eigenvalues are
\be
\begin{array}{c|c|c}
X & d_X & t_X \\
&& \\
\hline
&& \\
\l[2^N]   &1 & 1\\
&& \\
\l[42^{N-3}11]&\frac{[N-2][N-1][N+1][N+2]}{[2]^2}& -A^2\\
&& \\
\l[332^{N-3}]&\frac{[N-2][N-1][N+1][N+2]}{[2]^2}& -A^2 \\
&&\\
\l[332^{N-4}11]&\frac{[N-3][N]^2[N+1]}{[2]^2}& - q^{-2}A^2\\
&& \\
\l[432^{N-4}1]&\frac{[N-3][N-1]^2[N+1]^2[N+3]}{[3]^2}& A^3 \\
&& \\
\l[42^{N-2}] &\frac{[N-1][N]^2[N+3]}{[2]^2}& q^2A^2\\
&& \\
\hline
&& \\
\underline{\l[32^{N-2}1]}&\underline{[N-1][N+1]} & \underline{\pm A} \\
&& \\
\end{array}
\label{dims21a}
\ee

Only one (underlined) item $[32^{N-2}1]=[21^{N-2}]=Adj$ enters (\ref{21bar21})
with non-trivial multiplicity $m_{Adj}=2$,
however this makes the Racah matrices $10\times 10$, since $\sum_X m_X^2 = 6\cdot 1+1\cdot 2^2 = 10$.
They are explicitly found in \cite{GuJ}.
The pure antiparallel matrix $\bar S$ is

\begin{landscape}
\tiny{
\be
\left(\begin{array}{cccccc|cccc}
\frac{[3]}{D_{-1}D_0D_1}&{[3]\over [2]D_0}\sqrt{D_2D_{-2}\over D_1D_{-1}}&{[3]\over [2]D_0}\sqrt{D_2D_{-2}\over D_1D_{-1}}
&{[3]\over [2]D_{-1}}\sqrt{D_3\over D_1}&{\sqrt{D_3D_{-3}}\over D_0}&{[3]\over [2]D_1}\sqrt{D_3\over D_{-1}}&
{[3]\over D_0}{1\over\sqrt{D_1D_{-1}}}&{[3]\over D_0}{1\over\sqrt{D_1D_{-1}}}&0&0\\
&&&&&&&&&\\
{[3]\over [2]D_0}\sqrt{D_2D_{-2}\over D_1D_{-1}}&{[3]\over [2]^2D_0}&{[3]\over [2]^2D_0}&
-{[3]\over [2]^2D_0}\sqrt{D_{-3}D_2\over D_{-2}D_{-1}}&0&-{[3]\over [2]^2D_0}\sqrt{D_3D_{-2}\over D_2D_1}&\bar S_{[2,7]}&
-{\sqrt{D_2D_{-2}}\over [2]D_0}&{D_{-1}\over [2]D_0}&{D_1\over [2]D_0}\\
&&&&&&&&&\\
{[3]\over [2]D_0}\sqrt{D_2D_{-2}\over D_1D_{-1}}&{[3]\over [2]^2D_0}&{[3]\over [2]^2D_0}&-{[3]\over [2]^2D_0}
\sqrt{D_{-3}D_2\over D_{-1}D_{-2}}&0&-{[3]\over [2]^2D_0}
\sqrt{D_3D_{-2}\over D_1D_2}&\bar S_{[2,7]}&-{\sqrt{D_2D_{-2}}\over [2]D_0}&-{D_1\over [2]D_0}&-{D_{-1}\over [2]D_0}\\
&&&&&&&&&\\
{[3]\over [2]D_{-1}}\sqrt{D_3\over D_1}&-{[3]\over [2]^2D_0}\sqrt{D_{-3}D_2\over D_{-2}D_{-1}}&-{[3]\over [2]^2D_0}
\sqrt{D_{-3}D_2\over D_{-1}D_{-2}}&\bar S_{[4,4]}&{[3]\sqrt{D_1D_3}\over D_0D_2D_{-2}}&-{[3]^2D_0\over D_2D_{-2}}
\sqrt{D_3D_{-3}\over D_1D_{-1}}&\bar S_{[4,7]}&{D_2\over [2]D_0}\sqrt{D_{-3}\over D_{-1}}&
-{\sqrt{D_{-1}D_{-3}D_2}\over [2]D_0\sqrt{D_{-2}}}&{\sqrt{D_{-1}D_{-3}D_2}\over [2]D_0\sqrt{D_{-2}}}\\
&&&&&&&&&\\
{\sqrt{D_3D_{-3}}\over D_0}&0&0&{[3]\sqrt{D_1D_3}\over D_0D_2D_{-2}}&-{[3]\over D_{-2}D_0D_2}&
{[3]\sqrt{D_{-1}D_{-3}}\over D_{-2}D_0D_2}&-{[3]\sqrt{D_{-1}D_1D_{-3}D_3}\over D_{-2}D_0D_2}&0&0&0\\
&&&&&&&&&\\
{[3]\over [2]D_1}\sqrt{D_3\over D_{-1}}&-{[3]\over [2]^2D_0}\sqrt{D_3D_{-2}\over D_2D_1}&-{[3]\over [2]^2D_0}
\sqrt{D_3D_{-2}\over D_1D_2}&-{[3]^2D_0\over D_2D_{-2}}
\sqrt{D_3D_{-3}\over D_1D_{-1}}&{[3]\sqrt{D_{-1}D_{-3}}\over D_{-2}D_0D_2}&\bar S_{[6,6]}&\bar S_{[6,7]}&
{D_{-2}\over [2]D_0}\sqrt{D_3\over D_1}&{\sqrt{D_1D_3D_{-2}}\over [2]D_0\sqrt{D_{2}}}&
-{\sqrt{D_1D_3D_{-2}}\over [2]D_0\sqrt{D_{2}}}
\\&&&&&&&&&\\
\hline
&&&&&&&&&\\
{[3]\over D_0}{1\over\sqrt{D_1D_{-1}}}&\bar S_{[2,7]}&\bar S_{[2,7]}&\bar S_{[4,7]}&
-{[3]\sqrt{D_{-1}D_1D_{-3}D_3}\over D_{-2}D_0D_2}&\bar S_{[6,7]}&\bar S_{[7,7]}&{[4]\over [2]D_0}&
-{\{A^2\}\over \{q\}}{1\over D_0^2\sqrt{D_2D_{-2}}}&{\{A^2\}\over \{q\}}{1\over D_0^2\sqrt{D_2D_{-2}}}\\
&&&&&&&&&\\
{[3]\over D_0}{1\over\sqrt{D_1D_{-1}}}&-{\sqrt{D_2D_{-2}}\over [2]D_0}&-{\sqrt{D_2D_{-2}}\over [2]D_0}&
{D_2\over [2]D_0}\sqrt{D_{-3}\over D_{-1}}&0&{D_{-2}\over [2]D_0}\sqrt{D_3\over D_1}&{[4]\over [2]D_0}&-{1\over D_0}&
0&0\\
&&&&&&&&&\\
0&-{D_1\over [2]D_0}&{D_{-1}\over [2]D_0}&-{\sqrt{D_{-1}D_{-3}D_2}\over [2]D_0\sqrt{D_{-2}}}&0&
{\sqrt{D_1D_3D_{-2}}\over [2]D_0\sqrt{D_{2}}}&-{\{A^2\}\over \{q\}}{1\over D_0^2\sqrt{D_2D_{-2}}}&0&-{1\over D_0}&
{1\over D_0}\\
&&&&&&&&&\\
0&-{D_{-1}\over [2]D_0}&{D_1\over [2]D_0}&{\sqrt{D_{-1}D_{-3}D_2}\over [2]D_0\sqrt{D_{-2}}}&0&
-{\sqrt{D_1D_3D_{-2}}\over [2]D_0\sqrt{D_{2}}}&{\{A^2\}\over \{q\}}{1\over D_0^2\sqrt{D_2D_{-2}}}&0&{1\over D_0}&-{1\over D_0}
\end{array}\right)
\nn
\ee
}

\bigskip

\small{
\be
\bar S_{[2,7]}=
-{[2][3]^2D_{-2}D_0D_2-[2]D_2^2D_0D_{-2}^2-[3]^3D_0(D_2D_{-3}+D_3D_{-2})+[3]D_2D_{-2}(D_3+D_{-3})\over
[2]^2D_0^2D_1D_{-1}\sqrt{D_2D_{-2}}}
\nn
\ee
\be
\bar S_{[4,7]}=\sqrt{D_{-3}\over D_{-1}^3}
{[2]^2[3]^2+2[3]D_2D_{-2}(D_{-1}^2+D_1^2)-[3]^2D_0D_1(D_2D_{-3}+D_3D_{-2})-[2]D_{-1}D_{-2}^2D_2^3\over [2]^2D_0D_2D_{-2}D_1^2}
\label{Sa21}
\ee
\be
\bar S_{[6,7]}=\sqrt{D_{3}\over D_{1}^3}
{[2]^2[3]^2+2[3]D_2D_{-2}(D_{-1}^2+D_1^2)-[3]^2D_0D_{-1}(D_2D_{-3}+D_3D_{-2})-[2]D_{1}D_{2}^2D_{-2}^3\over [2]^2D_0D_2D_{-2}D_{-1}^2}
\nn
\ee
\be
\bar S_{[6,6]}={[2]^2(D_1+D_{-1})-D_3D_0D_{-4}\over [2]^2D_0D_1D_2D_{-2}}
\nn
\ee
\be
\bar S_{[7,7]}=
{[2]D_{-2}D_2(D_2D_{-1}^2+D_{-2}D_1^2)-2[3](D_{-1}^3+D_1^3))\over [2]D_0^2D_{-1}^2D_1^2}+
{[2][3]^2\{A^2\}^2\over \{q\}^2D_{-2}D_2D_0^3D_1^2D_{-1}^2(D_1D_{-2}+D_{-1}D_2)}+
{[3]^2(D_2D_{-3}+D_3D_{-2})\over D_{-2}D_0D_2(D_1D_{-2}+D_{-1}D_2)}
\nn
\ee}
\end{landscape}

For $S$ we additionally need the expansion\footnote{For comparison, for rectangular representations all multiplicities are unit:
$$
\l[2]\otimes [2]=[4]+[31]+[22],
$$
$$
\l[22]\otimes [22]=[44]+[431]+[422]+[3311]+[3221]+[2222],
$$
$$
\l[33]\otimes [33]=[66]+[651]+[642]+[633]+[5511]+[5421]+[5331]+[4422]+[4332]+[3333],
$$
$$
\ldots
$$
}
\be
\l[21]\otimes [21] = [42]+[411]+[33]+\underline{2\cdot[321]}+[222]+[3111]+[2211]
\ee
with
\be
\begin{array}{c|c|c}
X & d_X & t_X \\
&& \\
\hline
&& \\
\l[222]&\frac{[N-2][N-1]^2[N]^2[N+1]}{[2]^2[3]^2[4]}& q^{-3}A^{-3}\\
&& \\
\l[411] &\frac{[N-2][N-1][N][N+1][N+2][N+3]}{[2]^2[3][6]}& -q^3A^{-3}\\
&& \\
\l[33]&\frac{[N-1][N]^2[N+1]^2[N+2]}{[2]^2[3]^2[4]}& -q^3A^{-3}\\
&& \\
\l[2211]   &\frac{[N-3][N-2][N-1][N]^2[N+1]}{[2]^2[4][5]} & -q^{-5}A^{-3}\\
&&\\
 \l[3111]&\frac{[N-3][N-2][N-1][N][N+1][N+2]}{[2]^2[3][6]} & q^{-3}A^{-3} \\
&& \\
\l[42]&\frac{[N-1][N]^2[N+1][N+2][N+3]}{[2]^2[4][5]}&q^5A^{-3}   \\
&& \\
\hline
&& \\
\underline{\l[321]}&\underline{\frac{[N-2][N-1][N]^2[N+1][N+2]}{[3]^2[5]}}&\underline{\pm A^{-3}} \\
&&\\
\end{array}
\label{dims21p}
\ee
Here we again encounter one item with multiplicity two,
and the parallel $\rightarrow$ antiparallel Racah matrix $S$ is

\bigskip

\begin{landscape}
\small{
\be
\left(\begin{array}{cccccc|cccc}
\sqrt{\frac{ D_{-2}}{[2]^2[4]\,D_1 }} &\sqrt{\frac{[3]\,D_3D_2D_{-2}}{[2]^2[6]\,D_1D_0D_{-1} }}
&\sqrt{\frac{ D_2}{[2]^2[4]\,D_{-1} }}
&\sqrt{\frac{ [3]^2\,D_{-2}D_{-3}}{[2]^2[4][5]\,D_1D_{-1} }}
&\sqrt{\frac{ [3]\,D_2D_{-2}D_{-3}}{[2]^2[6]\,D_1D_0D_{-1} }}
&\sqrt{\frac{ [3]^2D_3D_2}{[2]^2[4][5]\,D_1D_{-1} }}
&\sqrt{\frac{D_2D_{-2}}{[5]\,D_1D_{-1} }}&\sqrt{\frac{D_2D_{-2}}{[5]\,D_1D_{-1}}}&0&0\\
&&&&&&&&&\\
{[3]\sqrt{D_{-1}D_2}\over [2]^2\sqrt{[4]}D_0}&{\sqrt{[3]D_3}\over [2]^2\sqrt{[6]D_0}}&{[3]\sqrt{D_1D_{-2}}\over [2]^2D_0\sqrt{[4]}}
&-{[3]\sqrt{D_2D_{-3}}\over [2]^2D_0\sqrt{[4][5]}}&{\sqrt{[3]D_{-3}}\over [2]^2\sqrt{D_0[6]}}&-{[3]\sqrt{D_3D_{-2}}\over
[2]^2D_0\sqrt{[4][5]}}&-{[4]D_0+[3](D_2-D_{-2})\over 2[2]^2\sqrt{[5]}D_0}
&-{[4]D_0-[3](D_2-D_{-2})\over 2[2]^2\sqrt{[5]}D_0}&\frac{1}{2}&\frac{1}{2}\\
&&&&&&&&&\\
{[3]\sqrt{D_{-1}D_2}\over [2]^2\sqrt{[4]}D_0}&{\sqrt{[3]D_3}\over [2]^2\sqrt{[6]D_0}}&{[3]\sqrt{D_1D_{-2}}\over [2]^2D_0\sqrt{[4]}}
&-{[3]\sqrt{D_2D_{-3}}\over [2]^2D_0\sqrt{[4][5]}}&{\sqrt{[3]D_{-3}}\over [2]^2\sqrt{D_0[6]}}&-{[3]\sqrt{D_3D_{-2}}\over
[2]^2D_0\sqrt{[4][5]}}&-{[4]D_0+[3](D_2-D_{-2})\over 2[2]^2\sqrt{[5]}D_0}
&-{[4]D_0-[3](D_2-D_{-2})\over 2[2]^2\sqrt{[5]}D_0}&-\frac{1}{2}&-\frac{1}{2}\\
&&&&&&&&&\\
{\sqrt{D_{-3}}\over [2]^2\sqrt{[4]}\sqrt{D_{-2}}}&{\sqrt{D_0D_3D_{-3}[3]^3}\over [2]^2\sqrt{[6]D_{-1}D_{-2}D_2}}&-{[3]\sqrt{D_1D_{-3}}
\over [2]^2\sqrt{[4]D_2D_{-1}}}
&-{[4]D_{-1}+[2][3]D_1\over [2]^2\sqrt{[4][5]D_{-1}D_{-2}}}&S_{[4,5]}&-{[3]^2D_3D_{-3}\over [2]^2\sqrt{[4][5]D_2D_{-1}}}
&{\sqrt{D_{-3}}(D_2+D_{-1})\over [2]\sqrt{[5]D_{-1}D_{-2}D_2}}&{\sqrt{D_{-3}}(D_2-D_{-1})\over [2]\sqrt{[5]D_{-1}D_{-2}D_2}}&0&0\\
&&&&&&&&&\\
{\sqrt{D_1D_3D_{-3}}\over [2]D_0\sqrt{[4]D_{-2}}}&-{\sqrt{[3]D_{-3}D_{-1}D_1}\over
 [2]\sqrt{[6]D_0D_2D_{-2}}}&-{\sqrt{D_{-3}D_{-1}D_3}\over [2]D_0\sqrt{[4]D_2}}
&-{\sqrt{[3]D_3D_1D_{-1}}\over [2]D_0\sqrt{[4][5]D_{-2}}}&{\sqrt{[3]D_1D_{-1}D_3}\over [2]\sqrt{[6]D_0D_2D_{-2}}}&
-{\sqrt{[3]D_1D_{-1}D_{-3}}\over [2]D_0\sqrt{[4][5]D_2}}
&{\sqrt{D_1D_{-1}D_3D_{-3}}\over D_0\sqrt{[5]D_2D_{-2}}}&{\sqrt{D_1D_{-1}D_3D_{-3}}\over D_0\sqrt{[5]D_2D_{-2}}}&0&0\\
&&&&&&&&&\\
{[3]\sqrt{D_{-1}D_3}\over [2]^2\sqrt{[4]D_1D_{-2}}}&S_{[6,2]}&-{\sqrt{D_3}\over [2]^2\sqrt{[4]D_2}}
&{[3]^2\sqrt{D_3D_{-3}}\over [2]^2\sqrt{[4][5]D_1D_{-2}}}&-{\sqrt{D_0D_3D_{-3}[3]^3}\over [2]^2\sqrt{[6]D_1D_2D_{-2}}}
&{[2][3]D_{-1}+[4]D_1\over [2]^3\sqrt{[4][5]D_1D_2}}
&-{\sqrt{D_3}(D_{-2}-D_1)\over [2]\sqrt{[5]D_1D_2D_{-2}}}&-{\sqrt{D_3}(D_{-2}+D_1)\over [2]\sqrt{[5]D_1D_2D_{-2}}}&0&0\\
&&&&&&&&&\\
\hline
&&&&&&&&&\\
-{\sqrt{D_{-1}^3}\over [2]D_0\sqrt{[4]D_{-2}}}&S_{[7,2]}&{\sqrt{D_{1}^3}\over [2]D_0\sqrt{[4]D_{2}}}
&-{\sqrt{D_{-3}}([2][3]D_{1}+[4]D_{-1})\over [2]^2D_0\sqrt{[4][5]D_{-2}}}&S_{[7,5]}
&{\sqrt{D_{3}}([2][3]D_{-1}+[4]D_{1}\over [2]^2D_0)\sqrt{[4][5]D_{2}}}&-{{\{A^2\}\over\{q\}}+1\over D_0\sqrt{[5]D_2D_{-2}}}
&-{{\{A^2\}\over\{q\}}-1\over D_0\sqrt{[5]D_2D_{-2}}}&0&0\\
&&&&&&&&&\\
-{\sqrt{D_{-1}D_{-2}}\over D_0\sqrt{[4]}}&0&{\sqrt{D_{1}D_{2}}\over D_0\sqrt{[4]}}
&{\sqrt{D_{-3}D_{-2}}\over D_0\sqrt{[4][5]}}&0&-{\sqrt{D_{3}D_{2}}\over D_0\sqrt{[4][5]}}
&{\sqrt{D_{2}D_{-2}}\over D_0\sqrt{[5]}}&-{\sqrt{D_{2}D_{-2}}\over D_0\sqrt{[5]}}&0&0\\
&&&&&&&&&\\
{\sqrt{D_{-1}D_2}\over [2]D_0\sqrt{[4]}}&-{\sqrt{[3]D_3}\over [2]\sqrt{[6]D_0}}&{\sqrt{D_1D_{-2}}\over [2]D_0\sqrt{[4]}}
&-{\sqrt{D_{-3}D_2}\over [2]D_0\sqrt{[4][5]}}&-{\sqrt{[3]D_{-3}}\over [2]\sqrt{[6]D_0}}&-{\sqrt{D_3D_{-2}}\over [2]D_0\sqrt{[4][5]}}
&{[6]D_0+[3](D_{-2}-D_2)\over 2[2][3]D_0\sqrt{[5]}}&{[6]D_0+[3](D_{2}-D_{-2})\over 2[2][3]D_0\sqrt{[5]}}&{1\over 2}&-{1\over 2}\\
&&&&&&&&&\\
{\sqrt{D_{-1}D_2}\over [2]D_0\sqrt{[4]}}&-{\sqrt{[3]D_3}\over [2]\sqrt{[6]D_0}}&{\sqrt{D_1D_{-2}}\over [2]D_0\sqrt{[4]}}
&-{\sqrt{D_{-3}D_2}\over [2]D_0\sqrt{[4][5]}}&-{\sqrt{[3]D_{-3}}\over [2]\sqrt{[6]D_0}}&-{\sqrt{D_3D_{-2}}\over [2]D_0\sqrt{[4][5]}}
&{[6]D_0+[3](D_{-2}-D_2)\over 2[2][3]D_0\sqrt{[5]}}&{[6]D_0+[3](D_{2}-D_{-2})\over 2[2][3]D_0\sqrt{[5]}}&-{1\over 2}&{1\over 2}
\end{array}\right)\nn
\ee

\bigskip

\be
S_{[4,5]}={[2][3]D_{-1}D_{-2}-[6]D_{-1}D_2-[3][4]D_1D_2\over [2]^2[4]\sqrt{[3][6]D_0D_{-1}D_{-2}D_2}}
\nn
\ee
\be\label{Sp21}
S_{[6,2]}=-{[2][3]D_{1}D_{2}-[6]D_{1}D_{-2}-[3][4]D_{-1}D_{-2}\over [2]^2[4]\sqrt{[3][6]D_0D_{1}D_{-2}D_2}}
\ee
\be
S_{[7,2]}=-\sqrt{D_3}{[2][3]D_{1}D_{2}-[6]D_{1}D_{-2}-[3][4]D_{-1}D_{-2}\over [2][4]\sqrt{[3][6]D_0^3D_{-2}D_2}}
\nn
\ee
\be
S_{[7,5]}=\sqrt{D_{-3}}{[2][3]D_{-1}D_{-2}-[6]D_{-1}D_2-[3][4]D_1D_2\over [2][4]\sqrt{[3][6]D_0^3D_{-2}D_2}}
\nn
\ee
}
\end{landscape}

\bigskip
\noindent
This matrix is obtained from that of \cite{GuJ} by transposition to
make it consistent with (\ref{firstline}) and thus with our
representation of knots.
Also minor (always allowed) conjugations are performed in
(\ref{Sa21}) and (\ref{Sp21}) to get rid
of unnecessary minuses and imaginary units.

Diagonalized ${\cal R}$-matrices $\bar T$ and $T$ are read off from the last columns of
(\ref{dims21a}) and (\ref{dims21p}):
\be
\overline T = \left(\begin{array}{cccccc|cccc}
1&&&&&&&&&\\
&-A^2&&&&&&&&\\
&&-A^2&&&&&&&\\
&&&-q^{-2}A^2&&&&&&\\
&&&&A^3&&&&&\\
&&&&&q^2A^2&&&&\\
\hline
&&&&&&A&&&\\
&&&&&&&-A&&\\
&&&&&&&&A&\\
&&&&&&&&& -A
\end{array}\right)
\label{Ta21}
\ee
and
\be
 T = \frac{1}{A^3}\left(\begin{array}{cccccc|cccc}
q^{-3}&&&&&&&&&\\
&-q^3&&&&&&&&\\
&&-q^3&&&&&&&\\
&&&-q^{-5}&&&&&&\\
&&&&q^{-3}&&&&&\\
&&&&&q^5&&&&\\
\hline
&&&&&&1&&&\\
&&&&&&&-1&&\\
&&&&&&&&-1&\\
&&&&&&&&& 1
\end{array}\right)
\label{Tp21}
\ee
In these $S$- and $T$-matrices the index $X$ runs from $1$ to $7$, and for $X=7$ there are
additional indices $a_i$ which take two values, which we substitute as $7_{11} \longrightarrow 7$, $7_{22}\longrightarrow 8$,
$7_{12}\longrightarrow 9$, $7_{21}\longrightarrow 10$, and $d_{10}=d_9=d_8=d_7$.
Then contractions in (\ref{basic}) for the simplest double-fat graph (\ref{pretzel})
can be rewritten as follows:
\be
2-{\rm finger\ case:}
&  \sum_{X=1}^7 \sum_{a,b=1}^{m_X} {A}_X^{ab} {B}_X^{ba}
&\longrightarrow\ \left(\sum_{X=1}^8 {\cal A}_X{\cal B}_X\right) +
\Big({\cal A}_9{\cal B}_{10} + {\cal A}_{10}{\cal B}_9\Big) \nn \\
&& \nn\\
3-{\rm finger\ case:}
&  \sum_{X=1}^7 \sum_{a,b,c =1}^{m_X}d_X^{-1/2}{A}_X^{ab} { B}_X^{bc}  { C}_X^{ca}
&\longrightarrow\ \left(\sum_{X=1}^8 d_X^{-1/2}{\cal A}_X{\cal B}_X{\cal C}_X\right) +
d_7^{-1/2}\Big({\cal A}_7{\cal B}_9{\cal C}_{10} + {\cal A}_8{\cal B}_{10}{\cal C}_9 + \nn \\
 && \ \ \ \ \ \ \ \ \ \ \  + {\cal B}_7{\cal C}_9{\cal A}_{10} + {\cal B}_8{\cal C}_{10}{\cal A}_9
 +{\cal C}_7{\cal A}_9{\cal B}_{10} + {\cal C}_8{\cal A}_{10}{\cal B}_9\Big)\nn \\
 && \nn \\
 \ldots \nn\\
{\rm in\ general}
&  \sum_{X=1}^7 d_X^{1-\frac{k}{2}}\prod^{\stackrel{k}{\leftarrow}}_{i=1} { A}_X^{(i)}
&\longrightarrow\ \left(\sum_{X=1}^8 d_X^{1-\frac{k}{2}}\prod_{i=1}^k {\cal A}_X^{(i)}\right) \ + \
\underline{d_7^{1-\frac{k}{2}} \sum_{i\neq j}^k{\cal A}^{(i)}_9{\cal A}^{(j)}_{10}\cdot\Big(\ldots\Big)}
 \label{acontras}
 \ee
It is the underlined term that breaks the permutation symmetry between different $A^{(i)}$
in the original product down to just cyclic symmetry, and it is the one that distinguishes mutants.
When all ${\cal A}^{(i)}_9$ or ${\cal A}^{(i)}_{10}$ are vanishing, the symmetry is preserved
and the mutants are not distinguished by the $[21]$-colored HOMFLY.

\section{$\sigma$-factors and other technicalities
\label{signs}}
We emphasize that the  eq.(\ref{basic}) is a kind of a conjecture/educated guess,
and was  well tested only for symmetric representations where no multiplicities
(indices $a,b,c,\ldots$) arise. Hence there is no reason to insist that contractions
(\ref{acontras}) are exactly correct.
In fact, they are not: for complicated enough knots, (\ref{basic}) with rule
(\ref{acontras}) does {\it not} produce {\it polynomials} in representation $[21]$
and, in result, fails to reproduce the known answers.
Both these problems are cured by switching to $\sigma$, which is allowed to take
the values $\pm 1$ only but not +1 always.

The procedure at the moment is to adjust these sign factors for a given knot so
that the answer for HOMFLY is a Laurent polynomial in $A$ and $q$.
The choice of sign is unique in some examples. We also observe that there
could more than one possible choice of sign giving same polynomial when one of the
fingers is pretzel of length one or  some matrix elements (${\cal A}^{(i)}_9$ or ${\cal A}^{(i)}_{10}$)
simply vanish.

No general rule is found yet for {\it a priori} determination of $\sigma$. We have
observed the sign pattern for some simple families of double fat diagrams we studied.
We present,  in the following section, the signs of $\sigma$ which we encountered
in obtaining polynomial invariants.

\subsection{Relevant choices}
\subsubsection{3 fingers}
In the simplest case of $3$ fingers we use the following notation:
\be
  \sum_{X=1}^7 \overline{\sum_{a,b,c =1}^{m_X}}d_X^{-1/2}{A}_X^{ab} { B}_X^{bc}  { C}_X^{ca}
 \longrightarrow  \ \  {\sum_{X=1}^6} d_X^{-1/2}{  A}_X{  B}_X{  C}_X +
d_7^{-1/2}{\sum_{a,b,c=1,2}} \sigma^{abc} \,{  A}_7^{ab}{  B}_7^{bc}{  C}_7^{ca}
\label{acontrasign}
 \ee
 First of all, it is always the case that $\sigma^{111}=\sigma^{222}=1$. Now,
it turns out that the two relevant sets of signs of the remaining $\sigma$'s,
appearing in the right formulas are either all equal to -1:
\be\label{Avariant}
\begin{array}{rl}
\hbox{variant \sl a}:&\ \ \ \ \ \sigma^{112}=\sigma^{211}=\sigma^{121}=\sigma^{122}=\sigma^{212}=\sigma^{221}=-1
\end{array}
\ee
or only two $\sigma^{abc}$ equal to -1. Depending on the order of fingers, it can be either $\sigma^{112}$ and $\sigma^{221}$ or two other pairs obtained by cyclic permutations from these. We will always choose the order in such a way that the first pattern is realized:
\be\label{Bvariant}
\begin{array}{rl}
\hbox{variant \sl b}:&\ \ \ \ \ \sigma^{112}=\sigma^{221}=-1\\
\end{array}
\ee
However, the choice between the two possibilities depends on the type of knot.
The following ideal option with all signs positive
\be
\hbox{variant \sl c}:~~~~~~\sigma^{112}=\sigma^{211}=\sigma^{121}=\sigma^{122}=\sigma^{212}=\sigma^{221}=+1
\label{allplus}
\ee
sometimes happens to provide right answers, but only when some of the
pretzel lengths are unit and identities (\ref{ramaid}) relate
(\ref{allplus}) to (\ref{Avariant})-(\ref{Bvariant}).

It deserves mentioning that (\ref{Bvariant}) explicitly breaks the cyclic symmetry
between $A$, $B$ and $C$. Thus one can expect that it is relevant
when such a symmetry is absent in 3-finger configuration. That is.,  if we have two parallel and
one antiparallel pretzel fingers. Having 3 parallel pretzel fingers is an impossible configuration to
obtain knots. However, we can have three antiparallel pretzel fingers
but  the off-diagonal matrix elements in  the second term of eq.\ref{acontrasign}
are  vanishing.

\subsubsection{More fingers\label{mf}}
For 4-finger parallel pretzel knots the relevant choice is
\be
\sigma^{2211} = \sigma^{1122}=\sigma^{2112} = \sigma^{1221} = -1, \ \ \ \
{\rm all \ other\ }\ \sigma^{abcd}=+1
\ee
For the 5-finger pretzel knot with one antiparallel finger (for the sake of definiteness, we take the first finger to be antiparallel pretzel ), the proper choice is
\be
\sigma^{21111}=\sigma^{22211}=\sigma^{22112}=\sigma^{21122}=\sigma^{21121}=\sigma^{21211}=\sigma^{12222}
=\sigma^{11122}=\sigma^{11221}=\nn\\=
\sigma^{12211}=\sigma^{12212}=\sigma^{12122}=-1,\ \ \ \ \ \ \ \
{\rm all \ other\ }\ \sigma^{abcd}=+1\hfill
\ee
Thus, we learn,
\begin{enumerate}
\item  always $\sigma^{11\dots 1}=\sigma^{22\ldots 2}=1$
\item  there is a symmetry of simultaneous replacement of  $1\leftrightarrow 2$ in all indices.
\item  Specifically for the parallel pretzel knots (which exist only for  even number of fingers),
there is also an additional symmetry: all sign factors obtained by a cyclic permutation are the same.
\end{enumerate}
\subsection{Sign dependence on knots
\label{signdep}}
This list of relevant sign choices
is not necessarily complete,
but it is necessary and sufficient for all the examples that we
studied so far.
With these sign prescriptions, $H_{[21]}$ for knots are {\it polynomials}
(otherwise denominators can occur) and coincide with the previously
known answers from \cite{Ano21,AnoMcabling,GuJ,MMM21}.

Among knots belonging to starfish configurations,  the 3-finger configurations we checked  are described by (\ref{Bvariant}) (with the  third finger chosen antiparallel but the first two fingers
taken parallel) except knot $10_{71}$ where  (\ref{Avariant}) choice is used. Similarly the 4 parallel finger knots
and 5-finger (one antiparallel + 4 parallel) configurations knots satisfy the rules described in sec.\ref{mf}.

As for the double sum configurations, due to propagators connecting two $k$-valent
vertices (see Appendix), $10_{152}$ is described by (\ref{Bvariant}) in the both sums, $10_{153}$ is described by (\ref{Bvariant}) in the sum over pretzel fingers and (\ref{Avariant}) in the second sum.
The knot $10_{154}$ is described by (\ref{Avariant}) in the both sums.

Finally,  the  double fat configurations involving three summation variables (\ref{Mpqmrs}) is described by (\ref{Avariant}) in the internal sum in (\ref{mutfam}) and (\ref{Bvariant}) in the two sums within the ppS-blocks.

\subsection{Explicit calculations in the pretzel case}
After the matrices $S$, $ T$ and the sign factors $\sigma$
are explicitly known, it is straightforward to insert them into
formulas of sec.\ref{exapols} and obtain the colored HOMFLY polynomials in representation $[21]$
for an enormous set of knots.
In fact, the answers themselves are huge, thus it makes no sense to list them all,
especially given the (large) number of examples.
Therefore, we  enumerate  the classes of diagrams, and in the Appendix we tabulate explicit examples.
Here, for illustrative purposes, we present some general discussion of the pretzel case. A vast set of concrete examples can be found again in the Appendix.

\subsubsection{Parallel}
The parallel pretzel finger is defined as
\be
{\cal A}^{\rm par}_{1X}({n}) = \Big(S {T}^n S^\dagger \Big)_{1X} \nn \\
\label{parfin21}
\ee
Unitarity of the $10\times 10$ matrix $S$ implies that
\be
{\cal A}^{\rm par}_{YX}(0) = \delta_{YX}, \ \ \ \ \
\ee
If instead of (\ref{parfin21}) we considered
$\widetilde{{\cal A}^{\rm par}}({n})_{1X} = \Big(S {T}^n \Lambda S^\dagger \Big)_{1X}
= \Big(S {T}^n S^\dagger\Lambda \Big)_{1X}$, where $\Lambda$ is the matrix of permutation of indices 9 and 10,
this property would be broken: $\widetilde{{\cal A}^{\rm par}}({0})_{YX} = \Lambda_{YX}$,
though the finger itself, at $Y=1$, would not be affected. This insertion of $\Lambda$ is just equivalent to the inverse order of indices of the right matrix $\bar S$: $T\Lambda \bar S=\sum_{X,c,d}T_X^{ab,cd}\bar S_{XZ}^{dc,ef}$ instead of $\sum_{X,c,d}T_X^{ab,cd}\bar S_{XZ}^{cd,ef}$.

Also vanishing are
\be
{\cal A}^{\rm par}_{1,9}(\pm 1) =  {\cal A}^{\rm par}_{1,10}(\pm 1) = 0
\ee
but only for unit braid length, $n=\pm 1$.

\bigskip

The simple checks and comments are now in order.

\bigskip

For the $2$-strand torus knots and links
\be
H_{[21]}^{[2,n]} = d_{[21]}\cdot {\cal A}^{\rm par}_{11}(-n)
\ee
and for the 4-parallel-finger pretzel knots, like
$\ \ 6_3\ (2,-3,1,1)$,\ \  $7_3\ (4,1,1,1)$,\ \  $7_5\ (3,2,1,1)$:
\be
H_{[21]}^{{\rm Pr}(n_1,n_2,n_3,n_4)} =
\overline{\sum_X} \frac{d_{[21]}^3}{d_X}\, {\cal A}_{1X}^{\rm par}(n_1)
{\cal A}_{1X}^{\rm par}(n_2){\cal A}_{1X}^{\rm par}(n_3){\cal A}_{1X}^{\rm par}(n_4)
\ee

In the latter (4-finger) case the off-diagonal terms in (\ref{acontras})
is non-vanishing, moreover, it is not a polynomial
and cancels the non-polynomial part of the diagonal terms.
For example, in the case of $6_3$ the off-diagonal contribution is
\be
{\rm off-diagonal\ contribution\ to}\ 6_3: \ \ \ \
-2\{q\}^4 \cdot\frac{[6]}{[2]}\frac{D_2D_{-2}}{D_0}\cdot
\left(\frac{[8]}{[2]}\,Aq \ +\  A^2q^4D_{-4}\right)
\ee
It is proportional to the fourth power of $\{q\}=q-q^{-1}$ and therefore vanishes
in the limit $q=1$ (for special polynomials \cite{DMMSS,IMMMfe}).
The same happens for  knot  $7_5$.
However, for knot $7_3$, we observe that  the off-diagonal contribution vanishes.

The simplest pair of parallel pretzel mutants is $Pr(3,3,3,2)$ and $Pr(3,3,2,3)$. We discuss this example, along with other mutants pairs of knots, in a separate section ss.\ref{Mutants}.

\subsubsection{Even antiparallel}
This finger depends on the even parameter $n$ and is given by the matrix element
\be
{\cal A}_{1X}^{\rm ea}({\bar n}) = \Big(\bar S \bar  T^n \bar S^\dagger\Big)_{1X}
\ee
Again, the unitarity ${\cal A}_{YX}(\bar 0) = \delta_{YX}$ makes this choice natural as compared with
those with additional insertions of $\Lambda$.

\bigskip
For $n=0$, this matrix element is just $\delta_{X,1}$.

For $n=\pm 2$, the matrix elements of ${\cal A}_{1X}^{\rm ea}$ with $X=9,10$ are very simple, similarly to the parallel case:
\be
X=9 & \pm \{q\}\cdot  \frac{[6]A^{\pm 4}}{[2]D_0}\sqrt{\frac{D_2D_{-2}}{D_1D_{-1}}} \nn\\
X=10 & \mp \{q\}\cdot  \frac{[6]A^{\pm 4}}{[2]D_0}\sqrt{\frac{D_2D_{-2}}{D_1D_{-1}}}
\ee

The simplest check to be made is for the Hopf link, which can be represented both
through parallel and antiparallel braid:
\be
H_{[21]}^{[2,2]} = {\cal A}_{11}^{\rm par}(2) = {\cal A}_{11}^{\rm ea}(-2)={1\over A^9q^{10}}{1\over [3]\{q\}^3}\Big(
A^6q^{20}-2A^6q^{18}-A^4q^{20}+3A^6q^{16}+A^4q^{18}-\nn\\-4A^6q^{14}-2A^4q^{16}
+5 A^6 q^{12}+3 A^4 q^{14}+A^2 q^{16}-5 A^6 q^{10}-4 A^4 q^{12}+5 A^6 q^8+3 A^4 q^{10}+A^2 q^{12}-\nn\\
-4 A^6 q^6-4 A^4 q^8-A^2 q^{10}+3 A^6 q^4+3 A^4 q^6+A^2 q^8-q^{10}-2 A^6 q^2-2 A^4 q^4+A^6+A^4 q^2+A^2q^4-A^4\Big)
\ee
This is the reduced HOMFLY invariant, and since it describes a link, it is not a polynomial.

Any combinations of even antiparallel fingers only provide links, not knots.
The simplest knot family involving such fingers is $Pr(n_1,n_2,\bar m)$, where
two of the three fingers are parallel. Then
\be
H_{[21]}^{Pr(n_1,n_2,\bar m)} = \overline{\sum_X} \frac{d_{[21]}^2}{\sqrt{d_X}}
{\cal A}^{\rm par}_{1X}(n_1){\cal A}_{1X}^{\rm par}(n_2){\cal A}_{1X}^{\rm ea}(\bar m)
\label{Prppa}
\ee
This family includes knots $\ \ 6_2 \ \ (3,1,\bar 2)$, \ \ $8_1\ \ (1,1,\bar 6)$, \ \ $8_2 \ \ (5,1,\bar 2)$,
$\ \ 8_4\ \ (3,1,\bar 4)$, $ \ \ 8_5\ \ (3,3,\bar 2)$, $\ \ 8_{19}\ \ (3,3,-\bar 2)$, $\ \ 8_{20}\ \ (3,-3,\bar 2)$, \ \ $\ldots$
which are all 3-strand and one can check that (\ref{Prppa}) reproduces the known
answers from \cite{AnoMcabling}.

Similarly, the five-finger family
\be
H_{[21]}^{Pr(n_1,n_2,n_3,n_4,\bar m)} = \overline{\sum_X} \frac{d_{[21]}^4}{{d_X}^{3/2}}
{\cal A}^{\rm par}_{1X}(n_1){\cal A}_{1X}^{\rm par}(n_2)
{\cal A}^{\rm par}_{1X}(n_3){\cal A}_{1X}^{\rm par}(n_4){\cal A}_{1X}^{\rm ea}(\bar m)
\label{Prppppa}
\ee
contains \ $7_6\ (1,1,-3,1,\bar 2)$,\  $8_3\ (1,1,1,1,\bar 4)$,\ $8_6\ (1,1,1,3,\bar 2)$, \ $\ldots$ \
At least for these three knots (\ref{Prppppa}), we
get  {\it polynomials} and they satisfy
the consistency checks putforth in sec.\ref{tests}.

\subsubsection{Odd antiparallel\label{oapr21}}
This finger depends on the odd parameter $n$ and is given by the matrix element
\be
{\cal A}^{\rm oa}_{1X}({\bar n}) = \Big(\bar S^\dagger \bar  T^n S\Big)_{1X}
\ee
This time $n$ is odd and cannot vanish, therefore there is no obstacle for inserting $\Lambda$ and writing   $ \Big(\bar S^\dagger T^n\Lambda S\Big)_{_{1X}}$ instead.  However, this does not actually affect the answers, at least in the cases discussed here.

\bigskip
Elementary calculation or application of identity (\ref{SbTbS}) gives that at $n=\pm 1$
\be
{\cal A}_{1X}^{\rm oa}(\pm 1) =  \left\{
\begin{array}{ccc} \frac{\sqrt{d_X}}{d_{[21]}}\cdot  T^{\,\mp 1}_{XX} &\ \ & X=1,\ldots,8 \\
0 && X=9,10 \end{array} \right.
\ee
where $X$ belongs to the parallel sector, $X\in R\otimes R$.
Once again, this immediately provides the answers for the trefoil:
\be
\sum_{X\in [21]\otimes [21]} \frac{d_{[21]}^2}{\sqrt{d_X}}\cdot \Big( {\cal A}_{1X}^{\rm oa}(1)\Big)^3 =
\frac{1}{d_{[21]}} \sum_{X\in [21]\otimes [21]} d_X^2  T_{XX}^{-3}
\ee
and for the unknot
\be
\sum_{X\in [21]\otimes [21]} \frac{d_{[21]}^2}{\sqrt{d_X}}\cdot
 {\cal A}_{1X}^{\rm oa}(-1) \Big( {\cal A}_{1X}^{\rm oa}(1)\Big)^2 =
\frac{1}{d_{[21]}} \sum_{X\in [21]\otimes [21]} d_X^2  T_{XX}^{-1} = 1
\ee
For other values of $n$ these matrix elements are multiplied by polynomials:
\be
{\cal A}^{\rm oa}_{1X}({\pm\bar n})  =
{\cal A}^{\rm oa}_{1X}({\pm\bar 1})  \cdot {\cal P}^{\rm oa}_X\big(n\big|A^{\mp 1},q^{\mp 1}\big)
\ee
In particular, this means that the odd antiparallel fingers vanish at $X=9,10$ for all values $n$,
which means that
{\it odd antiparallel pretzel mutants are not distinguished by the colored HOMFLY polynomials in representation $[21]$}.

The polynomials ${\cal P}$ themselves are, however, somewhat involved. For instance,
\be
{\cal P}_1(3) = {1\over q^{10}}\Big(
-A^4 q^{16}+A^4 q^{14}+A^6 q^{10}-A^2 q^{14}+2 A^4 q^{10}+2 A^2 q^{12}-A^4 q^8-\nn\\-A^2 q^{10}-q^{12}+2 A^2 q^8+2 q^{10}-A^4 q^4-3 A^2 q^6-q^8+A^2 q^4+q^6-A^2 q^2-2 q^4+A^2+q^2\Big)
\ee
The twist knots are pretzel $Tw_k = Pr(\bar 1,\bar 1, \overline{2k-1})$
i.e. are made from three odd antiparallel fingers:
\be
H_{[21]}^{Pr(\bar n_1,\bar n_2, \bar n_3)} =
\sum_{X=1}^8 \frac{d_{[21]}^2}{\sqrt{d_X}}\, {\cal A}^{\rm oa}_{1X}(\bar n_1)
{\cal A}^{\rm oa}_{1X}(\bar n_2)  {\cal A}^{\rm oa}_{1X}(\bar n_3)
\ee
The off-diagonal terms in (\ref{acontras}) vanish for odd antiparallel fingers implying that
there is no difference between $\sum_{X=1}^8$ and $\overline{\sum_X}$.
The answers for twist knots agree with  those in Refs. \cite{AnoMcabling,GuJ,MMM21}.
Other members of the same three-finger  family are: $\ \ 7_4 \ \ (\bar 3,\bar 3,\bar 1)$, $\ \ 9_5 \ \ (-\bar 1,-\bar 3,-\bar 5)$, $\ \ 9_{35} \ \ (\bar 3,\bar 3,\bar 3)$, $\ \ 9_{46} \ \ (\bar 3,-\bar 3,\bar 3)$, $\ \ 10_3 \ \ (\bar 1,\bar 5,-\bar 5)$, ...

\subsection{Checks
\label{tests}}
All answers for the HOMFLY polynomials in representation $[21]$ in this paper for the knots that have a three braid representation were compared with the results obtained by the cabling method of \cite{Ano21,AnoMcabling}. Besides, there are five types of self-consistency checks one can
putforth to  test our answers. The first four are true for all our examples,
the fifth one is more tedious and was only partly verified.
\subsubsection{Special polynomials}
As conjectured in \cite{DMMSS,IMMMfe} and proved in \cite{Zhu},
the "special" polynomials, i.e. reduced HOMFLY polynomials at $q=1$,
obey the factorization rule
\be
H_R^{\cal K}(q=1,A) = \Big(H_{_\Box}^{\cal K}(q=1,A)\Big)^{|R|}
\label{spepofac}
\ee
i.e. are unambiguously restored from those in the fundamental representation
$R=[1]= \phantom._\Box$.
The size $|R|$ of the Young diagram $R$ is just the
number of boxes, $|[21]|=3$.

For superpolynomials this factorization gets far more interesting,
see \cite{Che,Anton,GGS} and \cite{Ano21}.

\subsubsection{Alexander polynomials for 1-hook representations}

As conjectured in \cite{IMMMfe} a "dual" factorization property holds
at $A=1$, i.e. for the Alexander polynomials, but only for representations $R$ described by
single-hook diagrams:
\be
H_R^{\cal K}(A=1,q) = H_{_\Box}^{\cal K}(A=1,q^{|R|})
\ee
Representation $[21]$ belongs to this class.

Neither any meaning, nor generalizations of this
remarkable factorization property is currently available.

\subsubsection{Jones polynomials}
For $SU_q(2)$, i.e. for $A=q^2$ all two-line Young diagrams $[r_1r_2]$
get equivalent to the single-line $[r_1-r_2]$, in particular, $[21]$ gets
indistinguishable from $[1]$.
This means that
\be
H_{[21]}^{\cal K}(A=q^2) = H_{_\Box}^{\cal K}(A=q^2) = {\rm Jones}^{\cal K}
\ee
\subsubsection{The weak form of differential expansion}
Expressions for colored knot polynomials are extremely complicated,
but in fact they have a lot of hidden structure and satisfy a lot
of non-trivial relations.
Understanding of these structures nicknamed "differential expansions"
\cite{IMMMfe,evo,ArthMM}
because their first traces were observed in \cite{DGR} devoted
to the study of Khovanov-Rozansky "differentials" is still very poor.
In its weakest possible form, the differential expansion conjecture
implies for representation $[21]$ that \cite{Ano21,MMM21}
\be
H_{[21]}^{\cal K}(A,q) = 1 + F_1^{\cal K}(A^2,q^2)\Big(\{Aq^3\}\{A/q^3\}
+ \{Aq^2\}\{A\} + \{A\}\{A/q^2\}\Big) +   \{Aq^2\}\{A/q^2\}\cdot G_{[21]}^{\cal K}(A^2,q^2)
\label{diffexpan21}
\ee
with {\it some} (complicated) function $G_{[21]}^{\cal K}$,
where $F_1^{\cal K}(A^2,q^2)= F_1^{\cal K}(A^2,q^{-2})$ is the coefficient in
\be
H_{_\Box}^{\cal K}(A,q) = 1 + F_1^{\cal K}(A^2,q^2)\{Aq\}\{A/q\}
\ee
and $H_R$ are the {\it reduced} HOMFLY polynomials.
In other words certain linear combination of $H_{[21]}$ and $H_{[1]}$ should
vanish at $A=q^{\pm 2}$, i.e. in the case of $SU_q(2)$.
\subsubsection{Quasiclassical expansion and Vassiliev invariants}
Differential expansions from the previous paragraph are examples of
cleverly-structured quasiclassical expansions in parameters like $\hbar$,
where $q=e^{\hbar}$ and $A=q^{N\hbar}$ or $z=\{q\} = q-1/q$.
The most structured expansion of this type, known so far is
the Hurwitz-style formula \cite{Sle} for reduced colored HOMFLY:
\be
H_R^{\cal K}(q=e^\hbar,A) = \Big(H_{_\Box}(q=1,A)\Big)^{|R|}\cdot
\exp\left(\sum_{\Delta}\hbar^{|\Delta|+l(\Delta)-2+k}\Sigma^{{\cal K}}_{\Delta,k} (A)\varphi_R(\Delta) \right)
\ee
Here $\varphi_R(\Delta)$ are characters of the universal symmetric group
(defined as in \cite{MMN}), $l(\Delta)$ is the number of lines in the Young diagram $\Delta$
and the coefficients of generalized special polynomials $\Sigma^{{\cal K}}_{\Delta,k} (A)$
are made from the Vassiliev invariants.

This formula includes (\ref{spepofac}) as a particular case and is also closely
related to (\ref{diffexpan21}).
It imposes non-trivial restrictions on the coefficients of colored polynomials.
As already mentioned, we made only a few simple checks of our answers with this
formula, and it is very interesting to extend them in order to understand
what are the really independent "degrees of freedom" in $H_{[21]}$,
as compared to those already captured by the colored HOMFLY polynomials in symmetric representations.
This study can be also helpful for superpolynomial extensions of the $[21]$-colored
knot polynomials, which still remains a complete mystery.

\section{Mutants\label{Mutants}}
Of special interest among the newly available answers are those for
the pairs of mutants, i.e. knots related by the mutation transformation,
which are inseparable by knot polynomials in symmetric representations.

\subsection{Generalities}
Mutation in knot theory is the transformation of link diagram, when one cuts
a sub-diagram with exactly four external legs, rotate and glue it back to
the original position.
Within the Reshetikhin-Turaev approach, it is clear that cutting corresponds to decomposition of
knot polynomial in the channel $R_1\otimes R_2$ and mutation is a rotation
in the spaces of intertwining operators $W^Q_{R_1R_2}: R_1\otimes R_2\rightarrow Q$.
If these spaces are one-dimensional, like in the case of rectangular
representations $R_1=R_2$ or $\bar R_1=R_2$, the mutation does not affect
the corresponding HOMFLY polynomial.
Thus, mutants can be distinguished by colored HOMFLY polynomials in non-rectangular
representations, the first of which is $R=[21]$.
The difference
\be
\boxed{
H_{[21]}^{{\rm mutant}_1} -H_{[21]}^{{\rm mutant}_2} =
\{q\}^{11} \cdot D_3^2D_2D_0D_{-2}D_{-3}^2
\cdot A^{\gamma}\cdot M_{[21]}^{\rm mutant}(q^2)
= }
\label{mutdiff}
\ee
$$
= \{q\}^4\cdot \{Aq^3\}^2\{Aq^2\}\{A\}\{A/q^2\}\{A/q^3\}^2 \cdot
A^{\gamma}\cdot M_{[21]}^{\rm mutant}(q^2)
$$
has the universal prefactor. The differentials
$D_{\pm 2}$ and $D_0$ appear in it because the
Alexander and the Jones polynomials at $A=1$ and $A=q^2$ are not affected
by the mutation and, thus, the difference should vanish for $SU_q(N)$ with $N=0$ and $N=2$.
The Young diagram is symmetric under row-column interchange (also called rank-level duality
\cite{DMMSS,GS,IMMMfe}). Hence
the polynomial in variables $A$ and $q$ must remain same when we change $q\rightarrow
q^{-1}$ and, hence, the difference should vanish also at $A=q^{-2}$.
Vanishing for $SU_q(3)$ follows from a more involved argument of ref.\cite{Mort}.
There are no factors $D_1D_{-1}$, because we consider the {\it reduced} HOMFLY polynomial,
which is equal to the original unreduced one divided by $d_{21} = \frac{D_1D_0D_{-1}}{[3]}$.
The additional $A$-independent factor $\{q\}^4$ seems to be typical for all
non-diagonal terms considered in this paper, and this provides an explanation
for power $11$ in (\ref{mutdiff}).
In all examples, which we managed to analyze, $M^{\rm mutant}(q^2)$ is, indeed, a function of $q$ only,
with no $A$-dependence.
\subsection{Examples of mutants}
The mutant pairs appear for the first time
for knots with eleven intersections, where there are
$16$ pairs (n\&a in the knots denote alternating or non-alternating):
\be\label{mutlist}
11a19 & 11a25&  \nn \\
11a24 & 11a26&  \nn \\
11a44 & 11a47& {\bf Pr}(-3,3,2,1,1,-3)\ \& \ {\bf Pr}(3,-3,2,1,1,-3) \nn \\
11a57 & 11a231& {\bf Pr}(2,1,3,3,-3)\ \& \ {\bf Pr}(2,1,3,-3,3) \nn \\
11a251 & 11a253&  \nn \\
11a252 & 11a254&  \nn \\
\nn \\
11n34 & 11n42 &  (3, -2|2|-3, 2)\ \&\ (3,-2|2|2,-3) \ \ \ \ {\bf KTC \ mutants} \nn \\
11n35 & 11n43 &  \nn \\
11n36 & 11n44 & (3, 2|2|-3, 2)\ \&\ (3,2|2|2,-3)  \nn \\
11n39 & 11n45 & (3, -2|2|3, -2)\ \&\ (3,-2|2|-2,3) \nn \\
11n40 & 11n46 &  \nn \\
11n41 & 11n47 & (3, -2|2|3, 2)\ \&\ (3,-2|2|2,3) \nn \\
11n71 & 11n75 & {\bf Pr}(2,-3,3,-3,1)\ \& \ {\bf Pr}(2,3,-3,-3,1)  \nn \\
11n73 & 11n74 & {\bf Pr}(-2,3,3,-3) \ \& \ {\bf Pr}(-2,3,-3,3)  \nn \\
11n76 & 11n78 & {\bf Pr}(2,3,3,-3) \ \& \ {\bf Pr}(2,3,-3,3)   \nn \\
11n151 & 11n152 & (3, -2|-2|3, -2)\ \&\ (3,-2|-2|-2,3)
\ee
Of these, the KTC pair is most famous,  probably  because their Alexander polynomials are just unity.  The simplest mutants are the pretzel ones: there are five such pairs
in the above list, see also eq.(\ref{mu3332}) below. For enumeration of mutants with up to 16 intersections see \cite{stoi}.

\subsection{HOMFLY for the Pretzel mutants\label{pmu}}
First of all, we repeat that $H_{[21]}$ {\bf do not distinguish pretzel mutants
made from antiparallel pretzel fingers of odd lengths}, see sec.\ref{oapr21} above.
The simplest example of this kind are 12-intersection pair of mutant links $Pr(\bar3, \bar3,-\bar 3,-\bar 3)\  \& \ Pr(\bar 3,-\bar 3,\bar 3,-\bar 3)$ and 15-intersection pair of mutant knots  $Pr(\bar3, \bar3,\bar 3,-\bar 3,-\bar 3)\  \& \ Pr(\bar 3,\bar 3,-\bar 3,\bar 3,-\bar 3)$. In both cases the links/knots in the pair seem topologically different, but
\be
H_{[21]}^{Pr(\bar 3,\bar 3,\bar 3,-\bar 3,-\bar3)} - H_{[21]}^{Pr(\bar 3,\bar 3, -\bar 3,\bar3,-\bar 3)}
= 0
\ee
and so on.

\bigskip

For the pretzel parallel fingers the situation is better.
The simplest parallel finger pretzel mutants are made from four fingers of lengths $3,3,3,2$ and
therefore contain $11$ intersections.
Of eight potential mutant pairs in this set there are actually only two pairs with distinct polynomials.
In fact, pretzel knot  $Pr(3,3,3,\pm 2)=Pr(3,3,\pm 2,3)$ because of the cyclic symmetry, $Pr(3,3,-3,\pm 2|A,q) = Pr(3,-3,-3,\mp 2|A^{-1},q)$,
and, finally, $Pr(3,-3,-3,\pm 2|A,q) = Pr(3,3,-3,\mp 2|A^{-1},q^{-1})$ if one reads the sequence of
lengths in the opposite direction and applies the cyclic symmetry.
This leaves two pairs of 11-intersection mutants, which are pretzels with four parallel fingers:
$11n73 \ \&\ 11n74$ ($Pr(-2,3,3,-3) \ \& \ Pr(-2,3,-3,3)$)
and
$11n76\ \&\ 11n78$ ($Pr(2,3,3,-3) \ \& \ Pr(2,3,-3,3)$).
For them
\be
H_{[21]}^{Pr(-2,3,-3,3)}-H_{[21]}^{Pr(-2,3,3,-3)}
= A^{-3}\cdot\frac{[3]^2[8][14]}{[2]^2[7]}\cdot
\{q\}^4 \{Aq^3\}^3\{Aq^2\}\{A\}\{A/q^2\}\{A/q^3\}^2, \nn \\
H_{[21]}^{Pr(2,3,3,-3)} - H_{[21]}^{Pr(2,3,-3,3)}
=  A^{-15}\cdot\frac{[3]^2[8]}{[2]}\cdot
\{q\}^4 \{Aq^3\}^3\{Aq^2\}\{A\}\{A/q^2\}\{A/q^3\}^2
\label{mu3332}
\ee
One can easily analyze any other mutant pair of this kind, for example:
\be
H_{[21]}^{Pr(9,3,2,5)} \!\!- H_{[21]}^{Pr(9,3,5,2)} =
-A^{-57}\!\cdot\!\frac{[6]^2[8][10][11]}{[2]^3[4]}\left(\frac{[14]}{[2]}-\frac{[12]}{[6]}\right)
\cdot \{q\}^4 \{Aq^3\}^3\{Aq^2\}\{A\}\{A/q^2\}\{A/q^3\}^2
\label{mu9325}
\ee
In fact, the pretzel mutant pairs provide us with a much better set of examples than concrete knots like the KTC pair: one can easily
construct whole families of mutants. An example of such a family we consider in the next section.

Note that among the pretzel mutant pairs there are more knots with 11 intersections, though their pretzel presentation has more
intersections (12 and even 13). These are $11n71\ \&\ 11n75$ ($Pr(2,-3,3,-3,1)\ \& \ Pr(2,3,-3,-3,1)$),  $11a44\ \&\ 11a47$
($Pr(-3,3,2,1,1,-3)\ \& \ Pr(3,-3,2,1,1,-3)$) and $11a57\ \&\ 11a231$ ($Pr(2,1,3,3,-3)\ \& \ Pr(2,1,3,-3,3)$). The
differences of their HOMFLY are:
\be
H_{[21]}^{Pr(2,-3,3,-3,1)}-H_{[21]}^{Pr(2,3,-3,-3,1)}=A^{13}\frac{[3]^2[8]}{[2]}\cdot
\{q\}^4 \{Aq^3\}^3\{Aq^2\}\{A\}\{A/q^2\}\{A/q^3\}^2\nn\\
H_{[21]}^{Pr(2,1,3,-3,3)}-H_{[21]}^{Pr(2,1,3,3,-3)}= A^{-5}\frac{[3]^2[8][14]}{[2]^2[7]}\cdot
\{q\}^4 \{Aq^3\}^3\{Aq^2\}\{A\}\{A/q^2\}\{A/q^3\}^2\\
H_{[21]}^{Pr(-3,3,2,1,1,-3)}-H_{[21]}^{Pr(3,-3,2,1,1,-3)}=A \frac{[3]^2[8][14]}{[2]^2[7]}\cdot
\{q\}^4 \{Aq^3\}^3\{Aq^2\}\{A\}\{A/q^2\}\{A/q^3\}^2 \nn
\ee
\subsection{KTC mutants}
\label{KTCflas}

The Kinoshita-Terasaka ($11n42$) and Conway ($11n34$) knots (KTC mutants)
are respectively
$M(3,-2|2|-3,2)$ and $M(3-2|2|2,-3)$.
Both knots have representations with 11-intersection, but for our purposes the realization with 12 intersections
is more convenient, see Figure 3.

These diagrams are already easy to bring to the form (\ref{mutfam}) with $(p,q,r,s)$ = $(3,-2,-3,2)$ and $(3,-2,2,-3)$ for $11n42$ and $11n34$ respectively.
Their HOMFLY are:\footnote{
We use here the notation from \cite{NRZ2}.
Matrix lists
the coefficients of a polynomial in $A^2$ and $q^2$
by the following rule:

\setlength{\arraycolsep}{3pt}

\begin{equation}
\nonumber q^{10}A^{16} \left(\begin{array}{rr}
3 & 4 \\
& \\
1 & 2 \\
\end{array}\right)
= q^{10}A^{16}+2q^{12}A^{16}+3q^{10}A^{18}+4q^{12}A^{18}.
\end{equation}
}

\newpage

\begin{landscape}
\be
H_{[21]}^{11n34} =
\frac{1}{q^{36}A^{6}} \cdot
\ee

\bigskip

\centerline{{\tiny
$
\setlength{\arraycolsep}{1pt}
\left(\begin{array}{rrrrrrrrrrrrrrrrrrrrrrrrrrrrrrrrrrrrr}
0 & 0 & 0 & 0 & 0 & 0 & 0 & 0 & 1 & -2 & 5 & -9 & 14 & -17 & 22 & -25 &
29 & -29 & 30 & -29 & 29 & -25 & 22 & -17 & 14 & -9 & 5 & -2 & 1 & 0 & 0
& 0 & 0 & 0 & 0 & 0 & 0 \\
&&&&&&&&&&&&&&&&&&&&&&&&&&&&&&&&&&&& \\
0 & 0 & 0 & 0 & 0 & -1 & 0 & -2 & 0 & 3 & -18 & 28 & -61 & 94 & -144 &
178 & -226 & 245 & -264 & 245 & -226 & 178 & -144 & 94 & -61 & 28 & -18
& 3 & 0 & -2 & 0 & -1 & 0 & 0 & 0 & 0 & 0 \\
&&&&&&&&&&&&&&&&&&&&&&&&&&&&&&&&&&&& \\
0 & 0 & 0 & 2 & -2 & 9 & -13 & 29 & -30 & 57 & -52 & 84 & -77 & 121 &
-107 & 169 & -165 & 213 & -176 & 213 & -165 & 169 & -107 & 121 & -77 &
84 & -52 & 57 & -30 & 29 & -13 & 9 & -2 & 2 & 0 & 0 & 0 \\
&&&&&&&&&&&&&&&&&&&&&&&&&&&&&&&&&&&& \\
0 & -1 & 0 & -4 & 3 & -16 & 16 & -56 & 60 & -130 & 144 & -250 & 239 &
-356 & 327 & -431 & 351 & -452 & 368 & -452 & 351 & -431 & 327 & -356 &
239 & -250 & 144 & -130 & 60 & -56 & 16 & -16 & 3 & -4 & 0 & -1 & 0 \\
&&&&&&&&&&&&&&&&&&&&&&&&&&&&&&&&&&&& \\
1 & -1 & 6 & -8 & 24 & -21 & 55 & -49 & 105 & -79 & 183 & -157 & 307 &
-275 & 488 & -446 & 662 & -567 & 738 & -567 & 662 & -446 & 488 & -275 &
307 & -157 & 183 & -79 & 105 & -49 & 55 & -21 & 24 & -8 & 6 & -1 & 1 \\
&&&&&&&&&&&&&&&&&&&&&&&&&&&&&&&&&&&& \\
-1 & 1 & -6 & 8 & -24 & 22 & -54 & 46 & -105 & 80 & -185 & 168 & -347 &
332 & -574 & 547 & -798 & 701 & -888 & 701 & -798 & 547 & -574 & 332 &
-347 & 168 & -185 & 80 & -105 & 46 & -54 & 22 & -24 & 8 & -6 & 1 & -1 \\
&&&&&&&&&&&&&&&&&&&&&&&&&&&&&&&&&&&& \\
0 & 1 & 0 & 4 & -3 & 15 & -16 & 57 & -61 & 131 & -142 & 248 & -212 & 309
& -229 & 311 & -170 & 263 & -141 & 263 & -170 & 311 & -229 & 309 & -212
& 248 & -142 & 131 & -61 & 57 & -16 & 15 & -3 & 4 & 0 & 1 & 0 \\
&&&&&&&&&&&&&&&&&&&&&&&&&&&&&&&&&&&& \\
0 & 0 & 0 & -2 & 2 & -9 & 12 & -27 & 30 & -58 & 51 & -89 & 85 & -133 &
102 & -163 & 137 & -186 & 130 & -186 & 137 & -163 & 102 & -133 & 85 &
-89 & 51 & -58 & 30 & -27 & 12 & -9 & 2 & -2 & 0 & 0 & 0 \\
&&&&&&&&&&&&&&&&&&&&&&&&&&&&&&&&&&&& \\
0 & 0 & 0 & 0 & 0 & 1 & 0 & 2 & 1 & -4 & 19 & -32 & 67 & -95 & 142 &
-172 & 218 & -228 & 246 & -228 & 218 & -172 & 142 & -95 & 67 & -32 & 19
& -4 & 1 & 2 & 0 & 1 & 0 & 0 & 0 & 0 & 0 \\
&&&&&&&&&&&&&&&&&&&&&&&&&&&&&&&&&&&& \\
0 & 0 & 0 & 0 & 0 & 0 & 0 & 0 & -1 & 2 & -5 & 9 & -15 & 20 & -27 & 32 &
-38 & 40 & -42 & 40 & -38 & 32 & -27 & 20 & -15 & 9 & -5 & 2 & -1 & 0 &
0 & 0 & 0 & 0 & 0 & 0 & 0
\setlength{\arraycolsep}{6pt}
\end{array}\right)
$
}}

\bigskip

\be
H_{[21]}^{11n42} =
\frac{1}{q^{34}A^{6}}   \cdot
\ee

\bigskip

\centerline{{\tiny
$
\setlength{\arraycolsep}{1pt}
\left(\begin{array}{rrrrrrrrrrrrrrrrrrrrrrrrrrrrrrrrrrr}
0 & 0 & 0 & 0 & 0 & 0 & 0 & 1 & -2 & 5 & -9 & 14 & -17 & 22 & -25 & 29 &
-29 & 30 & -29 & 29 & -25 & 22 & -17 & 14 & -9 & 5 & -2 & 1 & 0 & 0 & 0
& 0 & 0 & 0 & 0 \\
&&&&&&&&&&&&&&&&&&&&&&&&&&&&&&&&&& \\
0 & 0 & 0 & 0 & -1 & 0 & -2 & 1 & -1 & -10 & 14 & -37 & 58 & -95 & 116 &
-154 & 165 & -180 & 165 & -154 & 116 & -95 & 58 & -37 & 14 & -10 & -1 &
1 & -2 & 0 & -1 & 0 & 0 & 0 & 0 \\
&&&&&&&&&&&&&&&&&&&&&&&&&&&&&&&&&& \\
0 & 0 & 2 & -2 & 7 & -6 & 17 & -11 & 27 & -13 & 38 & -26 & 73 & -64 &
131 & -138 & 193 & -156 & 193 & -138 & 131 & -64 & 73 & -26 & 38 & -13 &
27 & -11 & 17 & -6 & 7 & -2 & 2 & 0 & 0 \\
&&&&&&&&&&&&&&&&&&&&&&&&&&&&&&&&&& \\
-1 & 1 & -6 & 3 & -12 & 5 & -30 & 14 & -62 & 49 & -132 & 108 & -216 &
189 & -305 & 233 & -344 & 268 & -344 & 233 & -305 & 189 & -216 & 108 &
-132 & 49 & -62 & 14 & -30 & 5 & -12 & 3 & -6 & 1 & -1 \\
&&&&&&&&&&&&&&&&&&&&&&&&&&&&&&&&&& \\
3 & -3 & 8 & -1 & 13 & 14 & -3 & 59 & -37 & 142 & -113 & 254 & -201 &
385 & -310 & 492 & -375 & 540 & -375 & 492 & -310 & 385 & -201 & 254 &
-113 & 142 & -37 & 59 & -3 & 14 & 13 & -1 & 8 & -3 & 3 \\
&&&&&&&&&&&&&&&&&&&&&&&&&&&&&&&&&& \\
-3 & 3 & -8 & 1 & -12 & -13 & 0 & -59 & 38 & -144 & 124 & -294 & 258 &
-471 & 411 & -628 & 509 & -690 & 509 & -628 & 411 & -471 & 258 & -294 &
124 & -144 & 38 & -59 & 0 & -13 & -12 & 1 & -8 & 3 & -3 \\
&&&&&&&&&&&&&&&&&&&&&&&&&&&&&&&&&& \\
1 & -1 & 6 & -3 & 11 & -5 & 31 & -15 & 63 & -47 & 130 & -81 & 169 & -91
& 185 & -52 & 155 & -41 & 155 & -52 & 185 & -91 & 169 & -81 & 130 & -47
& 63 & -15 & 31 & -5 & 11 & -3 & 6 & -1 & 1 \\
&&&&&&&&&&&&&&&&&&&&&&&&&&&&&&&&&& \\
0 & 0 & -2 & 2 & -7 & 5 & -15 & 11 & -28 & 12 & -43 & 34 & -85 & 59 &
-125 & 110 & -166 & 110 & -166 & 110 & -125 & 59 & -85 & 34 & -43 & 12 &
-28 & 11 & -15 & 5 & -7 & 2 & -2 & 0 & 0 \\
&&&&&&&&&&&&&&&&&&&&&&&&&&&&&&&&&& \\
0 & 0 & 0 & 0 & 1 & 0 & 2 & 0 & 0 & 11 & -18 & 43 & -59 & 93 & -110 &
146 & -148 & 162 & -148 & 146 & -110 & 93 & -59 & 43 & -18 & 11 & 0 & 0
& 2 & 0 & 1 & 0 & 0 & 0 & 0 \\
&&&&&&&&&&&&&&&&&&&&&&&&&&&&&&&&&& \\
0 & 0 & 0 & 0 & 0 & 0 & 0 & -1 & 2 & -5 & 9 & -15 & 20 & -27 & 32 & -38
& 40 & -42 & 40 & -38 & 32 & -27 & 20 & -15 & 9 & -5 & 2 & -1 & 0 & 0 &
0 & 0 & 0 & 0 & 0
\end{array}\right)
\setlength{\arraycolsep}{6pt}
$
}}

\end{landscape}

\newpage

\begin{figure}[h!]
\centering\leavevmode
\includegraphics[width=2.2 cm]{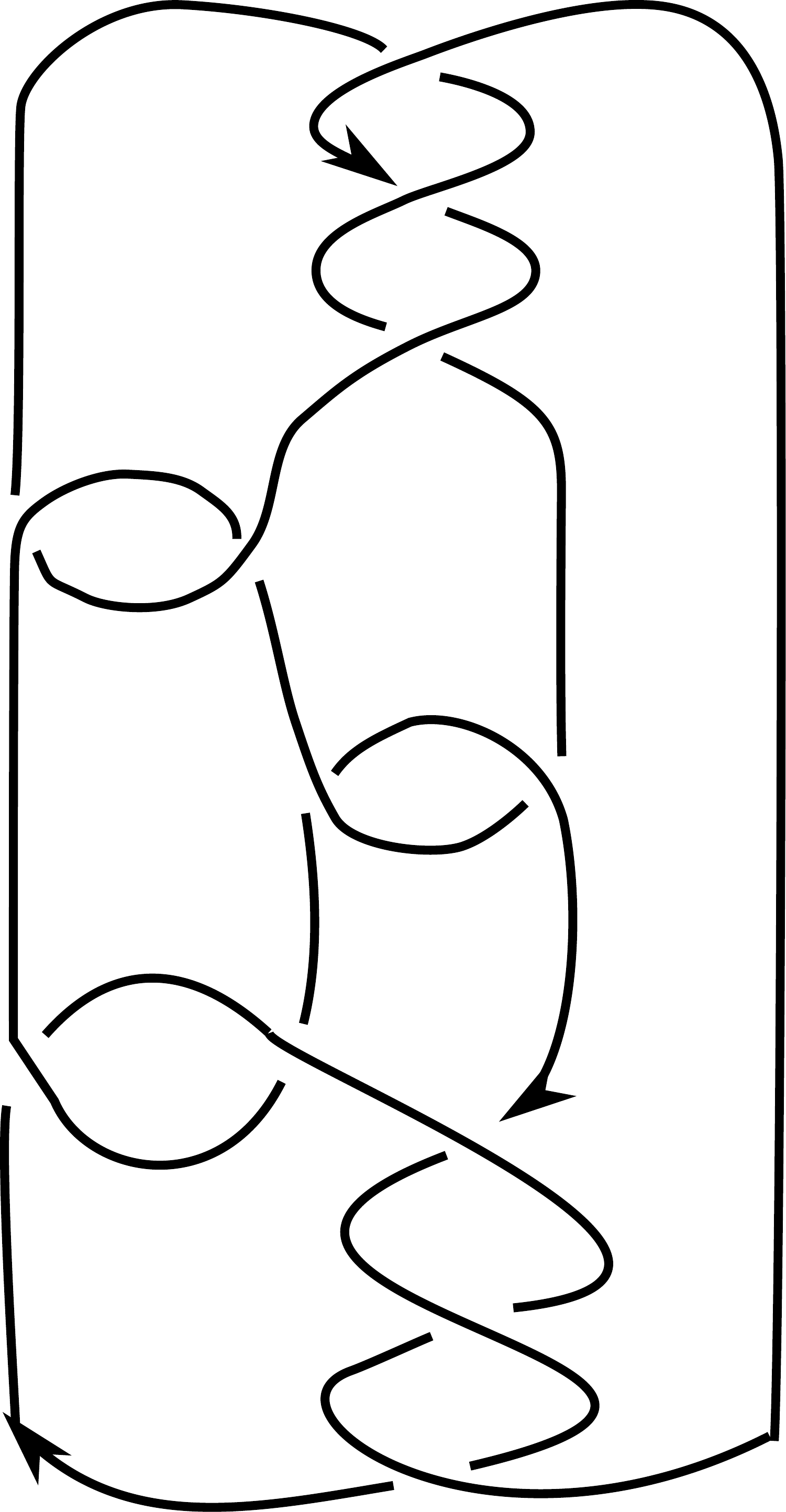}
\includegraphics[width=2.2 cm]{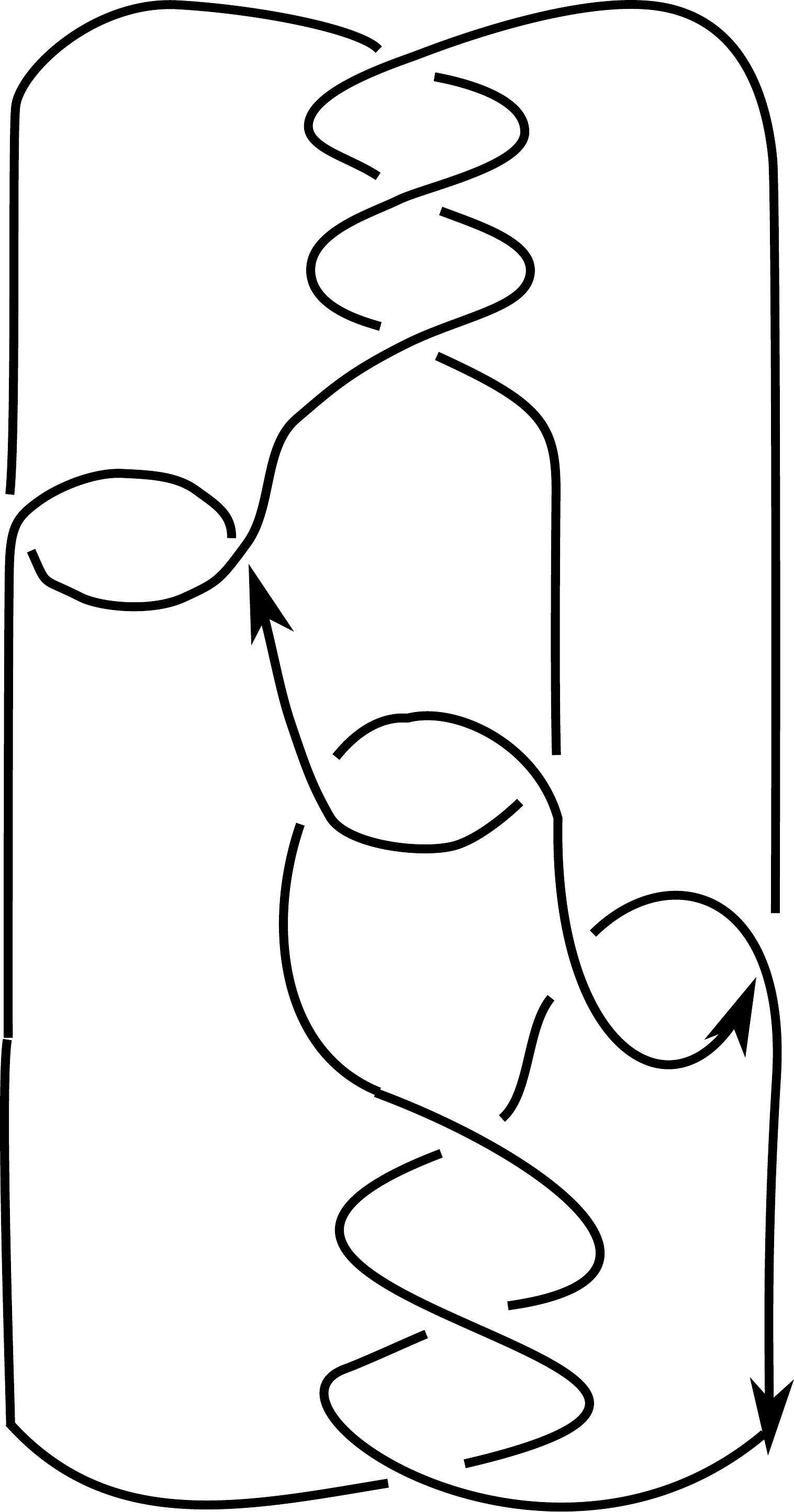}
\caption{(a) Kinoshita-Terasaka (b) Conway}
\end{figure}

The KTC knots can be viewed as two 3-fingers connected by propagator involving
horizontal braid with $m=2$ crossings.
Changing $m=2\longrightarrow m=-2$ simply switches between these two mutant
polynomials.

At $q=1$ both these polynomials are cubes of the special polynomial,
\be
H_{[21]}(q=1) = \Big(H_{[1]}(q=1)\Big)^3 = \left(\frac{2A^6-6A^4+7A^2-2}{A^2}   \right)^3
\ee
At $A=q^2$ both reproduce the same Jones
\be
H_{[21]}(A=q^2) = H_{[1]}(A=q^2) = q^{12}-2q^{10}+2q^8-2q^6+q^4+2q^{-2}-2q^{-4}+2q^{-6}-q^{-8}
\ee
and at $A=1$ the Alexander polynomial, which in this particular case is just unity,
\be
H_{[21]}(A=1,q) = H_{[1]}(A=1,q^3) = 1
\ee
The first terms of the differential expansion (see ss.\ref{tests}) in the classical limit,  $A=q^N$, $q=e^\hbar$, are
\be
H_{[21]} = 1 +   2\{A\} \Big(\{Aq^3\}\{A/q^3\} + \{Aq^2\}\{A\}+ \{A\}\{A/q^2\}\Big) + O(\hbar^4)
\ee
In accordance with \cite{Mort}, the difference between the two polynomials shows up in the order $\hbar^{11}$, which is related to the power 11 of $\{q\}$ in (\ref{mutdiff}).

The complete difference is
\be
\boxed{
H_{[21]}^{M(3,-2|2|-3,2)} -  H_{[21]}^{M(3,-2|2|2,-3)} =
-\rho^2 A^3 \cdot D_{3}^2D_{2} D_0 D_{-2}D_{-3}^2 \cdot z^{11}
}
\label{KTCdiff}
\ee
$$
= -\rho^2 A^3 \cdot \{q\}^4\{Aq^3\}^2\{Aq^2\}\{A\}\{A/q^2\}\{A/q^3\}^2
$$
with $\rho = \frac{[3][14]}{[2][7]}$ and $z = q-q^{-1}=\{q\}$.
It perfectly matches the result of \cite{Mura}. The difference could actually be calculated for three-cabled knots
(because symmetric and antisymmetric representations $[3]$ and $[111]$ do
not contribute to it), what allows one to make calculations with the help of the
ordinary skein relations.
However, the answers for individual  $H_{[21]}$ could not be obtained in that way.

\subsection{Other mutant pairs}
In the list (\ref{mutlist}) of 11-intersection mutant pairs there are four more
pairs, for which the fat tree description is already provided. In fact, from this point of view they belong exactly to the same class as the KTC mutants, just some intersection signs are different.
The differences are
\be
H_{[21]}^{11n36} -H_{[21]}^{11n44}=A^{-9}\frac{[3]^2 [14]}{[2][7]}\cdot
\{q\}^4 \{Aq^3\}^3\{Aq^2\}\{A\}\{A/q^2\}\{A/q^3\}^2\\
H_{[21]}^{11n39} -H_{[21]}^{11n45}= A^{-3}\frac{[3]^2 [14]^2}{[2]^2[7]^2}\cdot
\{q\}^4 \{Aq^3\}^3\{Aq^2\}\{A\}\{A/q^2\}\{A/q^3\}^2\\
H_{[21]}^{11n41} -H_{[21]}^{11n47}=A^{-15}\frac{[3]^2 [14]}{[2][7]}\cdot
\{q\}^4 \{Aq^3\}^3\{Aq^2\}\{A\}\{A/q^2\}\{A/q^3\}^2\\
H_{[21]}^{11n151} -H_{[21]}^{11n152}= A^{-9}\frac{[3]^2 [14]^2}{[2]^2[7]^2}\cdot
\{q\}^4 \{Aq^3\}^3\{Aq^2\}\{A\}\{A/q^2\}\{A/q^3\}^2
\ee
Somewhat surprisingly,  some differences are the same for different pairs: for 11n36/11n44 and 11n41/11n47, as well as for 11n39/11n45 and 11n151/11n152. The same phenomenon one could observe for the pretzel mutants: the differences are the same for pairs: 11n71/11n75 and 11n76/11n78;
11a47/11a44, 11a57/11a231 and 11n73/11n74 (in the latter case the three differences coincide), see ss.\ref{pmu}.
The entire HOMFLY are, of course quite different, see the Appendix.

\bigskip

It remains to be seen, if  the remaining pairs in (\ref{mutlist}) possess a {\it double} fat tree description.

\section{Evolution method in application to mutant families}

Of course, most interesting are not just particular knots, but entire families,
depending on various parameters.
The way to study this kind of problems is provided by the evolution method of
\cite{DMMSS,evo}.
For its application to knot polynomials in representation $[21]$ see \cite{MMM21},
but there only the simplest family of twist knots was considered
(the torus  knots were described in this way in arbitrary representation \cite{RJ,AS,DMMSS,Che},
but this is a much simpler exercise, because of the clear algebraic nature of the torus family).
We briefly remind this story and then extend consideration to the first mutant-containing family.

\subsection{Twist knots \cite{AnoMcabling,GuJ,MMM21}}

The $[21]$-colored HOMFLY for generic twist knots was obtained by the
evolution method in \cite{MMM21}:
\be
H^{(k)}_{[21]} = u_3A^{6k} + \Big(u_{2p}q^{4k}+u_{20}+u_{2m}q^{-4k}\Big)A^{4k} +u_1A^{2k} + u_0
 \nn \\ \nn \\
u_3   = -A^3\frac{\{Aq^3\}\{A/q^{3}\}\{Aq\}\{A/q\}}{\{A\}}
\nn \\
u_{2p}= \frac{[3]}{[2]^2} A^3\frac{\{Aq^3\}\{A\}}{\{Aq\}}
\Big([2]A^2q^{-3}-(q^4+1-q^{-2}+q^{-4})\Big) \nn \\
u_2  = -2\{q\}^2 A^3\left(\frac{[3]}{[2] }\right)^2\frac{\{Aq^2\}\{A/q^2\}}{\{A\}}
\nn\\
u_{2m}= \frac{[3]}{[2]^2}A^3\frac{\{Aq^{-3}\}\{A\}}{\{Aq^{-1}\}}
\Big([2]A^2q^3 -(q^4+1-q^{2}+q^{-4})\Big)
\nn\\
u_1 = -\frac{[3]A^3}{\{A\}}\Big(A^4 -(q^6+q^{-6})A^2 + (2q^6-4q^4+4q^2-3+4q^{-2}-4q^{-4}+2q^{-6})\Big)
\nn\\
u_0 = \frac{A^3}{\{Aq\}\{A\}\{A/q\}}\left(A^6-\frac{[3][10]}{[2][5]}A^4
+ \frac{[3][10]^2}{[2]^2[5]^2}A^2-\frac{[10][14]}{[2]^2[5][7]}\right)
\ee
This example is already quite interesting for the study of differential expansion
\cite{ArthMM} beyond (anti)symmetric representations. However, this family does not
contain mutant pairs and, hence we need to look for knots beyond this family.

\subsection{4-finger pretzel knots}
The simplest {\it multi-parametric family} containing mutant pairs is the four-parallel-finger
pretzels $Pr(n_1,n_2,n_3,n_4)$ with even $n_1$
and $n_2,n_3,n_4$ odd.
In this case, one has
\be
\frac{H^{Pr(n_1,n_2,n_3,n_4)}_{[21]}-H^{Pr(n_1,n_2,n_4,n_3)}_{[21]}}
{A^{-3(n_1+n_2+n_3+n_4)}\{q\}^4\{Aq\}^3\{Aq^2\}\{A\}\{A/q^2\}\{A/q^3\}} \ = \
\frac{1}{ \{q\}^{11}}\cdot F(n_1|n_2)\cdot G(n_3,n_4)
\label{4fprmut}
\ee
with
\be
G(n_3,n_4) =
\Big(q^{4(n_3+n_4)}+q^{-4(n_3+n_4)}\Big)\Big(q^{n_4-n_3}-q^{n_3-n_4}\Big)
+ \frac{[4]^2}{[2]^2}
\Big(q^{n+3+n_4}+q^{-n_3-n_4}\Big)\Big(q^{4(n_4-n_3)}-q^{4(n_3-n_4)}\Big) + \nn \\
+\frac{[3][6]}{[2][5]}\Big(q^{5(n_3-n_4)}-q^{5(n_4-n_3)}\Big)
+ \frac{[5]}{[3]}(q^4+3+q^{-4})\Big(q^{3(n_3-n_4)}-q^{3(n_4-n_3)}\Big) + \nn
\ee
\vspace{-0.5cm}
\be
+\frac{[2][4][6]}{[3][5]}\Big(q^{5n_3}-q^{-5n_3} - q^{5n_4}+q^{-5n_4}\Big)
+ \frac{[2][4][6]}{[3]^2}\Big(-q^{3n_3}+q^{-3n_3} + q^{3n_4}-q^{-3n_4}\Big)
\ee
and
\be
F(n_1=2\big|n_2)= \{q\}^2 \frac{[3] }{[2]^4[4]^3[5]^2[6]^2}\left([6][5n_2-3] - [2][5][3n_2-5] -\frac{[2]^2[4][6]}{[3]}
  + \{q\}^2 [2][5][3n_2-1]\right)
\ee
For generic even values of $n_1$ expression is more complicated:
\be
F(n_1|n_2) = \frac{[3]}{[2]^4[4]^4[5]^2[6]^2}
\left(\begin{array}{ccccc}q^{-5n_1}&q^{-3n_1}&1&q^{3n_1}&q^{5n_1}\end{array}\right)\cdot
\nn
\ee
\be
\cdot
\left(\begin{array}{ccccc}
0& 1 & -\frac{[2][4][6]}{[3][5]}& \frac{[4]^2}{[2]^2} & -\frac{[3][6]}{[2][5]} \\ \\
1& 0 & -\frac{[2][4][6]}{[3]^2} & \frac{[5]}{[3]}(q^4+3+q^{-4}) & -\frac{[4]^2}{[2]^2} \\ \\
\frac{[4][6]}{[5]} & -\frac{[2][4]}{[3]}(q^4+3+q^{-4})& 2\cdot\frac{[2]^2[4]^2[6]}{[3]^2[5]}
& -\frac{[2][4]}{[3]}(q^4+3+q^{-4}) & \frac{[4][6]}{[5]} \\ \\
-\frac{[4]^2}{[2]^2} & \frac{[5]}{[3]}(q^4+3+q^{-4}) & -\frac{[2][4][6]}{[3]^2}&0&1\\ \\
-\frac{[3][6]}{[2][5]} & \frac{[4]^2}{[2]^2} & -\frac{[2][4][6]}{[3][5]} & 1 & 0
\end{array}\right)
\left(\begin{array}{c}q^{-5n_2}\\ \\ q^{-3n_2}\\ \\ 1 \\ \\
q^{3n_2} \\ \\ q^{5n_2}\end{array}\right)
\ee
The RHS. of (\ref{4fprmut}) is actually a Laurent polynomial in $q$.
This formula is obtained by the evolution method of \cite{evo} with respect to  the variables $n_2,n_3,n_4$, and it has the structure predicted by the general expression (\ref{mutdiff}) for mutants. Also it vanishes whenever any of $n_2,n_3,n_4$ is unity. This is because the pretzel knots with unit length
always commutes with any other length due to identities like (\ref{ramaid}).
Eqs.(\ref{mu3332})-(\ref{mu9325}) in ss.\ref{pmu} are particular cases of (\ref{4fprmut}).

\section{Conclusion
\label{conc}}
In this paper we continued the study of the double fat tree realization of
link diagrams, for which HOMFLY polynomials are described by an
absolutely new and impressively effective formula (\ref{basic}).
This formula requires redrawing  knots as  double fat tree diagram.

In the present paper,  we showed that  eqn. (\ref{basic}),
actually  simplifies  evaluation of  knot polynomials of many knots
presentable as two-bridge knots.
It goes without saying that this overcomes most calculational
complexities in knot/link theory.
As an immediate illustration, we calculate a number of $[21]$-colored
polynomials, and actually can now do this for many knots.

Particularly, this methodology helped  us to explicitly illustrate/validate old
expectations about the mutant pairs. Specifically, we explained the
conditions when mutants are distinguishable or indistinguishable by
$[21]$-colored polynomials. For instance, the KTC mutant pair {\bf can} be resolved by these polynomials, but the pretzel mutants made from antiparallel fingers of odd lengths {\bf cannot}
be distinguished.

Of greatest importance for knot theory are now three questions:\\

$\bullet$ How large is the double fat tree family: what are the knots/links
beyond it (if any?), and what is the way to find a double fat tree realization of
a given knot/link?

$\bullet$ What is the origin of (\ref{basic}) and its
effective field theory interpretation?
What could be the meaning of quantization and loop diagrams
in this effective theory.

$\bullet$ What is the $\beta$-deformation of (\ref{basic}), i.e.
can this powerful formula be lifted to the super-
and Khovanov-Rozansky polynomials?

Added to these could be two obvious {\it technical} next steps:

$\bullet$ to extend the double fat trees made from 4-strand braids  propagators to arbitrary braids.
The {\it need} for this for knot theory depends on the answer to
the very first question. However, the problem itself can have its own
value, especially because of its relation to multi-point
conformal blocks. This approach will give complete solution provided we
know the Racah matrices for arbitrary representations in the multi-point conformal blocks.


$\bullet$ to develop a systematic calculus for colored knots in
arbitrary representations, beyond symmetric and $[21]$. This requires
 calculating the simplest Racah matrices describing the 2-bridge knots.
We hope some of the non-rectangular representations may distinguish
odd antiparallel pretzel mutants.


\bigskip
Calculations in the case of representation $[21]$ are quite
involved and can be performed in slightly different ways.
In particular, such technically independent exercise for
the KTC mutants is reported in a parallel paper \cite{Satoshi}.

\section*{Acknowledgements}

We are indebted to S.Nawata for participation at the early stage of the project and for critically reading the final version of manuscript. We also highly appreciate considerable help in calculations and discussions with S.Arthamonov and A.Sleptsov.
We would like to thank the hospitality of the Indian Institute of Technology Bombay, where a part of this work was done.

Our work is partly supported by the Laboratory of Quantum Topology of Chelyabinsk State University
through Russian Federation government grant 14.Z50.31.0020 (And.Mor.), by grants NSh-1500.2014.2, by RFBR grants 13-02-00457 (A.Mir.), 13-02-00478 (A.Mor.), 14-02-00627 (An.Mor.), by the joint grants 15-51-52031-NSC-a (A.M.'s), 14-01-92691-Ind-a (A.M.'s), INT/RFBR/P-162 (P.R. and V.K.S.), and by 15-52-50041-YaF (A.M.'s). Also we are partly supported by the Brazilian Ministry of Science, Technology and Innovation through the National Counsel of Scientific and Technological Development (A.Mor.).

\newpage

\section{Appendix. Examples of $[21]$-colored HOMFLY}

Here we list examples of knots from the Rolfsen table of \cite{katlas},
for which the $[21]$-colored HOMFLY are now available.

\subsection{A list of knots with up to 10
intersections}

We start from the table which contains some relevant information about the knots up to 10 intersections.

The left part of the table lists the previously known cases:
\begin{itemize}
\item For the torus knots, the arbitrary colored HOMFLY polynomial is given by the Rosso-Jones formula \cite{RJ}.
\item For {\it three}-strand knots, the $[21]$-colored HOMFLY polynomials can be calculated
by the cabling method of \cite{AnoMcabling} on ordinary computers.
When the number of strands is three, we explicitly give a braid word,
only instead of $\tau_1^{a_1}\tau_2^{a_2}\tau_1^{a_3}\ldots$
we write the sequence $a_1,a_2,a_3,\ldots$
\item For {\it two}-bridge knots the knowledge of Racah matrices $S$ and $\bar S$ from \cite{GuJ}
is sufficient: eq.(\ref{basic}) in this case reduces to an obvious matrix element ${\cal A}_{11}$.
The corresponding answers are available from \cite{GuJ}.
When the number of bridges is two, we explicitly give an $S-T$ word.
\end{itemize}

\bigskip

The right part of the table lists the cases which are available only now,
by the method of the present paper, these knots are boldfaced in the first column.
Numerous intersections between the left and right parts of the table
are important for checking our conjecture (\ref{basic}).

An exhaustive list of the pretzel knots up to 10 intersections
is borrowed from the third paper of ref.\cite{GMMMS}.
They are distinguished from generic double fat tree knots only
by simplicity of computer calculations, what is actually quite important.

The starfish cases are also usually simple enough: they still involve just one
sum over representations
(all pretzel knots are automatically starfish).
The cases with two and three propagators involve two and three such sums
and are considerably more difficult for computers.
A search for maximally simple representations are therefore important
from this point of view, and hopefully the table can be significantly
improved.

The right part of the table is currently incomplete, but it seems
all the knots with 10 or less intersections to fit into it (with the possible
exception of $10_{161}$, which is anyhow present in the left part of the table).

In addition to this table, answers are available for some
explicitly identified mutants with $11$ intersections.

\bigskip

\centerline{
$
\begin{array}{c|c|c|c||c|c|c|c|  }
\multicolumn{3}{c}{}&
\multicolumn{5}{|||c|||}{\textbf{Implications of the new topological theory (\ref{basic})}}\\
\multicolumn{3}{c}{}&\multicolumn{5}{|||c|||}{}\\
\hline
&&&&&&&\\
{\rm knot} & {\rm torus} & \#\ {\rm of\ strands} & \#\ {\rm of\ bridges}
& {\rm pretzel} &{\rm starfish}
& \text{double-sum}&
\\
&&&&&&&\\ \hline &&&&&&&\\
3_1 & [2,3] &   (1,1)^2 &d_R\cdot(ST^3S^\dagger)_{_{11}} & (3,0) &&&\\
&&&&&&&\\ \hline &&&&&&&\\
4_1 & &  (1,-1)^2 &d_R\cdot(\bS\bT^{-2}\bS\bT^{2}\bS)_{_{11}}& (1,\bar 2,1) &&&\\
&&&&&&&\\ \hline &&&&&&&\\
5_1 & [2,5] & 2 &d_R\cdot(ST^5S^\dagger)_{_{11}}& (5,0)&&&\\
5_2 & &  (3,1,-1,1) &d_R\,(\bS\bT^{4}\bS\bT^{2}\bS)_{_{11}}
\!\! = d_R\, (\bar S\bar T^3 ST^{-2}S^\dagger)_{_{11}}
& (\bar 3,\bar 1,\bar 1)&&& \\
&&&&&&&\\ \hline &&&&&&&\\
6_1 && 4  &d_R\cdot(\bS\bT^{-4}\bS\bT^{2}\bS)_{_{11}} & (\bar 5,-\bar 1,-\bar 1) &&&\\
6_2 &&   (3,-1,1,-1) &d_R\cdot(\bS\bT^2\bS\bT^{-1}ST^3S^\dagger)_{_{11}}
& (3,\bar 2,1) &&& \\
6_3 &&   ( 2,-1,1,-2) &d_R\cdot( ST^{-2}S^\dagger\bT\bS\bT^{-1}ST^2S^\dagger)_{_{11}}
& (2,-3,1,1) &&& \\
&&&&&&&\\ \hline  &&&&&&&\\
7_1 & [2,7] & 2 &d_R\cdot(ST^7S^\dagger)_{_{11}}& (7,0) &&& \\
{7_2} &&4&d_R\,(\bS\bT^{6}\bS\bT^{2}\bS)_{_{11}}
\!\! = d_R\, (\bar S\bar T^5 ST^{-2}S^\dagger)_{_{11}}& (\bar 5,\bar 1,\bar 1)  &&& \\
7_3 &&    (5,1,-1,1) &d_R\cdot(ST^4S^\dagger \bT^{-3}\bS)_{_{11}}& (4,1,1,1) &&&\\
{ 7_4} &&4&
d_R\cdot(\bS \bT^{-3}STS^\dagger \bT^{-2}STS^\dagger)_{_{11}}
&(\bar 3,\bar 3,\bar 1) &&& \\
7_5 &&   (4,1,-1,2) &d_R\cdot (ST^3S^\dagger\bT^{-2}ST^2S^\dagger )_{_{11}}
& (3,2,1,1) &&& \\
{7_6} &&4&d_R\cdot(\bS\bT^2\bS\bT^{-2}\bS\bT ST^{-2}S^\dagger)_{_{11}} & (-3,1,\bar 2,1,1)  &&&  \\
{7_7} &&4&d_R\cdot(\bS\bT^{-2}\bS\bT ST^{-1}S^\dagger\bT\bS\bT^{-2}\bS )_{_{11}}
&(-\bar 3,\bar 1,-\bar 3,\bar 1,\bar 1) &&&  \\
&&&&&&&\\ \hline  &&&&&&&\\
%
%
{8_1} &&5&d_R\cdot(\bS\bT^{-6}\bS\bT^{2}\bS)_{_{11}}& (1,\bar 6,1)&&& \\
{8_2} &&5,1,-1,1&d_R\cdot(STS^{\dagger}\bar{T}^{-1}\bar{S}\bar{T}ST^{-5}S^{\dagger})_{_{11}}& (5,\bar 2,1)  &&&\\
{8_3} &&5&d_R\cdot(\bar{S}\bar{T}^4\bar{S}\bar{T}^{-4}\bar{S})_{_{11}}& (1,1,\bar 4,1,1)  &&& \\
 {8_4} &&4&d_R\cdot(\bar{S}\bar{T}^4\bar{S}\bar{T}^{-1}ST^3S^{\dagger})_{_{11}}& (3,\bar 4,1)  &&&        \\
{ 8_5} && 3,-1,3,-1&3& (3,\bar 2,3)  &&& \\
 {8_6} &&4&d_R\cdot(STS^{\dagger}\bar{T}^{-1}\bar{S}\bar{T}^3ST^{-3}S^{\dagger})_{_{11}}& (1,3,\bar 2,1,1)  &&&  \\
 { 8_7} &&4,-1,1,-2&d_R\cdot(ST^2S^{\dagger}\bar{T}^{-1}\bar{S}\bar{T}ST^{-4}S^{\dagger})_{_{11}}&(4, -3, 1, 1)  &&& \\
{8_8} &&4&d_R\cdot(ST^{-2}S^{\dagger}\bar{T}\bar{S}\bar{T}^{-3}ST^2S^{\dagger})_{_{11}}
& (2, -3, 1^4)  &&& \\
 { 8_9} && 3,-1,1,-3&d_R\cdot(ST^{-3}S^{\dagger}\bar{T}\bar{S}\bar{T}^{-1}ST^3S^{\dagger})_{_{11}}&  (4, -3, -1, -1)  &&& \\
{ 8_{10}} && 3,-1,2,-2&3& (2, -3, 1, 3)  &&& \\
{8_{11}} &&4&d_R\cdot(STS^{\dagger}\bar{T}^{-1}\bar{S}\bar{T}ST^{-2}S^{\dagger}\bar{T}^3\bar{S})_{_{11}}& (-\bar 3, \bar 1, \bar 1, \bar 3, \bar 1)  &&&\\
{8_{12}} &&5&d_R\cdot(\bS\bT^2\bS\bT^{-2}\bS\bT^2\bS\bT^{-2}\bS)_{_{11}}&    &&&  \\
{8_{13}} &&4&
d_R\,(\bS\bT^3ST^{-1}S^\dagger\bT\bS\bT^{-1}ST^2S^\dagger)_{_{11}}
&(-\bar 4,-3, 1, 1, 1)  &&& \\
{8_{14}} &&4&d_R\cdot(ST^3S^\dagger\bT\bS\bT^{-1}STS^\dagger\bT^{-1}\bS\bT^2\bS)_{_{11}}&&&   &  \\
{\bf 8_{15}} &&4&3& (2,3,3, -1^3)& (\ref{815})
&& \\
 { 8_{16}} && 2,-1,2,-1,1,-1&3&& &&    \\
  { 8_{17}} && 2,-1,1,-1,1,-2 &3&&&&     \\
  { 8_{18}} && (1,-1)^4 &3&&&& \\
{ 8_{19}} &[3,4]& 3,1,3,1 &3& (3,-\bar 2,3)  &&& \\
{ 8_{20}} & &  3,-1,-3,-1&3& (3,\bar 2,-3) &&&\\
{ 8_{21}} && 3,-1,-3,-1 &3& (2, -3, 1, -3) &&&    \\
&&&&&&&\\ \hline  &&&&&&&\\
\end{array}
$
}

\newpage

\newpage

\centerline{
$
\begin{array}{c|c|c|c||c|c|c|c|  }
&&&&&&&\\
{9_1} & [2,9] & (9,0)&d_R\cdot(ST^9S^\dagger)_{_{11}} &&&& \\
{9_2} &&5&d_R(\bS\bT^{8}\bS\bT^{2}\bS)_{_{11}}
\!\! = d_R(\bar S\bar T^7 ST^{-2}S^\dagger)_{_{11}}& (\bar 1,\bar 7,\bar 1) &&& \\
{9_3} &&7,1,-1,1&d_R\cdot(\bS\bT^{-3}ST^{6}S^{\dagger})_{_{11}}& (6, 1, 1, 1)  &&& \\
{9_4} &&4&d_R\cdot(ST^{-1}S^{\dagger}\bar{T}^{4}ST^{-4}S^{\dagger})_{_{11}}& (4,1^5) &&& \\
{9_5} &&5&d_R\cdot(\bar{S}\bar{T}^{5}ST^{-1}S^{\dagger}\bar{T}^{2}ST^{-1}S^{\dagger})_{_{11}}& (-\bar 1, -\bar 3, -\bar 5) &&& \\
{9_6} && 6,1,-1,2 &d_R\cdot(ST^2S^{\dagger}\bar{T}^{-2}ST^{5}S^{\dagger})_{_{11}}& (2, 1, 5, 1) &&& \\
{9_7} &&4&d_R\cdot(ST^2S^{\dagger}\bar{T}^{-4}ST^{3}S^{\dagger})_{_{11}}& (2,3,1^4)  &&&\\
{9_8} &&5&d_R\cdot(\bar{S}\bar{T}^{-2}\bar{S}\bar{T}^4\bar{S}\bar{T}^{-1}ST^2S^{\dagger})_{_{11}}& (-2, -3, 1^6) &&&\\
{9_9} && 5,1,-1,3 &d_R\cdot(\bar{S}\bar{T}^{-1}ST^2S^{\dagger}\bar{T}^{-2}ST^4S^{\dagger})_{_{11}}
& (-4, 1, -5, 1)  &&&\\
{9_{10}} &&4&d_R\cdot(\bar{S}\bar{T}^{3}ST^{-3}S^{\dagger}\bar{T}^{-2}ST^{-1}S^{\dagger})_{_{11}}
& (\bar 3, \bar 3, \bar 1^3)  &&&\\
{9_{11}} &&4&d_R\cdot(\bar{S}\bar{T}^{-2}\bar{S}\bar{T}^2\bar{S}\bar{T}^{-1}ST^4S^{\dagger})_{_{11}}
& (-5,-2, 1^4) &&& \\
{9_{12}} &&5&d_R\cdot(\bar{S}\bar{T}^{4}\bar{S}\bar{T}^{-2}\bar{S}\bar{T}ST^{-2}S^{\dagger})_{_{11}}
& (-3, 1^3, \bar 4) &&& \\
{9_{13}} &&4&d_R\cdot(ST^{3}S^{\dagger}\bar{T}^{-2}STS^{\dagger}\bar{T}^{-3}\bar{S})_{_{11}}& (1, 3, 1, 1, -\bar 4) &&& \\
{9_{14}} &&5&d_R\cdot (\bar S \bar T^4 \bar S \bar T^{-1}STS^\dagger \bar T^{-1}\bar S\bar T^2\bar S)_{_{11}}
& (-\bar 5, -\bar 3, \bar 1^3) &&& \\
{9_{15}} &&5&d_R\cdot(\bar S\bar T^2\bar S\bar T^{-2}\bar S\bar T^3ST^{-2}S^\dagger)_{_{11}}& &&& \\
{ 9_{16}} &&4,2,-1,3&3& (\bar 2, 1, 3, 3) & &&\\
{9_{17}} &&4&d_R\cdot(\bar S\bar T^2 \bar S\bar T^{-1}ST^3S^\dagger\bar T^{-1}\bar S\bar T^2\bar S)_{_{11}}
& (\bar 3, \bar 3, -\bar 1^5) &&& \\
{9_{18}} &&4&d_R\cdot(\bar S\bar T^{-3}ST^2S^\dagger \bar T^{-2}ST^2S^\dagger)_{_{11}}& &&& \\
{9_{19}} &&5&d_R\cdot(\bar S\bar T^2\bar S\bar T^{-1}STS^\dagger\bar T^{-3}\bar S\bar T^2\bar S)_{_{11}}
& &&& \\
{9_{20}} &&4&d_R\cdot(ST^2S^\dagger \bar T^{-1}\bar S \bar T^2 \bar S\bar T^{-1}ST^3S^\dagger)_{_{11}}
& (4, 3, -1^4) &&& \\
{9_{21}} &&5&d_R\cdot(\bar S\bar T^3 ST^{-1}S^\dagger \bar T\bar S \bar T^{-2}\bar S\bar T^2\bar S)_{_{11}}
& &&&  \\
{\bf 9_{22}} &&4&3& &&&  \\
{9_{23}} &&4&d_R\cdot(ST^2S^\dagger \bar T^{-2}STS^\dagger \bar T^{-2}ST^2S^\dagger)_{_{11}}& &&&  \\
{\bf 9_{24}} &&4&3& (-2, -3, 3, 1^3)  &&& \\
 {\bf 9_{25}} &&5&3&   &&& \\
 {9_{26}} &&4
 &d_R\cdot(ST^3S^\dagger \bar T^{-1}\bar S\bar T ST^{-1}S^\dagger \bar T \bar S \bar T^{-2}\bar S)_{_{11}}
 &  &&&  \\
 {9_{27}} &&4&
 d_R\cdot(ST^2S^\dagger \bar T^{-1}\bar S\bar T ST^{-2}S^\dagger \bar T \bar S \bar T^{-2}\bar S)_{_{11}}
 & &&& \\
 {\bf 9_{28}} &&4&3& (2, -3, -3, 1^3)  &&& \\
{\bf 9_{29}} &&4&3& &&& \\
{\bf 9_{30}} &&4&3   &&&& \\
{9_{31}} &&4&\!d_R\,(ST^2S^\dagger \bar T^{-1}\!\bar S\bar T ST^{-1}\!S^\dagger
\bar T\bar S\bar T^{-1}\!ST^2S^\dagger)_{_{11}}\!\!&&&& \\
{\bf 9_{32}} &&4&3   &&&& \\
{\bf 9_{33}} &&4&3   &&&&   \\
{\bf 9_{34}} &&4&3   &&&& \\
{\bf 9_{35}} &&5&3&  (\bar 3, \bar 3, \bar 3)  &&& \\
{\bf 9_{36}} &&4&3&   &&&  \\
{\bf 9_{37}} &&5&3& (-\bar 3^2, \bar 3, \bar 1^2)  &&&  \\
{\bf 9_{38}} &&4&3&   &&&  \\
{\bf 9_{39}} &&5&3&   &&&  \\
 {\bf 9_{40}} &&4&3&   &&   &      \\
{\bf 9_{41}} &&5&3&&&& \\
{\bf 9_{42}} &&4&3&&     \overline{\sum\limits_{ X\in R\otimes \bar R}^{{\phantom.}}} \!\!
\frac{d_R\cdot(ST^3S^\dagger)_{_{1X}}(ST^{-2}S^\dagger)_{_{1X}}(STS^\dagger\bT^{-1}\bS\bT^2\bS)_{_{1X}}}
{\bar S_{_{1X}}^{\phantom{.^5}}}&&\\
{\bf 9_{43}} &&4&3&&&&     \\
{\bf 9_{44}} &&4&3&&&&       \\
{\bf 9_{45}} &&4&3&&&&    \\
{\bf 9_{46}} &&4&3& (\bar 3, -\bar 3, \bar 3)&&& \\
{\bf 9_{47}} &&4&3&&&&      \\
{\bf 9_{48}} &&4&3& (-\bar 3^3,   \bar 1^2) &&&     \\
{\bf 9_{49}} &&4&3& &&&      \\
\end{array}
$}


\centerline{
$
\begin{array}{c|c|c|c||c|c|c|c|  }
{10_1} &&&(\bS\bT^{-8}\bS\bT^{2}\bS)_{_{11}}& (\bar 1, \bar 7, -\bar 3)  &&&   \\
{10_2} & &7,-1,1,-1 &
d_R\,( \bS\bT^2\bS\bT^{-1}ST^7S^\dagger )_{_{11}} & (2, -7, -1, -1)&&&     \\
{10_3} &&&
d_R\,( \bS\bT^6\bS\bT^{-4}\bS)_{_{11}}& (\bar 1, \bar 5, -\bar 5)  &&& \\
{ 10_4} &&&
d_R\,( ST^3S^\dagger \bT^{-1}\bS\bT^6\bS )_{_{11}}& (-\bar 7,\bar 1,\bar 1,\bar 1,\bar 1)  &&& \\
{ 10_5} &&6,-1,1,-2 &
d_R\,( ST^6S^\dagger \bT^{-1}\bS\bT ST^{-2}S^\dagger )_{_{11}}& (-2, 7, -1, -1)  &&&    \\
{ 10_6} &&&
d_R\,( ST^5S^\dagger \bT^{-3}\bS\bT^2\bS )_{_{11}}& (-5, -1, -1, -1, -\bar 2)&&&\\
{ 10_7} &&&
d_R\,( \bS\bT^5S T^{-2}S^\dagger \bT\bS\bT^{-2}\bS )_{_{11}}& (-\bar 3,\bar 1,\bar 5,\bar 1,\bar 1)  & &&\\
{ 10_8} &&&
d_R\,( \bS\bT^4\bS\bT^{-1}ST^5S^\dagger )_{_{11}}& (-6, 1^5)   &&&\\
{ 10_9} &&5,-1,1,-3 &
d_R\,( ST^5S^\dagger\bT^{-1}\bS\bT ST^{-3}S^\dagger )_{_{11}}& (6, -3, -1, -1) &&&\\
{ 10_{10}} &&&
d_R\,( \bS\bT^5 ST^{-1}S^\dagger \bT\bS\bT^{-1}ST^2S^\dagger )_{_{11}}& &&& \\
{ 10_{11}} &&&
d_R\,( \bS\bT^4\bS\bT^{-3}ST^3S^\dagger )_{_{11}}& (3, 1, 1, 1, \bar 4)  &&&\\
{ 10_{12}} &&&
d_R\,( ST^{-4}S^\dagger\bT^3\bS\bT^{-1}ST^2S^\dagger  )_{_{11}}& (4, -3, 1^4)  &&& \\
{ 10_{13}} &&&
d_R\,( \bS\bT^{-4}\bS\bT^2\bS\bT^{-2}\bS\bT^2\bS )_{_{11}}& &&& \\
{ 10_{14}} &&&
d_R\,( ST^4S^\dagger \bT^{-2}STS^\dagger\bT^{-1}\bS\bT^2\bS )_{_{11}}& &&& \\
{ 10_{15}} &&&
d_R\,( ST^{-4}S^\dagger \bT\bS\bT^{-3}ST^2S^\dagger )_{_{11}}& (-2, -1, 5, -1^3) &&&\\
{ 10_{16}} &&&
d_R\,( \bS\bT^4 \bS\bT^{-1}ST^2S^\dagger\bT^{-3}\bS)_{_{11}}& (-\bar 3, -\bar 1, \bar 5, -\bar 1, -\bar 1) &&& \\
{ 10_{17}} &&4,-1,1,-4 &
d_R\,( ST^4S^\dagger \bT^{-1}\bS\bT ST^{-4}S^\dagger )_{_{11}}& (4, -5, 1, 1) & &&\\
{ 10_{18}} &&&
d_R\,( \bS\bT^4\bS\bT^{-1}STS^\dagger\bT^{-2}ST^2S^\dagger )_{_{11}}& &&& \\
{ 10_{19}} &&&
\!\!d_R\,( ST^4S^\dagger\bT^{-1}\bS\bT ST^{-1}S^\dagger \bT^3\bS )_{_{11}}\!\!& (\bar 4, 5,-1,-1,-1) &&& \\
 { 10_{20}} &&&
 d_R\,( ST^3S^\dagger \bT^{-5}\bS\bT^2\bS )_{_{11}}& (-2, 1, 3, 1^5) &&& \\
 { 10_{21}} &&&
 d_R\,( \bS\bT^{-3}ST^4S^\dagger\bT^{-1}\bS\bT^2\bS )_{_{11}}& (-\bar 3,\bar 1,\bar 1,\bar 3,\bar 1,\bar 1,\bar 1) &&&\\
 { 10_{22}} &&&
 d_R\,( ST^3S^\dagger \bT^{-3}\bS\bT ST^{-3}S^\dagger )_{_{11}}& (-4, 1, 1, 3, 1, 1) &&& \\
 { 10_{23}} &&&
 d_R\,( ST^2S^\dagger \bT^{-1}\bS\bT ST^{-3}S^\dagger\bT^3\bS )_{_{11}}& &&& \\
 { 10_{24}} &&&
 d_R\,( \bS\bT^{-3}ST^2S^\dagger \bT^{-3}\bS\bT^2\bS )_{_{11}}& &&& \\
 { 10_{25}} &&&
 d_R\,( \bS\bT^2\bS\bT^{-1}ST^2S^\dagger\bT^{-2}ST^3S^\dagger )_{_{11}}& &&& \\
 { 10_{26}} &&&
 d_R\,( ST^3S^\dagger\bT^{-1}\bS\bT ST^{-2}S^\dagger\bT^3\bS )_{_{11}}& &&& \\
 { 10_{27}} &&&
 d_R\,(ST^{2}S^\dagger \bT^{-1}\bS\bT^{1}ST^{-1}S^\dagger\bT^{2}ST^{-3}S^\dagger )_{_{11}}& &&& \\
 { 10_{28}} &&&
 d_R\,( ST^{2}S^\dagger \bT^{-1}\bS\bT^{3}ST^{-1}S^\dagger\bT^{2}ST^{-1}S^\dagger )_{_{11}}& (\bar 4, 3,-1^5)&&& \\
 { 10_{29}} &&&
 d_R\,( \bS\bT^2\bS\bT^{-2}\bS\bT^2\bS\bT^{-1}ST^3S^\dagger )_{_{11}}& &&& \\
 { 10_{30}} &&&
 d_R\,( \bS\bT^{-3}STS^\dagger\bT^{-2}STS^\dagger\bT^{-1}\bS\bT^2\bS )_{_{11}}& &&& \\
 { 10_{31}} &&&
 d_R\,( \bS\bT^3ST^{-1}S^\dagger\bT\bS\bT^{-3}ST^2S^\dagger )_{_{11}}& &&& \\
 { 10_{32}} &&&
 d_R\,(ST^2S^\dagger \bT^{-2}STS^\dagger \bT^{-1}\bS\bT ST^{-3}S^\dagger )_{_{11}}& &&& \\
 { 10_{33}} &&&
 d_R\,( \bS\bT^3 ST^{-1}S^\dagger \bT\bS\bT^{-1} STS^\dagger \bT^{-3}\bS )_{_{11}}& &&& \\
 { 10_{34}} &&&
 d_R\,( ST^2S^\dagger\bT^{-1}\bS\bT^5 ST^{-2}S^\dagger)_{_{11}}& (2, -3, 1^6) &&& \\
 { 10_{35}} &&&
 d_R\,( \bS\bT^2\bS\bT^{-2}\bS\bT^4\bS\bT^{-2}\bS )_{_{11}}& &&& \\
 { 10_{36}} &&&
 d_R\,( ST^2S^\dagger \bT^{-4}STS^\dagger\bT^{-1}\bS\bT^2\bS )_{_{11}}& &&& \\
 { 10_{37}} &&&
 d_R\,( ST^2S^\dagger \bT^{-3}\bS\bT^3 ST^{-2}S^\dagger)_{_{11}}& &&& \\
 { 10_{38}} &&&
 d_R\,( ST^2S^\dagger \bT^{-2}ST^2S^\dagger\bT^{1}\bS\bT^{-2}\bS\bT^2\bS )_{_{11}}& &&& \\
 { 10_{39}} &&&
 d_R\,( ST^2S^\dagger \bT^{-2} ST^3S^\dagger \bT^{-1}\bS\bT^2\bS)_{_{11}}& &&& \\
 { 10_{40}} &&&
 d_R\,( ST^{-2}S^\dagger\bT\bS\bT^{-1}ST^2S^\dagger\bT^{-2}ST^2S^\dagger )_{_{11}}& &&& \\
 { 10_{41}} &&&
 d_R\,( \bS\bT^2\bS\bT^{-1}ST^2S^\dagger\bT^{-1}\bS\bT^2\bS\bT^{-2}\bS )_{_{11}}& &&& \\
 { 10_{42}} &&&
 d_R\,(STS^\dagger \bT^{-1}\bS\bT^2\bS\bT^{-1}ST^2S^\dagger \bT^2ST^{-2}S^\dagger )_{_{11}}& &&& \\
 { 10_{43}} &&&
 d_R\,(ST^2S^\dagger\bT^{-1}\bS\bT^2\bS\bT^{-2}\bS\bT ST^{-2}S^\dagger )_{_{11}}& &&& \\
 { 10_{44}} &&&
 d_R\,(\bS\bT^2\bS\bT^{-1}STS^\dagger\bT^{-1}\bS\bT^2\bS\bT^{-1}ST^2S^\dagger )_{_{11}}& &&& \\
 { 10_{45}} &&&
 \!d_R\,(\bS\bT^2\bS\bT^{-1}STS^\dagger\bT^{-1}\bS\bT ST^{-1}S^\dagger\bT\bS\bT^{-2}\bS )_{_{11}}\!\!
 & &&& \\
 { 10_{46}} &&5,-1,3,-1 &3& (-2, 3, 5, 1) &&& \\
 { 10_{47}} &&5,-1,2,-2 &3&  (2, -3, 5, 1) &&& \\
  { 10_{48}} &&4,-2,1,-3 &3&  (2, -5, 1, 3) &&& \\
  {\bf 10_{49}} &&&3&  (2,3,5,-1^3)\bigvee(\bar 2,-5,-3,1^2)&&& \\
  {\bf 10_{50}} &&&3& &&& \\
  {\bf 10_{51}} &&&3& &&& \\
  {\bf 10_{52}} &&&3& &&& \\
  {\bf 10_{53}} &&&3& &&& \\
  {\bf 10_{54}} &&&3& &&& \\
  {\bf 10_{55}} &&&3& &&& \\
  &&&& &&&
  \end{array}
  $
  }

  \newpage

  \centerline{
  $
  \begin{array}{|c|c|c|c||c|c|c|c|}
  {\bf 10_{56}} &&&3& &&& \\
  {\bf 10_{57}} &&&3& &&& \\
  {\bf 10_{58}} &&&3& &&& \\
  {\bf 10_{59}} &&&3& &&& \\
  {\bf 10_{60}} &&&3& &&& \\
  {\bf 10_{61}} &&&3& (3, 3, \bar 4) &&& \\
  { 10_{62}} &&4,-1,3,-2 &3& (4, -3, 1, 3)  &&& \\
  {\bf 10_{63}} &&&3& (\bar 4,-3,-3,1,1)&&& \\
  { 10_{64}} &&3,-1,3,-3  &3&  (-4, 3, 3, 1)  &&& \\
  {\bf 10_{65}} &&&3& (\bar 4,3,-3,-1,-1)&&& \\
  {\bf 10_{66}} &&&3& &&& \\
  {\bf 10_{67}} &&&3& &&& \\
  {\bf 10_{68}} &&&3& &&& \\
  {\bf 10_{69}} &&&3& (4,3,3,-1^3)&&& \\
  {\bf 10_{70}} &&&3& &&& \\
{\bf 10_{71}} &&&3& &     \overline{\sum\limits_{ X\in R\otimes \bar R}^{{\phantom.}}} \!\!
\frac{d_R\cdot(\bS\bT^2\bS\bT^{-2}\bS)_{_{1X}}(ST^2S^\dagger T^{-1}\bS)_{_{1X}}(ST^{-3}S^\dagger\bT^{-1}\bS)_{_{1X}}}
{\bar S_{_{1X}}^{\phantom{.^5}}}&& \\
{\bf 10_{72}} &&&3& &&& \\
{\bf 10_{73}} &&&3& &&& \\
{\bf 10_{74}} &&&3& (-\bar 3,\bar 1,\bar 3,\bar 3,\bar 1)  &&& \\
{\bf 10_{75}} &&&3& &&& \\
{\bf 10_{76}} &&&3&  (1, 3, 3, 1, \bar 2)  &&&\\
{\bf 10_{77}} &&&3& (2, -3, 1, 3, 1, 1) &&& \\
{\bf 10_{78}} &&&3& (\bar 2,-3,-3,1^4)&&& \\
10_{79} &&3,-2,2,-3 &3&&&& \\
{\bf 10_{80}} &&&3& &&& \\
{\bf 10_{81}} &&&3& &&& \\
10_{82} &&4,-1,1,-1,1,-2&3&&&& \\
{\bf 10_{83}} &&&3& &&& \\
{\bf 10_{84}} &&&3& &&& \\
10_{85} &&4,-1,2,-1,1,-1&3& &&& \\
{\bf 10_{86}} &&&3& &&& \\
{\bf 10_{87}} &&&3& &&& \\
{\bf 10_{88}} &&&3& &&& \\
{\bf 10_{89}} &&&3& &&& \\
{\bf 10_{90}} &&&3& &&& \\
10_{91} &&3,-1,1,-2,1,-2 &3& &&& \\
{\bf 10_{92}} &&&3& &&& \\
{\bf 10_{93}} &&&3& &&& \\
10_{94} &&3,-1,2,-2,1,-1&3&&&& \\
{\bf 10_{95}} &&&3& &&& \\
{\bf 10_{96}} &&&3& &&& \\
{\bf 10_{97}} &&&3& &&& \\
{\bf 10_{98}} &&&3& &&& \\
10_{99} &&2,-1,2,-2,1,-2&3& &&& \\
10_{100}&&3,-1,2,-1,2,-1&3& &&& \\
{\bf 10_{101}} &&&3& &&& \\
{\bf 10_{102}} &&&3& &&& \\
{\bf 10_{103}} &&&3& &&& \\
10_{104}&&3,-2,1,-1,1,-2&3& &&& \\
{\bf 10_{105}} &&&3& &&& \\
10_{106}&&3,-1,1,-1,2,-2&3& &&& \\
{\bf 10_{107}} &&&3& &&& \\
{\bf 10_{108}} &&&3& &&& \\
10_{109}&&2,-1,1,-2,2,-2&3& &&& \\
{\bf 10_{110}} &&&3& &&& \\
&&&& &&&
\end{array}
$
}

\newpage

\centerline{
$
  \begin{array}{|c|c|c|c||c|c|c|c|}
{\bf 10_{111}} &&&3& &&& \\
10_{112}&&3,-1,1,-1,1,-1,1,-1&3& &&& \\
{\bf 10_{113}} &&&3& &&& \\
{\bf 10_{114}} &&&3& &&& \\
{\bf 10_{115}} &&&3& &&& \\
10_{116}&&2,-1,2,-1,1,-1,1,-1&3& &&& \\
{\bf 10_{117}} &&&3& &&& \\
{\bf 10_{118}} &&&3& &&& \\
{\bf 10_{119}} &&&3& &&& \\
{\bf 10_{120}} &&&3& &&& \\
{\bf 10_{121}} &&&3& &&& \\
{\bf 10_{122}} &&&3& &&& \\
10_{123}&&(1,-1)^5&3& &&&\\
{10_{124}} &[3,5] &5,1,3,1 &3&  (2, -1, 5, 3)\bigvee (\bar 2,-5,-3) &
\overline{\sum\limits_{ X\in R\otimes \bar R}^{{\phantom.}}} \!\!
\frac{d_R\cdot(ST^5S^\dagger)_{_{1X}}(ST^2S^\dagger)_{_{1X}}(ST^2S^\dagger\bT^{-1}\bS)_{_{1X}}}
{\bar S_{_{1X}}^{\phantom{.^5}}} &&\\
{10_{125}} && 5,-1,-3,-1 &3& (2, -5, -1, 3)\bigvee (\bar 2,5,-3)  &  &&    \\
{10_{126}} && 5,1,-3,1&3& (-2, 3, -5, 1)\bigvee (\bar 2,-5,3) &  &&    \\
{10_{127}} &&5,1,-2,2&3& (2, -5, -3, 1)\bigvee (\bar 2,5,3) &&& \\
{\bf 10_{128}} &&&3& &      \overline{\sum\limits_{ X\in R\otimes \bar R}^{{\phantom.}}} \!\!
\frac{d_R\cdot(\bS\bT^{-2}\bS)_{_{1X}}(\bS\bT^{-1}ST^2S^\dagger)_{_{1X}}(\bS\bT^{-3}ST^2S^\dagger)_{_{1X}}}
{\bar S_{_{1X}}^{\phantom{.^5}}}&& \\
{\bf 10_{129}} &&&3& (2, 1, 1, -3, 1, 1)  &&&\\
{\bf 10_{130}} &&&3& &&& \\
{\bf 10_{131}} &&&3& &&& \\
{\bf 10_{132}} &&&3& &    \overline{\sum\limits_{ X\in R\otimes \bar R}^{{\phantom.}}} \!\!
\frac{d_R\cdot (ST^{-2}S^\dagger\bT^3\bS)_{_{1X}}(ST^{-2}S^\dagger)_{_{1X}}(\bS\bT^{-1}ST^2S^\dagger)_{_{1X}}}
{\bar S_{_{1X}}^{\phantom{.^5}}}&& \\
{\bf 10_{133}} &&&3& &&& \\
{\bf 10_{134}} &&&3& &&& \\
{\bf 10_{135}} &&&3& &&& \\
{\bf 10_{136}} &&&3& &  \overline{\sum\limits_{ X\in R\otimes \bar R}^{{\phantom.}}} \!\! \frac{d_R\cdot
(\bS\bT^{-2}\bS\bT^2\bS)_{_{1X}}(ST^2S^\dagger)_{_{1X}}(\bS\bT^{-3}ST^{-2}S^\dagger)_{_{1X}}}
{\bar S_{_{1X}}^{\phantom{.^5}}}&& \\
{\bf 10_{137}} &&&3& &&& \\
{\bf 10_{138}} &&&3& &&& \\
{10_{139}} &&4,1,3,2 &3&  (4, -1, 3, 3)   &&&    \\
{\bf 10_{140}} &&&3&  (-3, 3,\bar 4) &&&  \\
{10_{141}} &&4,-1,-3,-2 &3&  (4, -3, -3, 1) &&&  \\
{\bf 10_{142}} &&&3&  (3, 3, -\bar 4)  &&&  \\
{ 10_{143}} &&4,1,-3,2 &3&  (-4, 3, 1, -3) &&&  \\
{\bf 10_{144}} &&&3& (\bar 4,3,3,-1,-1)&&& \\
{\bf 10_{145}} &&&3& & \sum\limits_{ X\in R\otimes R} \!\! \frac{d_R\cdot
(ST^{-2}S^\dagger\bT^2S)_{_{1X}}(\bS\bT^3S)_{_{1X}}(ST^{-1}S^\dagger\bT\bS\bT^{-1}S)_{_{1X}} }{S_{_{1X}}}
&& \\
{\bf 10_{146}} &&&3& &&& \\
{\bf 10_{147}} &&&3& &&& \\
10_{148}&&4,1,-2,1,-1,1 &3& &&& \\
10_{149}&&4,1,-1,1,-1,2 &3& &&&\\
{\bf 10_{150}} &&&3& &&& \\
{\bf 10_{151}} &&&3& &&& \\
10_{152}&&3,2,2,3 &3& &&(\ref{10152})&\\
{\bf 10_{153}} &&&3& &&(\ref{10153})& \\
{\bf 10_{154}} &&&3& &&(\ref{10154})& \\
10_{155}&&3,1,-2,1,-2,1 &3& &&&\\
{\bf 10_{156}} &&&3& &&& \\
10_{157}&&3,2,-1,1,-1,2 &3& &&&\\
{\bf 10_{158}} &&&3& &&& \\
10_{159}&&3,1,-1,1,-2,2 &3& &&&\\
{\bf 10_{160}} &&&3& &&& \\
10_{161}&&3,1,-1,1,2,2 &3& &&&  \\
{\bf 10_{162}} &&&3& &&& \\
{\bf 10_{163}} &&&3& &&& \\
{\bf 10_{164}} &&&3& &&& \\
{\bf 10_{165}} &&&3& &&& \\
&&&&&&&
\end{array}
$
}

\subsection{Some thin knots}

A lot of explicit examples with less than nine intersections can be found in
\cite{AnoMcabling} and \cite{GuJ}.
Here we add the only 8-intersection knot $8_{15}$ which was not present in those lists
(it is pretzel, but possesses also a simpler triple-finger realization),
\be
H_{R}^{8_{15}} =Pr(-2,-3,-3, 1, 1, 1) =
\overline{\sum_{X\in R\otimes \bar R}}
\frac{d_R}{\bar S_{_{1X}}}
(\bS\bT^{2}\bS)_{_{1X}}(ST^{-3}S^\dagger)_{_{1X}}(ST^{-2}S^\dagger\bT^{1}\bS\bT^{-1})_{_{1X}}
\label{815}
\ee
and a couple of 10-intersection knots:
\be
H_{R}^{10_{71}} =
\overline{\sum_{X\in R\otimes \bar R}}
\frac{d_R}{\bar S_{_{1X}}}
(\bar S \bar T^{2}\bar S\bar T^{-2}\bar S )_{_{1X}}
(S T^{2}S^\dagger \bar T^{-1}\bar S )_{_{1X}}
(S T^{-3}S^\dagger\bar T^{-1}\bar S  )_{_{1X}}
\ee
and
\be
H_{R}^{10_{125}} = \overline{\sum_{X\in R\otimes \bar R}}
\frac{d_R}{\bar S_{_{1X}}^2}
(S T^{-5}S^\dagger  )_{_{1X}}
(S T^{-1}S^\dagger  )_{_{1X}}
(S T^{3}S^\dagger  )_{_{1X}}
(S T^{2}S^\dagger  )_{_{1X}}
\ee

\newpage

\begin{landscape}
\be
H_{[21]}^{8_{15}} =
\frac{1}{q^{24}A^{30}}\cdot
\ee

\centerline{{\tiny
$
\setlength{\arraycolsep}{1pt}
\left(\begin{array}{rrrrrrrrrrrrrrrrrrrrrrrrr}
0 & 0 & 1 & -4 & 10 & -20 & 35 & -54 & 76 & -98 & 117 & -130 & 135 &
-130 & 117 & -98 & 76 & -54 & 35 & -20 & 10 & -4 & 1 & 0 & 0 \\
&&&&&&&&&&&&&&&&&&&&&&&& \\
0 & 2 & -8 & 26 & -61 & 117 & -192 & 291 & -402 & 510 & -598 & 662 &
-685 & 662 & -598 & 510 & -402 & 291 & -192 & 117 & -61 & 26 & -8 & 2 &
0 \\
&&&&&&&&&&&&&&&&&&&&&&&& \\
3 & -11 & 37 & -85 & 163 & -275 & 430 & -609 & 795 & -975 & 1133 & -1228
& 1259 & -1228 & 1133 & -975 & 795 & -609 & 430 & -275 & 163 & -85 & 37
& -11 & 3 \\
&&&&&&&&&&&&&&&&&&&&&&&& \\
-5 & 16 & -45 & 87 & -159 & 256 & -382 & 516 & -672 & 808 & -919 & 992 &
-1028 & 992 & -919 & 808 & -672 & 516 & -382 & 256 & -159 & 87 & -45 &
16 & -5 \\
&&&&&&&&&&&&&&&&&&&&&&&& \\
2 & -8 & 15 & -28 & 50 & -78 & 105 & -143 & 170 & -203 & 219 & -239 &
234 & -239 & 219 & -203 & 170 & -143 & 105 & -78 & 50 & -28 & 15 & -8 &
2 \\
&&&&&&&&&&&&&&&&&&&&&&&& \\
0 & 1 & 0 & 5 & 2 & 0 & 17 & -17 & 56 & -59 & 94 & -87 & 123 & -87 & 94
& -59 & 56 & -17 & 17 & 0 & 2 & 5 & 0 & 1 & 0 \\
&&&&&&&&&&&&&&&&&&&&&&&& \\
0 & 0 & 0 & -1 & -5 & 0 & -13 & 10 & -25 & 8 & -42 & 19 & -35 & 19 & -42
& 8 & -25 & 10 & -13 & 0 & -5 & -1 & 0 & 0 & 0 \\
&&&&&&&&&&&&&&&&&&&&&&&& \\
0 & 0 & 0 & 0 & 0 & 0 & 3 & 6 & 2 & 4 & -2 & 13 & 5 & 13 & -2 & 4 & 2 &
6 & 3 & 0 & 0 & 0 & 0 & 0 & 0 \\
&&&&&&&&&&&&&&&&&&&&&&&& \\
0 & 0 & 0 & 0 & 0 & 0 & 0 & 0 & 0 & -3 & -2 & -2 & 2 & -2 & -2 & -3 & 0
& 0 & 0 & 0 & 0 & 0 & 0 & 0 & 0 \\
&&&&&&&&&&&&&&&&&&&&&&&& \\
0 & 0 & 0 & 0 & 0 & 0 & 0 & 0 & 0 & 0 & 0 & 0 & 1 & 0 & 0 & 0 & 0 & 0 &
0 & 0 & 0 & 0 & 0 & 0 & 0
\end{array}\right)
$
}}

\bigskip

\be
H_{[21]}^{10_{71}} =
\frac{1}{q^{30}A^{12}}\cdot
\label{1071nonpre}
\ee

\centerline{{\tiny
$
\setlength{\arraycolsep}{1pt}
\left(\begin{array}{rrrrrrrrrrrrrrrrrrrrrrrrrrrrrrr}
0 & 0 & 0 & 0 & 0 & 0 & 0 & 0 & 0 & 0 & -1 & 2 & -3 & 4 & -5 & 5 & -5 &
4 & -3 & 2 & -1 & 0 & 0 & 0 & 0 & 0 & 0 & 0 & 0 & 0 & 0 \\
&&&&&&&&&&&&&&&&&&&&&&&&&&&&&& \\
0 & 0 & 0 & 0 & 0 & 0 & 0 & 2 & -3 & 5 & -8 & 13 & -15 & 20 & -21 & 23 &
-21 & 20 & -15 & 13 & -8 & 5 & -3 & 2 & 0 & 0 & 0 & 0 & 0 & 0 & 0 \\
&&&&&&&&&&&&&&&&&&&&&&&&&&&&&& \\
0 & 0 & 0 & 0 & -1 & -1 & 4 & -11 & 20 & -39 & 56 & -84 & 104 & -131 &
140 & -150 & 140 & -131 & 104 & -84 & 56 & -39 & 20 & -11 & 4 & -1 & -1
& 0 & 0 & 0 & 0 \\
&&&&&&&&&&&&&&&&&&&&&&&&&&&&&& \\
0 & 0 & 2 & -5 & 12 & -18 & 27 & -26 & 27 & -4 & -18 & 70 & -105 & 155 &
-175 & 206 & -175 & 155 & -105 & 70 & -18 & -4 & 27 & -26 & 27 & -18 &
12 & -5 & 2 & 0 & 0 \\
&&&&&&&&&&&&&&&&&&&&&&&&&&&&&& \\
-1 & 4 & -16 & 40 & -89 & 164 & -290 & 451 & -681 & 938 & -1251 & 1545 &
-1865 & 2085 & -2267 & 2301 & -2267 & 2085 & -1865 & 1545 & -1251 & 938
& -681 & 451 & -290 & 164 & -89 & 40 & -16 & 4 & -1 \\
&&&&&&&&&&&&&&&&&&&&&&&&&&&&&& \\
4 & -18 & 59 & -144 & 315 & -594 & 1046 & -1679 & 2548 & -3575 & 4795 &
-6035 & 7245 & -8187 & 8874 & -9074 & 8874 & -8187 & 7245 & -6035 & 4795
& -3575 & 2548 & -1679 & 1046 & -594 & 315 & -144 & 59 & -18 & 4 \\
&&&&&&&&&&&&&&&&&&&&&&&&&&&&&& \\
-6 & 28 & -92 & 224 & -486 & 923 & -1624 & 2593 & -3902 & 5473 & -7290 &
9111 & -10884 & 12287 & -13258 & 13545 & -13258 & 12287 & -10884 & 9111
& -7290 & 5473 & -3902 & 2593 & -1624 & 923 & -486 & 224 & -92 & 28 & -6 \\
&&&&&&&&&&&&&&&&&&&&&&&&&&&&&& \\
4 & -18 & 62 & -152 & 333 & -632 & 1117 & -1784 & 2691 & -3770 & 5036 &
-6301 & 7539 & -8514 & 9205 & -9398 & 9205 & -8514 & 7539 & -6301 & 5036
& -3770 & 2691 & -1784 & 1117 & -632 & 333 & -152 & 62 & -18 & 4 \\
&&&&&&&&&&&&&&&&&&&&&&&&&&&&&& \\
-1 & 4 & -17 & 42 & -96 & 178 & -314 & 491 & -741 & 1008 & -1330 & 1641
& -1960 & 2166 & -2351 & 2395 & -2351 & 2166 & -1960 & 1641 & -1330 &
1008 & -741 & 491 & -314 & 178 & -96 & 42 & -17 & 4 & -1 \\
&&&&&&&&&&&&&&&&&&&&&&&&&&&&&& \\
0 & 0 & 2 & -5 & 13 & -19 & 29 & -27 & 22 & 10 & -46 & 121 & -179 & 246
& -283 & 322 & -283 & 246 & -179 & 121 & -46 & 10 & 22 & -27 & 29 & -19
& 13 & -5 & 2 & 0 & 0 \\
&&&&&&&&&&&&&&&&&&&&&&&&&&&&&& \\
0 & 0 & 0 & 0 & -1 & -1 & 4 & -12 & 22 & -44 & 66 & -98 & 122 & -153 &
165 & -176 & 165 & -153 & 122 & -98 & 66 & -44 & 22 & -12 & 4 & -1 & -1
& 0 & 0 & 0 & 0 \\
&&&&&&&&&&&&&&&&&&&&&&&&&&&&&& \\
0 & 0 & 0 & 0 & 0 & 0 & 0 & 2 & -3 & 5 & -8 & 13 & -14 & 18 & -19 & 21 &
-19 & 18 & -14 & 13 & -8 & 5 & -3 & 2 & 0 & 0 & 0 & 0 & 0 & 0 & 0 \\
&&&&&&&&&&&&&&&&&&&&&&&&&&&&&& \\
0 & 0 & 0 & 0 & 0 & 0 & 0 & 0 & 0 & 0 & -1 & 2 & -3 & 4 & -5 & 5 & -5 &
4 & -3 & 2 & -1 & 0 & 0 & 0 & 0 & 0 & 0 & 0 & 0 & 0 & 0
\end{array}\right)
$
}}

\bigskip

\be
H_{[21]}^{10_{125}} =
\frac{1}{q^{32}A^{6}}\cdot
\label{10125pre}
\ee

\centerline{{\tiny
$
\setlength{\arraycolsep}{1pt}
\left(\begin{array}{rrrrrrrrrrrrrrrrrrrrrrrrrrrrrrrrr}
0 & 0 & 0 & 0 & 0 & -7 & 7 & -21 & 28 & -49 & 42 & -70 & 63 & -84 & 42 &
-70 & 49 & -70 & 42 & -84 & 63 & -70 & 42 & -49 & 28 & -21 & 7 & -7 & 0
& 0 & 0 & 0 & 0 \\
&&&&&&&&&&&&&&&&&&&&&&&&&&&&&&&& \\
0 & 0 & 7 & 0 & 14 & 0 & 28 & 28 & 21 & 70 & 0 & 133 & 7 & 154 & 21 &
154 & 49 & 154 & 21 & 154 & 7 & 133 & 0 & 70 & 21 & 28 & 28 & 0 & 14 & 0
& 7 & 0 & 0 \\
&&&&&&&&&&&&&&&&&&&&&&&&&&&&&&&& \\
-7 & 7 & -28 & 21 & -84 & 35 & -140 & 28 & -203 & -21 & -245 & -84 &
-301 & -126 & -343 & -147 & -378 & -147 & -343 & -126 & -301 & -84 &
-245 & -21 & -203 & 28 & -140 & 35 & -84 & 21 & -28 & 7 & -7 \\
&&&&&&&&&&&&&&&&&&&&&&&&&&&&&&&& \\
7 & 0 & 28 & -7 & 84 & -7 & 161 & 7 & 252 & 49 & 336 & 112 & 399 & 189 &
462 & 210 & 483 & 210 & 462 & 189 & 399 & 112 & 336 & 49 & 252 & 7 & 161
& -7 & 84 & -7 & 28 & 0 & 7 \\
&&&&&&&&&&&&&&&&&&&&&&&&&&&&&&&& \\
0 & -7 & -7 & -21 & -21 & -42 & -63 & -70 & -119 & -105 & -182 & -154 &
-224 & -196 & -252 & -231 & -266 & -231 & -252 & -196 & -224 & -154 &
-182 & -105 & -119 & -70 & -63 & -42 & -21 & -21 & -7 & -7 & 0 \\
&&&&&&&&&&&&&&&&&&&&&&&&&&&&&&&& \\
0 & 0 & 0 & 7 & 7 & 21 & 14 & 35 & 35 & 49 & 63 & 56 & 84 & 70 & 98 & 77
& 91 & 77 & 98 & 70 & 84 & 56 & 63 & 49 & 35 & 35 & 14 & 21 & 7 & 7 & 0
& 0 & 0 \\
&&&&&&&&&&&&&&&&&&&&&&&&&&&&&&&& \\
0 & 0 & 0 & 0 & 0 & 0 & -7 & 0 & -14 & 7 & -28 & 7 & -28 & 7 & -28 & 7 &
-35 & 7 & -28 & 7 & -28 & 7 & -28 & 7 & -14 & 0 & -7 & 0 & 0 & 0 & 0 & 0 & 0
\end{array}\right)
$
}}

\end{landscape}

\subsection{The first thick knots}

Thick are the knots, for which the fundamental superpolynomials are {\it not} obtained
from HOMFLY by a change of variables and Khovanov homologies have non-trivial entries
off critical diagonals (marked in red in \cite{katlas}).
In most cases, these superpolynomials have more terms than the HOMFLY polynomial
(though this discrepancy can often be eliminated by switching to differential
expansion {\it a la} \cite{ArthMM}).

The first thick knots in the Rolfsen table of \cite{katlas} are
\be
8_{19}, \ \ 9_{42},\ \  10_{124}, \ 10_{128},\ 10_{132},\ 10_{136},\ 10_{139},
10_{145},\ 10_{152},\ 10_{153},\  10_{154},\  10_{161}
\ee
The next have eleven and more intersections.

\subsubsection{3-strand cases
\label{3strthick}}

Five of these are 3-strand and therefore the answers for $H_{[21]}$ are easily
available by the methods of \cite{AnoMcabling}:

\be
H_{[21]}^{8_{19}} =
\frac{1}{q^{30}A^{30}}\cdot
\label{819threestr}
\ee

\centerline{{\tiny
$
\setlength{\arraycolsep}{1pt}
\left(\begin{array}{rrrrrrrrrrrrrrrrrrrrrrrrrrrrrrr}
1 & 0 & 2 & 1 & 3 & 1 & 5 & 3 & 6 & 3 & 8 & 4 & 10 & 3 & 10 & 5 & 10 & 3
& 10 & 4 & 8 & 3 & 6 & 3 & 5 & 1 & 3 & 1 & 2 & 0 & 1 \\
&&&&&&&&&&&&&&&&&&&&&&&&&&&&&& \\
-1 & -1 & -3 & -3 & -6 & -6 & -10 & -10 & -16 & -14 & -21 & -16 & -25 &
-19 & -27 & -19 & -27 & -19 & -25 & -16 & -21 & -14 & -16 & -10 & -10 &
-6 & -6 & -3 & -3 & -1 & -1 \\
&&&&&&&&&&&&&&&&&&&&&&&&&&&&&& \\
0 & 1 & 1 & 3 & 4 & 8 & 9 & 13 & 15 & 20 & 22 & 26 & 27 & 31 & 29 & 32 &
29 & 31 & 27 & 26 & 22 & 20 & 15 & 13 & 9 & 8 & 4 & 3 & 1 & 1 & 0 \\
&&&&&&&&&&&&&&&&&&&&&&&&&&&&&& \\
0 & 0 & 0 & -1 & -1 & -3 & -4 & -7 & -8 & -13 & -13 & -18 & -17 & -22 &
-18 & -25 & -18 & -22 & -17 & -18 & -13 & -13 & -8 & -7 & -4 & -3 & -1 &
-1 & 0 & 0 & 0 \\
&&&&&&&&&&&&&&&&&&&&&&&&&&&&&& \\
0 & 0 & 0 & 0 & 0 & 0 & 1 & 1 & 3 & 3 & 5 & 5 & 7 & 8 & 8 & 8 & 8 & 8 &
7 & 5 & 5 & 3 & 3 & 1 & 1 & 0 & 0 & 0 & 0 & 0 & 0 \\
&&&&&&&&&&&&&&&&&&&&&&&&&&&&&& \\
0 & 0 & 0 & 0 & 0 & 0 & 0 & 0 & 0 & 0 & -1 & -1 & -2 & -1 & -2 & -1 & -2
& -1 & -2 & -1 & -1 & 0 & 0 & 0 & 0 & 0 & 0 & 0 & 0 & 0 & 0 \\
&&&&&&&&&&&&&&&&&&&&&&&&&&&&&& \\
0 & 0 & 0 & 0 & 0 & 0 & 0 & 0 & 0 & 0 & 0 & 0 & 0 & 0 & 0 & 1 & 0 & 0 &
0 & 0 & 0 & 0 & 0 & 0 & 0 & 0 & 0 & 0 & 0 & 0 & 0
\end{array}\right)
$
}}

\bigskip

\be
H_{[21]}^{10_{124}} = \frac{1}{q^{40}A^{36}}\cdot
\ee
{\tiny
\be
\hspace{-1cm}
\setlength{\arraycolsep}{1pt}
\left(\begin{array}{rrrrrrrrrrrrrrrrrrrrrrrrrrrrrrrrrrrrrrrrr}
1 & 0 & 2 & 1 & 3 & 3 & 4 & 5 & 7 & 7 & 8 & 10 & 11 & 13 & 12 & 15 & 13 & 18 & 13 & 18 & 15 & 18 & 13 & 18 & 13 & 15 & 12 & 13 & 11 & 10 & 8 & 7 & 7 & 5 & 4 & 3 & 3 & 1 & 2 & 0 & 1 \\
&&&&&&&&&&&&&&&&&&&&&&&&&&&&&&&&&&&&&&&& \\
-1 & -1 & -3 & -4 & -6 & -9 & -11 & -16 & -18 & -25 & -27 & -35 & -35 & -45 & -43 & -55 & -49 & -61 & -52 & -65 & -54 & -65 & -52 & -61 & -49 & -55 & -43 & -45 & -35 & -35 & -27 & -25 & -18 & -16 & -11 & -9 & -6 & -4 & -3 & -1 & -1 \\
&&&&&&&&&&&&&&&&&&&&&&&&&&&&&&&&&&&&&&&& \\
0 & 1 & 1 & 4 & 4 & 10 & 11 & 19 & 21 & 31 & 35 & 46 & 50 & 62 & 65 & 76 & 76 & 87 & 85 & 92 & 86 & 92 & 85 & 87 & 76 & 76 & 65 & 62 & 50 & 46 & 35 & 31 & 21 & 19 & 11 & 10 & 4 & 4 & 1 & 1 & 0 \\
&&&&&&&&&&&&&&&&&&&&&&&&&&&&&&&&&&&&&&&& \\
0 & 0 & 0 & -1 & -1 & -4 & -5 & -9 & -13 & -17 & -24 & -29 & -38 & -41 & -53 & -53 & -64 & -63 & -72 & -67 & -76 & -67 & -72 & -63 & -64 & -53 & -53 & -41 & -38 & -29 & -24 & -17 & -13 & -9 & -5 & -4 & -1 & -1 & 0 & 0 & 0 \\
&&&&&&&&&&&&&&&&&&&&&&&&&&&&&&&&&&&&&&&& \\
0 & 0 & 0 & 0 & 0 & 0 & 1 & 1 & 4 & 4 & 9 & 8 & 15 & 14 & 23 & 21 & 29 & 26 & 33 & 29 & 34 & 29 & 33 & 26 & 29 & 21 & 23 & 14 & 15 & 8 & 9 & 4 & 4 & 1 & 1 & 0 & 0 & 0 & 0 & 0 & 0 \\
&&&&&&&&&&&&&&&&&&&&&&&&&&&&&&&&&&&&&&&& \\
0 & 0 & 0 & 0 & 0 & 0 & 0 & 0 & 0 & 0 & -1 & -1 & -3 & -3 & -4 & -5 & -5 & -8 & -6 & -9 & -6 & -9 & -6 & -8 & -5 & -5 & -4 & -3 & -3 & -1 & -1 & 0 & 0 & 0 & 0 & 0 & 0 & 0 & 0 & 0 & 0 \\
&&&&&&&&&&&&&&&&&&&&&&&&&&&&&&&&&&&&&&&& \\
0 & 0 & 0 & 0 & 0 & 0 & 0 & 0 & 0 & 0 & 0 & 0 & 0 & 0 & 0 & 1 & 0 & 2 & -1 & 2 & 0 & 2 & -1 & 2 & 0 & 1 & 0 & 0 & 0 & 0 & 0 & 0 & 0 & 0 & 0 & 0 & 0 & 0 & 0 & 0 & 0
\end{array}\right)
\label{10124threestr}
\ee
}

\bigskip

\be
H_{[21]}^{10_{139}} = \frac{A^{12}}{q^{22}}\cdot
\ee
{\tiny
\be
\nn
\setlength{\arraycolsep}{1pt}
\left(\begin{array}{rrrrrrrrrrrrrrrrrrrrrrr}
0 & 0 & 0 & 0 & 0 & 0 & 0 & 0 & 0 & 0 & 0 & -1 & 0 & 0 & 0 & 0 & 0 & 0 & 0 & 0 & 0 & 0 & 0 \\
&&&&&&&&&&&&&&&&&&&&&& \\
0 & 0 & 0 & 0 & 0 & 0 & 0 & 0 & 1 & 0 & 1 & -1 & 1 & 0 & 1 & 0 & 0 & 0 & 0 & 0 & 0 & 0 & 0 \\
&&&&&&&&&&&&&&&&&&&&&& \\
0 & 0 & 0 & 0 & 0 & 0 & -1 & 0 & -2 & 2 & -3 & 2 & -3 & 2 & -2 & 0 & -1 & 0 & 0 & 0 & 0 & 0 & 0 \\
&&&&&&&&&&&&&&&&&&&&&& \\
0 & 0 & 0 & 1 & 0 & 3 & -2 & 3 & -2 & 5 & -3 & 3 & -3 & 5 & -2 & 3 & -2 & 3 & 0 & 1 & 0 & 0 & 0 \\
&&&&&&&&&&&&&&&&&&&&&& \\
0 & -1 & 0 & -1 & 2 & -5 & 4 & -9 & 10 & -13 & 12 & -16 & 12 & -13 & 10 & -9 & 4 & -5 & 2 & -1 & 0 & -1 & 0 \\
&&&&&&&&&&&&&&&&&&&&&& \\
0 & -1 & 2 & -1 & 2 & -1 & 1 & 3 & 0 & 4 & -1 & 5 & -1 & 4 & 0 & 3 & 1 & -1 & 2 & -1 & 2 & -1 & 0 \\
&&&&&&&&&&&&&&&&&&&&&& \\
-1 & 2 & -3 & 4 & -8 & 8 & -9 & 7 & -10 & 5 & -8 & 1 & -8 & 5 & -10 & 7 & -9 & 8 & -8 & 4 & -3 & 2 & -1 \\
&&&&&&&&&&&&&&&&&&&&&& \\
1 & -1 & 2 & -3 & 3 & -1 & 1 & 4 & -6 & 12 & -11 & 16 & -11 & 12 & -6 & 4 & 1 & -1 & 3 & -3 & 2 & -1 & 1 \\
&&&&&&&&&&&&&&&&&&&&&& \\
0 & 0 & -1 & 0 & 1 & -4 & 7 & -12 & 16 & -24 & 24 & -26 & 24 & -24 & 16 & -12 & 7 & -4 & 1 & 0 & -1 & 0 & 0 \\
&&&&&&&&&&&&&&&&&&&&&& \\
0 & 1 & 0 & 0 & 0 & 1 & -1 & 4 & -6 & 9 & -11 & 14 & -11 & 9 & -6 & 4 & -1 & 1 & 0 & 0 & 0 & 1 & 0
\end{array}\right)
\label{10139threestr}
\ee}

\begin{landscape}

\be
H_{[21]}^{10_{152}} =
\frac{A^{24}}{q^{40}}\cdot
\ee
{\tiny
\be
\nn
\setlength{\arraycolsep}{1pt}
\left(\begin{array}{rrrrrrrrrrrrrrrrrrrrrrrrrrrrrrrrrrrrrrrrr}
0 & 0 & 0 & 0 & 0 & 0 & 0 & 0 & 0 & 0 & 0 & 0 & 0 & 0 & 8 & -4 & 12 & -8 & 22 & -14 & 22 & -14 & 22 & -8 & 12 & -4 & 8 & 0 & 0 & 0 & 0 & 0 & 0 & 0 & 0 & 0 & 0 & 0 & 0 & 0 & 0 \\
&&&&&&&&&&&&&&&&&&&&&&&&&&&&&&&&&&&&&&&& \\
0 & 0 & 0 & 0 & 0 & 0 & 0 & 0 & 0 & 0 & -10 & -6 & -18 & -18 & -14 & -46 & -6 & -74 & 10 & -92 & 8 & -92 & 10 & -74 & -6 & -46 & -14 & -18 & -18 & -6 & -10 & 0 & 0 & 0 & 0 & 0 & 0 & 0 & 0 & 0 & 0 \\
&&&&&&&&&&&&&&&&&&&&&&&&&&&&&&&&&&&&&&&& \\
0 & 0 & 0 & 0 & 0 & 0 & 4 & 8 & 16 & 26 & 36 & 42 & 74 & 66 & 114 & 94 & 134 & 138 & 130 & 172 & 124 & 172 & 130 & 138 & 134 & 94 & 114 & 66 & 74 & 42 & 36 & 26 & 16 & 8 & 4 & 0 & 0 & 0 & 0 & 0 & 0 \\
&&&&&&&&&&&&&&&&&&&&&&&&&&&&&&&&&&&&&&&& \\
0 & 0 & 0 & -4 & -2 & -18 & -16 & -38 & -50 & -66 & -108 & -92 & -202 & -94 & -324 & -76 & -444 & -54 & -530 & -40 & -564 & -40 & -530 & -54 & -444 & -76 & -324 & -94 & -202 & -92 & -108 & -66 & -50 & -38 & -16 & -18 & -2 & -4 & 0 & 0 & 0 \\
&&&&&&&&&&&&&&&&&&&&&&&&&&&&&&&&&&&&&&&& \\
0 & 4 & 0 & 14 & 12 & 34 & 30 & 72 & 58 & 128 & 96 & 190 & 166 & 214 & 276 & 196 & 416 & 146 & 532 & 104 & 576 & 104 & 532 & 146 & 416 & 196 & 276 & 214 & 166 & 190 & 96 & 128 & 58 & 72 & 30 & 34 & 12 & 14 & 0 & 4 & 0 \\
&&&&&&&&&&&&&&&&&&&&&&&&&&&&&&&&&&&&&&&& \\
-2 & -4 & -4 & -16 & -10 & -32 & -22 & -64 & -28 & -124 & -14 & -206 & 8 & -284 & -4 & -316 & -42 & -308 & -96 & -294 & -116 & -294 & -96 & -308 & -42 & -316 & -4 & 0 & 0 & 0 & 0 & 0 & 0 & -64 & -22 & -32 & -10 & -16 & -4 & -4 & -2 \\
&&&&&&&&&&&&&&&&&&&&&&&&&&&&&&&&&&&&&&&& \\
2 & 0 & 4 & 6 & 0 & 16 & 4 & 22 & 6 & 36 & 0 & 70 & -28 & 116 & -58 & 152 & -70 & 168 & -68 & 164 & -60 & 164 & -68 & 168 & -70 & 152 & -58 & 116 & -28 & 70 & 0 & 36 & 6 & 22 & 4 & 16 & 0 & 6 & 4 & 0 & 2
\end{array}\right)
\label{10152threestr}
\ee}

\bigskip

\be
H_{[21]}^{10_{161}} = \frac{A^{18}}{q^{30}}\cdot
\ee
{\tiny
\be
\setlength{\arraycolsep}{1pt}
\nn
\left(\begin{array}{rrrrrrrrrrrrrrrrrrrrrrrrrrrrrrr}
0 & 0 & 0 & 0 & 0 & 0 & 0 & 0 & 0 & -2 & 4 & -4 & 8 & -16 & 16 & -14 & 16 & -16 & 8 & -4 & 4 & -2 & 0 & 0 & 0 & 0 & 0 & 0 & 0 & 0 & 0 \\
&&&&&&&&&&&&&&&&&&&&&&&&&&&&&& \\
0 & 0 & 0 & 0 & 0 & 0 & 0 & -2 & -2 & 2 & -2 & 4 & -8 & 8 & -8 & 10 & -8 & 8 & -8 & 4 & -2 & 2 & -2 & -2 & 0 & 0 & 0 & 0 & 0 & 0 & 0 \\
&&&&&&&&&&&&&&&&&&&&&&&&&&&&&& \\
0 & 0 & 0 & 2 & 0 & 4 & -4 & 8 & -10 & 16 & -14 & 12 & -12 & 8 & -4 & 0 & -4 & 8 & -12 & 12 & -14 & 16 & -10 & 8 & -4 & 4 & 0 & 2 & 0 & 0 & 0 \\
&&&&&&&&&&&&&&&&&&&&&&&&&&&&&& \\
0 & -2 & 2 & -4 & 6 & -6 & 4 & -2 & 4 & 2 & -2 & 6 & -4 & 14 & -8 & 14 & -8 & 14 & -4 & 6 & -2 & 2 & 4 & -2 & 4 & -6 & 6 & -4 & 2 & -2 & 0 \\
&&&&&&&&&&&&&&&&&&&&&&&&&&&&&& \\
-2 & 4 & -4 & 2 & -4 & 2 & 2 & -10 & 10 & -12 & 10 & -16 & 6 & -2 & -2 & -4 & -2 & -2 & 6 & -16 & 10 & -12 & 10 & -10 & 2 & 2 & -4 & 2 & -4 & 4 & -2 \\
&&&&&&&&&&&&&&&&&&&&&&&&&&&&&& \\
0 & -2 & 2 & -4 & 0 & -2 & -2 & 2 & -4 & -6 & 8 & -22 & 28 & -44 & 44 & -50 & 44 & -44 & 28 & -22 & 8 & -6 & -4 & 2 & -2 & -2 & 0 & -4 & 2 & -2 & 0 \\
&&&&&&&&&&&&&&&&&&&&&&&&&&&&&& \\
2 & 0 & 0 & 4 & -2 & 2 & 2 & 4 & 2 & 0 & -4 & 20 & -22 & 32 & -38 & 50 & -38 & 32 & -22 & 20 & -4 & 0 & 2 & 4 & 2 & 2 & -2 & 4 & 0 & 0 & 2
\end{array}\right)
\label{10161threestr}
\ee}

\end{landscape}

\bigskip

\noindent
Moreover, two of these are torus knots, $8_{19}=Torus[3,4]$ and $10_{124}=Torus[3,5]$,
and therefore {\it arbitrary} colored HOMFLY polynomials for them are available:
provided by the Rosso-Jones formula \cite{RJ,DMMSS}.

\subsubsection{4-parallel pretzel finger cases}

Three of the above thick knots are of the pretzel type
and are described by the simple formulas:
\be
H_R^{Pr(n_1,n_2,\overline{n_3})} = \overline{\sum_{X\in R\otimes\bar R}}
\underbrace{\frac{d_R^2}{\sqrt{\bar d_X }}}_{\frac{d_R}{\bar S_{1X} }}
\bar{\cal A}^{\rm par}_{1X}(n_1)\bar{\cal A}^{\rm par}_{1X}(n_2)
\bar{\cal A}^{\rm ea}_{1X}(\overline{n_3})
\ee
and
\be
H_R^{Pr(n_1,n_2,n_3,n_4)} = \overline{\sum_{X\in R\otimes\bar R}}
\underbrace{\frac{d_R^3}{{\bar d_X }}}_{\frac{d_R}{\bar S_{1X}^2 }}
\bar{\cal A}^{\rm par}_{1X}(n_1)\bar{\cal A}^{\rm par}_{1X}(n_2)
\bar{\cal A}^{\rm par}_{1X}(n_3)\bar{\cal A}^{\rm par}_{1X}(n_4)
\ee

These three cases are 3-strand (and thus already known)
\be
8_{19} = Torus[3,4]=Pr(3,3,-\bar 2)\ \ \ \Longrightarrow \ \ \
H^{8_{19}}= (\ref{819threestr})
\ee
\be
10_{124}=Torus[3,5]=Pr(5,3,-\bar 2)=Pr(2,-1, 5, 3)\ \ \ \Longrightarrow \ \ \
H^{10_{124}}= (\ref{10124threestr})
\ee
and
\be
10_{139}=Pr(4, -1, 3, 3)
\ \ \ \Longrightarrow \ \ \
H^{10_{139}}=  (\ref{10139threestr})
\ee

\subsubsection{Realizations from \cite{NRZ2}}

According to \cite{Kaul} and \cite{NRZ2}, seven thick knots from \cite{DGR} can be
realized just as triple-finger starfish diagrams:

\be
H_R^{8_{19}} = \overline{\sum_{X\in R\otimes \bar R}}
\frac{d_R}{\bar S_{_{1X}}}(S T^{3}S^\dagger  )_{_{1X}}(S T^{3}S^\dagger  )_{_{1X}}
(\bar S \bar T^{-2}\bar S )_{_{1X}}
\ \ \Longrightarrow \ \  H_R^{8_{19}} = (\ref{819threestr})
\ee

\be
H_R^{9_{42}} = \overline{\sum_{X\in R\otimes\bar R}}\ \frac{d_R}{\bar S_{_{1X}} }
(S T^{3}S^\dagger  )_{_{1X}}(S T^{-2}S^\dagger  )_{_{1X}}
(S TS^\dagger \bar T^{-1}\bar S \bar T^2 \bar S)_{_{1X}}
\label{trefinger942}
\ee

\be
H_{[21]}^{9_{42}} =
\frac{1}{q^{22}A^{8}}\cdot
\ee
{\tiny
\be
\setlength{\arraycolsep}{1pt}
\nn
\left(\begin{array}{rrrrrrrrrrrrrrrrrrrrrrr}
0 & 0 & 0 & 0 & 0 & 0 & 1 & 0 & 2 & -1 & 2 & 0 & 2 & -1 & 2 & 0 & 1 & 0
& 0 & 0 & 0 & 0 & 0 \\
&&&&&&&&&&&&&&&&&&&&&& \\
0 & 0 & 0 & -1 & -1 & -2 & -1 & -2 & -3 & -3 & -4 & -2 & -4 & -3 & -3 &
-2 & -1 & -2 & -1 & -1 & 0 & 0 & 0 \\
&&&&&&&&&&&&&&&&&&&&&& \\
0 & 1 & 1 & 1 & 3 & 2 & 6 & 2 & 8 & 4 & 9 & 4 & 9 & 4 & 8 & 2 & 6 & 2 &
3 & 1 & 1 & 1 & 0 \\
&&&&&&&&&&&&&&&&&&&&&& \\
-1 & 1 & -5 & 3 & -9 & 4 & -15 & 5 & -19 & 5 & -21 & 5 & -21 & 5 & -19 &
5 & -15 & 4 & -9 & 3 & -5 & 1 & -1 \\
&&&&&&&&&&&&&&&&&&&&&& \\
1 & -2 & 4 & -3 & 9 & -6 & 14 & -7 & 20 & -10 & 24 & -10 & 24 & -10 & 20
& -7 & 14 & -6 & 9 & -3 & 4 & -2 & 1 \\
&&&&&&&&&&&&&&&&&&&&&& \\
0 & 0 & 0 & 0 & -1 & 0 & -4 & 1 & -8 & 4 & -12 & 4 & -12 & 4 & -8 & 1 &
-4 & 0 & -1 & 0 & 0 & 0 & 0 \\
&&&&&&&&&&&&&&&&&&&&&& \\
0 & 0 & 0 & 0 & -1 & 1 & -1 & 0 & 3 & -1 & 5 & -4 & 5 & -1 & 3 & 0 & -1
& 1 & -1 & 0 & 0 & 0 & 0 \\
&&&&&&&&&&&&&&&&&&&&&& \\
0 & 0 & 0 & 0 & 0 & 0 & 0 & 1 & -1 & 2 & -3 & 2 & -3 & 2 & -1 & 1 & 0 &
0 & 0 & 0 & 0 & 0 & 0
\end{array}\right)
\nn
\ee}

\bigskip

\be
H_R^{10_{124}} = \overline{\sum_{X\in R\otimes \bar R}}
\frac{d_R}{\bar S_{_{1X}}} (ST^5S^\dagger)_{_{1X}} (ST^{2}S^\dagger)_{_{1X}}
(ST^{2}S^\dagger \bar T^{-1}\bar S)_{_{1X}}
\ \ \Longrightarrow \ \  H_R^{10_{124}} = (\ref{10124threestr})
\ee

\bigskip

\be
H_R^{10_{128}} = \overline{\sum_{X\in R\otimes \bar R}}
\frac{d_R}{\bar S_{_{1X}}}
(\bar S\bar T^{-2}\bar S)_{_{1X}}
(\bar S\bar T^{-1} S T^{2}S^\dagger)_{_{1X}}
 (\bar S\bar T^{-3} ST^{2}S^\dagger)_{_{1X}}
\ee

\be
 H_{[21]}^{10_{128}} =
 \frac{1}{q^{34}A^{36}}\cdot
 \ee
{\tiny
\be
\setlength{\arraycolsep}{1pt}
\nn
\left(\begin{array}{rrrrrrrrrrrrrrrrrrrrrrrrrrrrrrrrrrr}
0 & 0 & 1 & -2 & 3 & -4 & 7 & -10 & 12 & -12 & 15 & -17 & 17 & -16 & 19 & -19 & 19 & -18 & 19 & -19 & 19 & -16 & 17 & -17 & 15 & -12 & 12 & -10 & 7 & -4 & 3 & -2 & 1 & 0 & 0 \\
&&&&&&&&&&&&&&&&&&&&&&&&&&&&&&&&&& \\
0 & 1 & -2 & 5 & -8 & 14 & -19 & 25 & -28 & 35 & -36 & 38 & -39 & 44 & -41 & 43 & -43 & 46 & -43 & 43 & -41 & 44 & -39 & 38 & -36 & 35 & -28 & 25 & -19 & 14 & -8 & 5 & -2 & 1 & 0 \\
&&&&&&&&&&&&&&&&&&&&&&&&&&&&&&&&&& \\
1 & -2 & 4 & -6 & 10 & -15 & 18 & -21 & 25 & -29 & 27 & -32 & 31 & -33 & 31 & -35 & 31 & -34 & 31 & -35 & 31 & -33 & 31 & -32 & 27 & -29 & 25 & -21 & 18 & -15 & 10 & -6 & 4 & -2 & 1 \\
&&&&&&&&&&&&&&&&&&&&&&&&&&&&&&&&&& \\
-1 & 0 & -2 & 0 & -4 & 0 & -4 & 0 & -4 & -3 & -3 & -5 & 2 & -8 & 5 & -12 & 7 & -12 & 7 & -12 & 5 & -8 & 2 & -5 & -3 & -3 & -4 & 0 & -4 & 0 & -4 & 0 & -2 & 0 & -1 \\
&&&&&&&&&&&&&&&&&&&&&&&&&&&&&&&&&& \\
0 & 1 & -1 & 3 & -2 & 7 & -4 & 11 & -8 & 16 & -13 & 22 & -19 & 27 & -25 & 34 & -30 & 34 & -30 & 34 & -25 & 27 & -19 & 22 & -13 & 16 & -8 & 11 & -4 & 7 & -2 & 3 & -1 & 1 & 0 \\
&&&&&&&&&&&&&&&&&&&&&&&&&&&&&&&&&& \\
0 & 0 & 0 & 0 & 1 & -1 & 2 & -2 & 4 & -3 & 7 & -3 & 9 & -6 & 15 & -6 & 18 & -10 & 18 & -6 & 15 & -6 & 9 & -3 & 7 & -3 & 4 & -2 & 2 & -1 & 1 & 0 & 0 & 0 & 0 \\
&&&&&&&&&&&&&&&&&&&&&&&&&&&&&&&&&& \\
0 & 0 & 0 & 0 & 0 & -1 & 0 & -3 & 0 & -5 & 1 & -9 & -1 & -12 & -1 & -14 & 0 & -16 & 0 & -14 & -1 & -12 & -1 & -9 & 1 & -5 & 0 & -3 & 0 & -1 & 0 & 0 & 0 & 0 & 0 \\
&&&&&&&&&&&&&&&&&&&&&&&&&&&&&&&&&& \\
0 & 0 & 0 & 0 & 0 & 0 & 0 & 0 & 1 & 1 & 2 & 3 & 1 & 5 & 0 & 9 & 0 & 10 & 0 & 9 & 0 & 5 & 1 & 3 & 2 & 1 & 1 & 0 & 0 & 0 & 0 & 0 & 0 & 0 & 0 \\
&&&&&&&&&&&&&&&&&&&&&&&&&&&&&&&&&& \\
0 & 0 & 0 & 0 & 0 & 0 & 0 & 0 & 0 & 0 & 0 & 0 & -1 & -1 & -2 & 0 & -2 & 0 & -2 & 0 & -2 & -1 & -1 & 0 & 0 & 0 & 0 & 0 & 0 & 0 & 0 & 0 & 0 & 0 & 0 \\
&&&&&&&&&&&&&&&&&&&&&&&&&&&&&&&&&& \\
0 & 0 & 0 & 0 & 0 & 0 & 0 & 0 & 0 & 0 & 0 & 0 & 0 & 0 & 0 & 0 & 0 & 1 & 0 & 0 & 0 & 0 & 0 & 0 & 0 & 0 & 0 & 0 & 0 & 0 & 0 & 0 & 0 & 0 & 0
\end{array}\right)
\nn
\ee}

\bigskip

\be
H_R^{10_{132}} = \overline{\sum_{X\in R\otimes \bar R}}
\frac{d_R}{\bar S_{_{1X}}} (ST^{-2}S^\dagger \bar T^3\bar S)_{_{1X}}
 (ST^{-2}S^\dagger)_{_{1X}}(\bar S\bar T^{-1}ST^2S^\dagger)_{_{1X}}
\ee

\be
H_{[21]}^{10_{132}} = \frac{A^{4}}{q^{20}}\cdot
\ee
{\tiny
\be
\setlength{\arraycolsep}{1pt}
\nn
\left(\begin{array}{rrrrrrrrrrrrrrrrrrrrr}
0 & 0 & 0 & 0 & 0 & -1 & 0 & -2 & 1 & -2 & 0 & -2 & 1 & -2 & 0 & -1 & 0 & 0 & 0 & 0 & 0 \\
&&&&&&&&&&&&&&&&&&&& \\
0 & 0 & 1 & 1 & 2 & 1 & 2 & 3 & 3 & 4 & 2 & 4 & 3 & 3 & 2 & 1 & 2 & 1 & 1 & 0 & 0 \\
&&&&&&&&&&&&&&&&&&&& \\
-1 & 0 & -2 & 1 & -5 & 1 & -8 & 1 & -10 & 1 & -10 & 1 & -10 & 1 & -8 & 1 & -5 & 1 & -2 & 0 & -1 \\
&&&&&&&&&&&&&&&&&&&& \\
0 & 0 & 2 & -3 & 5 & -3 & 7 & -3 & 5 & 0 & 7 & 0 & 5 & -3 & 7 & -3 & 5 & -3 & 2 & 0 & 0 \\
&&&&&&&&&&&&&&&&&&&& \\
0 & 1 & -1 & 1 & -1 & 2 & -1 & 0 & -1 & 0 & 0 & 0 & -1 & 0 & -1 & 2 & -1 & 1 & -1 & 1 & 0 \\
&&&&&&&&&&&&&&&&&&&& \\
1 & -1 & 1 & -2 & 4 & -5 & 6 & -8 & 10 & -11 & 10 & -11 & 10 & -8 & 6 & -5 & 4 & -2 & 1 & -1 & 1 \\
&&&&&&&&&&&&&&&&&&&& \\
0 & 0 & -1 & 2 & -4 & 6 & -9 & 13 & -15 & 17 & -18 & 17 & -15 & 13 & -9 & 6 & -4 & 2 & -1 & 0 & 0 \\
&&&&&&&&&&&&&&&&&&&& \\
0 & 0 & 0 & 0 & 0 & -1 & 3 & -5 & 7 & -9 & 10 & -9 & 7 & -5 & 3 & -1 & 0 & 0 & 0 & 0 & 0
\end{array}\right)
\nn
\ee}

\bigskip

\be
H_R^{10_{136}} = \overline{\sum_{X\in R\otimes \bar R}}
\frac{d_R}{\bar S_{_{1X}}} (\bar S\bar T^{-2}\bar S \bar T^2\bar S)_{_{1X}}
 (ST^{2}S^\dagger)_{_{1X}}(\bar S\bar T^{-3}ST^{-2}S^\dagger)_{_{1X}}
\ee

\be
H_{[21]}^{10_{136}} = \frac{1}{q^{22}A^{12}}\cdot
\ee
{\tiny
\be
\setlength{\arraycolsep}{1pt}
\nn
\left(\begin{array}{rrrrrrrrrrrrrrrrrrrrrrr}
0 & 0 & 0 & 0 & 0 & 0 & 1 & -2 & 3 & -4 & 5 & -5 & 5 & -4 & 3 & -2 & 1 & 0 & 0 & 0 & 0 & 0 & 0 \\
&&&&&&&&&&&&&&&&&&&&&& \\
0 & 0 & 0 & -1 & 1 & -2 & 3 & -4 & 3 & -5 & 4 & -4 & 4 & -5 & 3 & -4 & 3 & -2 & 1 & -1 & 0 & 0 & 0 \\
&&&&&&&&&&&&&&&&&&&&&& \\
0 & 1 & 0 & -1 & 4 & -4 & 11 & -17 & 24 & -24 & 33 & -33 & 33 & -24 & 24 & -17 & 11 & -4 & 4 & -1 & 0 & 1 & 0 \\
&&&&&&&&&&&&&&&&&&&&&& \\
-2 & 5 & -14 & 21 & -34 & 46 & -67 & 74 & -92 & 98 & -110 & 103 & -110 & 98 & -92 & 74 & -67 & 46 & -34 & 21 & -14 & 5 & -2 \\
&&&&&&&&&&&&&&&&&&&&&& \\
4 & -9 & 17 & -21 & 37 & -43 & 59 & -62 & 84 & -82 & 99 & -91 & 99 & -82 & 84 & -62 & 59 & -43 & 37 & -21 & 17 & -9 & 4 \\
&&&&&&&&&&&&&&&&&&&&&& \\
-2 & 2 & -3 & 0 & -6 & -1 & -7 & 1 & -16 & 10 & -26 & 12 & -26 & 10 & -16 & 1 & -7 & -1 & -6 & 0 & -3 & 2 & -2 \\
&&&&&&&&&&&&&&&&&&&&&& \\
0 & 1 & 0 & 3 & -2 & 5 & 3 & 2 & 14 & -10 & 24 & -14 & 24 & -10 & 14 & 2 & 3 & 5 & -2 & 3 & 0 & 1 & 0 \\
&&&&&&&&&&&&&&&&&&&&&& \\
0 & 0 & 0 & -1 & 0 & -2 & -5 & 10 & -21 & 23 & -38 & 35 & -38 & 23 & -21 & 10 & -5 & -2 & 0 & -1 & 0 & 0 & 0 \\
&&&&&&&&&&&&&&&&&&&&&& \\
0 & 0 & 0 & 0 & 0 & 0 & 2 & -2 & 5 & -5 & 8 & -7 & 8 & -5 & 5 & -2 & 2 & 0 & 0 & 0 & 0 & 0 & 0 \\
&&&&&&&&&&&&&&&&&&&&&& \\
0 & 0 & 0 & 0 & 0 & 0 & 0 & 0 & 0 & -1 & 1 & -1 & 1 & -1 & 0 & 0 & 0 & 0 & 0 & 0 & 0 & 0 & 0
\end{array}\right)
\nn
\ee}

\bigskip

\be
H_R^{10_{145}} = {\sum_{X\in R\otimes  R}}
\frac{d_R}{S_{_{1X}}} (ST^{-2}S^\dagger \bar T^2 S)_{_{1X}}
 (\bar S\bar T^{3}S)_{_{1X}}(S T^{-1}S^{\dagger}\bar T \bar S \bar T^{-1} S)_{_{1X}}
\ee
$10_{145}$ is the only example in our thick-knot set, where   summation is
over representations $X\in R\otimes R$, i.e. in the {\it parallel} sector.
{\it This} summation is {\it insensitive} to the choice of parameters $\xi$ in sec.\ref{signs},
thus there is no overline in this case.
\be
H_{[21]}^{10_{145}} = \frac{A^{12}}{q^{22}}\cdot
\ee
{\tiny
\be
\setlength{\arraycolsep}{1pt}
\nn
\left(\begin{array}{rrrrrrrrrrrrrrrrrrrrrrr}
0 & 0 & 0 & 0 & 0 & 0 & 0 & 0 & 0 & 0 & 0 & -1 & 0 & 0 & 0 & 0 & 0 & 0 & 0 & 0 & 0 & 0 & 0 \\
&&&&&&&&&&&&&&&&&&&&&& \\
0 & 0 & 0 & 0 & 0 & 0 & 0 & 0 & 1 & 0 & 1 & -1 & 1 & 0 & 1 & 0 & 0 & 0 & 0 & 0 & 0 & 0 & 0 \\
&&&&&&&&&&&&&&&&&&&&&& \\
0 & 0 & 0 & 0 & 0 & 0 & -1 & 0 & -2 & 2 & -3 & 2 & -3 & 2 & -2 & 0 & -1 & 0 & 0 & 0 & 0 & 0 & 0 \\
&&&&&&&&&&&&&&&&&&&&&& \\
0 & 0 & 0 & 1 & 0 & 3 & -2 & 3 & -2 & 5 & -3 & 3 & -3 & 5 & -2 & 3 & -2 & 3 & 0 & 1 & 0 & 0 & 0 \\
&&&&&&&&&&&&&&&&&&&&&& \\
0 & -1 & 0 & -1 & 2 & -5 & 4 & -9 & 10 & -13 & 12 & -16 & 12 & -13 & 10 & -9 & 4 & -5 & 2 & -1 & 0 & -1 & 0 \\
&&&&&&&&&&&&&&&&&&&&&& \\
0 & -1 & 2 & -1 & 2 & -1 & 1 & 3 & 0 & 4 & -1 & 5 & -1 & 4 & 0 & 3 & 1 & -1 & 2 & -1 & 2 & -1 & 0 \\
&&&&&&&&&&&&&&&&&&&&&& \\
-1 & 2 & -3 & 4 & -8 & 8 & -9 & 7 & -10 & 5 & -8 & 1 & -8 & 5 & -10 & 7 & -9 & 8 & -8 & 4 & -3 & 2 & -1 \\
&&&&&&&&&&&&&&&&&&&&&& \\
1 & -1 & 2 & -3 & 3 & -1 & 1 & 4 & -6 & 12 & -11 & 16 & -11 & 12 & -6 & 4 & 1 & -1 & 3 & -3 & 2 & -1 & 1 \\
&&&&&&&&&&&&&&&&&&&&&& \\
0 & 0 & -1 & 0 & 1 & -4 & 7 & -12 & 16 & -24 & 24 & -26 & 24 & -24 & 16 & -12 & 7 & -4 & 1 & 0 & -1 & 0 & 0 \\
&&&&&&&&&&&&&&&&&&&&&& \\
0 & 1 & 0 & 0 & 0 & 1 & -1 & 4 & -6 & 9 & -11 & 14 & -11 & 9 & -6 & 4 & -1 & 1 & 0 & 0 & 0 & 1 & 0
\end{array}\right)
\nn
\ee}

\bigskip

Three knots are more complicated, their HOMFLY polynomials are represented by double sums, at best:
\be
H_R^{10_{152}} = \overline{\sum_{X,Y\in R\otimes\bar R}}\ \frac{d_R}{\bar S_{_{1X}}\bar S_{_{1Y}}}
(\bar S \bar T^2 \bar S)_{_{1X}}(  S T^{-3}S^\dagger )_{_{1X}}
 \Big(\bar T\bar S\Big)_{_{YX}}
(S T^{-3}S^\dagger  )_{_{1Y}}
( \bar S\bar T S T^{-1}S^\dagger)_{_{1Y}}
\label{10152}
\ee

\be
H_{[21]}^{10_{152}} =
\frac{A^{24}}{q^{40}}\cdot
\ee
{\tiny
\be
\hspace{-1.6cm}
\setlength{\arraycolsep}{0.6pt}
\nn
\left(\begin{array}{rrrrrrrrrrrrrrrrrrrrrrrrrrrrrrrrrrrrrrrrr}
0 & 0 & 0 & 0 & 0 & 0 & 0 & 0 & 0 & 0 & 0 & 0 & 0 & 0 & 4 & -2 & 6 & -4
& 11 & -7 & 11 & -7 & 11 & -4 & 6 & -2 & 4 & 0 & 0 & 0 & 0 & 0 & 0 & 0 &
0 & 0 & 0 & 0 & 0 & 0 & 0 \\
&&&&&&&&&&&&&&&&&&&&&&&&&&&&&&&&&&&&&&&& \\
0 & 0 & 0 & 0 & 0 & 0 & 0 & 0 & 0 & 0 & -5 & -3 & -9 & -9 & -7 & -23 &
-3 & -37 & 5 & -46 & 4 & -46 & 5 & -37 & -3 & -23 & -7 & -9 & -9 & -3 &
-5 & 0 & 0 & 0 & 0 & 0 & 0 & 0 & 0 & 0 & 0 \\
&&&&&&&&&&&&&&&&&&&&&&&&&&&&&&&&&&&&&&&& \\
0 & 0 & 0 & 0 & 0 & 0 & 2 & 4 & 8 & 13 & 18 & 21 & 37 & 33 & 57 & 47 &
67 & 69 & 65 & 86 & 62 & 86 & 65 & 69 & 67 & 47 & 57 & 33 & 37 & 21 & 18
& 13 & 8 & 4 & 2 & 0 & 0 & 0 & 0 & 0 & 0 \\
&&&&&&&&&&&&&&&&&&&&&&&&&&&&&&&&&&&&&&&& \\
0 & 0 & 0 & -2 & -1 & -9 & -8 & -19 & -25 & -33 & -54 & -46 & -101 & -47
& -162 & -38 & -222 & -27 & -265 & -20 & -282 & -20 & -265 & -27 & -222
& -38 & -162 & -47 & -101 & -46 & -54 & -33 & -25 & -19 & -8 & -9 & -1 &
-2 & 0 & 0 & 0 \\
&&&&&&&&&&&&&&&&&&&&&&&&&&&&&&&&&&&&&&&& \\
0 & 2 & 0 & 7 & 6 & 17 & 15 & 36 & 29 & 64 & 48 & 95 & 83 & 107 & 138 &
98 & 208 & 73 & 266 & 52 & 288 & 52 & 266 & 73 & 208 & 98 & 138 & 107 &
83 & 95 & 48 & 64 & 29 & 36 & 15 & 17 & 6 & 7 & 0 & 2 & 0 \\
&&&&&&&&&&&&&&&&&&&&&&&&&&&&&&&&&&&&&&&& \\
-1 & -2 & -2 & -8 & -5 & -16 & -11 & -32 & -14 & -62 & -7 & -103 & 4 &
-142 & -2 & -158 & -21 & -154 & -48 & -147 & -58 & -147 & -48 & -154 &
-21 & -158 & -2 & -142 & 4 & -103 & -7 & -62 & -14 & -32 & -11 & -16 &
-5 & -8 & -2 & -2 & -1 \\
&&&&&&&&&&&&&&&&&&&&&&&&&&&&&&&&&&&&&&&& \\
1 & 0 & 2 & 3 & 0 & 8 & 2 & 11 & 3 & 18 & 0 & 35 & -14 & 58 & -29 & 76 &
-35 & 84 & -34 & 82 & -30 & 82 & -34 & 84 & -35 & 76 & -29 & 58 & -14 &
35 & 0 & 18 & 3 & 11 & 2 & 8 & 0 & 3 & 2 & 0 & 1
\end{array}\right)
\nn
\ee}

\bigskip

\be
H_R^{10_{153}} = \overline{\sum_{X,Y\in R\otimes\bar R}}\ \frac{d_R}{\bar S_{_{1X}}\bar S_{_{1Y}}}
(  S T^{3}S^\dagger )_{_{1X}}
(S T^2  S^\dagger)_{_{1X}} \bar S_{_{XY}}
(S T^{-2}S^\dagger \bar T\bar S)_{_{1Y}}
( ST^{-1}S^\dagger \bar T\bar S)_{_{1Y}}
\label{10153}
\ee

\be
H_{[21]}^{10_{153}} =
\frac{1}{q^{32}A^{6}}\cdot
\ee
{\tiny
\be
\setlength{\arraycolsep}{1pt}
\nn
\left(\begin{array}{rrrrrrrrrrrrrrrrrrrrrrrrrrrrrrrrr}
0 & 0 & 0 & 0 & 0 & 0 & 0 & 0 & 0 & 0 & 0 & -1 & 2 & -3 & 4 & -5 & 5 &
-5 & 4 & -3 & 2 & -1 & 0 & 0 & 0 & 0 & 0 & 0 & 0 & 0 & 0 & 0 & 0 \\
&&&&&&&&&&&&&&&&&&&&&&&&&&&&&&&& \\
0 & 0 & 0 & 0 & 0 & 0 & 0 & 0 & 0 & 0 & -1 & 2 & -5 & 8 & -11 & 13 & -15
& 13 & -11 & 8 & -5 & 2 & -1 & 0 & 0 & 0 & 0 & 0 & 0 & 0 & 0 & 0 & 0 \\
&&&&&&&&&&&&&&&&&&&&&&&&&&&&&&&& \\
0 & 0 & 0 & 0 & 0 & 0 & 0 & 0 & 2 & -3 & 5 & -8 & 13 & -12 & 15 & -14 &
19 & -14 & 15 & -12 & 13 & -8 & 5 & -3 & 2 & 0 & 0 & 0 & 0 & 0 & 0 & 0 &
0 \\
&&&&&&&&&&&&&&&&&&&&&&&&&&&&&&&& \\
0 & 0 & 1 & -1 & 3 & -3 & 6 & -3 & 6 & -1 & 2 & 2 & 2 & -3 & 5 & -8 & 10
& -8 & 5 & -3 & 2 & 2 & 2 & -1 & 6 & -3 & 6 & -3 & 3 & -1 & 1 & 0 & 0 \\
&&&&&&&&&&&&&&&&&&&&&&&&&&&&&&&& \\
-1 & 1 & -4 & 4 & -11 & 8 & -17 & 12 & -23 & 14 & -26 & 20 & -31 & 23 &
-35 & 29 & -34 & 29 & -35 & 23 & -31 & 20 & -26 & 14 & -23 & 12 & -17 &
8 & -11 & 4 & -4 & 1 & -1 \\
&&&&&&&&&&&&&&&&&&&&&&&&&&&&&&&& \\
1 & 0 & 2 & -1 & 6 & -4 & 11 & -11 & 17 & -15 & 15 & -15 & 10 & -14 & 3
& -12 & 5 & -12 & 3 & -14 & 10 & -15 & 15 & -15 & 17 & -11 & 11 & -4 & 6
& -1 & 2 & 0 & 1 \\
&&&&&&&&&&&&&&&&&&&&&&&&&&&&&&&& \\
0 & 0 & 1 & -1 & 3 & 1 & 5 & 5 & 6 & 11 & 15 & 9 & 26 & 4 & 43 & -4 & 49
& -4 & 43 & 4 & 26 & 9 & 15 & 11 & 6 & 5 & 5 & 1 & 3 & -1 & 1 & 0 & 0 \\
&&&&&&&&&&&&&&&&&&&&&&&&&&&&&&&& \\
0 & -1 & 0 & -2 & -2 & -4 & -6 & -6 & -12 & -11 & -16 & -15 & -26 & -12
& -33 & -8 & -43 & -8 & -33 & -12 & -26 & -15 & -16 & -11 & -12 & -6 &
-6 & -4 & -2 & -2 & 0 & -1 & 0 \\
&&&&&&&&&&&&&&&&&&&&&&&&&&&&&&&& \\
0 & 0 & 0 & 1 & 1 & 2 & 2 & 4 & 6 & 4 & 9 & 5 & 13 & 7 & 13 & 8 & 12 & 8
& 13 & 7 & 13 & 5 & 9 & 4 & 6 & 4 & 2 & 2 & 1 & 1 & 0 & 0 & 0 \\
&&&&&&&&&&&&&&&&&&&&&&&&&&&&&&&& \\
0 & 0 & 0 & 0 & 0 & 0 & -1 & 0 & -2 & 1 & -4 & 1 & -4 & 1 & -4 & 1 & -5
& 1 & -4 & 1 & -4 & 1 & -4 & 1 & -2 & 0 & -1 & 0 & 0 & 0 & 0 & 0 & 0
\end{array}\right)
\nn
\ee}

\bigskip

\be
H_R^{10_{154}} = \overline{\sum_{X, Y\in R\otimes\bar R}}\ \frac{d_R}{\bar S_{_{1X}}\bar S_{_{1Y}}}
(\bar S \bar T S T^{-1}S^\dagger \bar T \bar S)_{_{1X}}
(\bar S \bar T^2 \bar S)_{_{1X}} \bar S_{_{XY}}
(S T^{-2}S^\dagger \bar T\bar S)_{_{1Y}}
( ST^{-1}S^\dagger \bar T\bar S)_{_{1Y}}
\label{10154}
\ee

\be
H_{[21]}^{10_{154}} =
\frac{1}{q^{30}A^{36}}\cdot
\ee
{\tiny
\be
\setlength{\arraycolsep}{1pt}
\nn
\left(\begin{array}{rrrrrrrrrrrrrrrrrrrrrrrrrrrrrrr}
1 & 0 & 0 & 4 & -2 & 2 & 5 & -6 & 20 & -24 & 33 & -29 & 35 & -24 & 26 &
-18 & 26 & -24 & 35 & -29 & 33 & -24 & 20 & -6 & 5 & 2 & -2 & 4 & 0 & 0
& 1 \\
&&&&&&&&&&&&&&&&&&&&&&&&&&&&&& \\
0 & -2 & 2 & -3 & -5 & 8 & -23 & 29 & -48 & 48 & -59 & 44 & -47 & 34 &
-35 & 18 & -35 & 34 & -47 & 44 & -59 & 48 & -48 & 29 & -23 & 8 & -5 & -3
& 2 & -2 & 0 \\
&&&&&&&&&&&&&&&&&&&&&&&&&&&&&& \\
-2 & 3 & -4 & 2 & -2 & 0 & 4 & -10 & 9 & -10 & 4 & -7 & -5 & 8 & -17 & 6
& -17 & 8 & -5 & -7 & 4 & -10 & 9 & -10 & 4 & 0 & -2 & 2 & -4 & 3 & -2 \\
&&&&&&&&&&&&&&&&&&&&&&&&&&&&&& \\
1 & -2 & 3 & -5 & 12 & -12 & 16 & -13 & 20 & -8 & 13 & 9 & 2 & 24 & -8 &
32 & -8 & 24 & 2 & 9 & 13 & -8 & 20 & -13 & 16 & -12 & 12 & -5 & 3 & -2
& 1 \\
&&&&&&&&&&&&&&&&&&&&&&&&&&&&&& \\
0 & 1 & -1 & 3 & -1 & 4 & -1 & 1 & 3 & -9 & 21 & -41 & 52 & -79 & 84 &
-98 & 84 & -79 & 52 & -41 & 21 & -9 & 3 & 1 & -1 & 4 & -1 & 3 & -1 & 1 &
0 \\
&&&&&&&&&&&&&&&&&&&&&&&&&&&&&& \\
0 & 0 & 0 & -1 & -2 & -2 & -2 & -2 & -8 & 5 & -17 & 21 & -37 & 37 & -47
& 50 & -47 & 37 & -37 & 21 & -17 & 5 & -8 & -2 & -2 & -2 & -2 & -1 & 0 &
0 & 0 \\
&&&&&&&&&&&&&&&&&&&&&&&&&&&&&& \\
0 & 0 & 0 & 0 & 0 & 0 & 2 & 1 & 4 & -2 & 5 & 2 & 0 & -2 & 0 & 8 & 0 & -2
& 0 & 2 & 5 & -2 & 4 & 1 & 2 & 0 & 0 & 0 & 0 & 0 & 0 \\
&&&&&&&&&&&&&&&&&&&&&&&&&&&&&& \\
0 & 0 & 0 & 0 & 0 & 0 & 0 & 0 & 0 & 0 & 0 & 1 & -2 & 2 & -1 & 6 & -1 & 2
& -2 & 1 & 0 & 0 & 0 & 0 & 0 & 0 & 0 & 0 & 0 & 0 & 0 \\
&&&&&&&&&&&&&&&&&&&&&&&&&&&&&& \\
0 & 0 & 0 & 0 & 0 & 0 & 0 & 0 & 0 & 0 & 0 & 0 & -2 & 0 & -2 & 2 & -2 & 0
& -2 & 0 & 0 & 0 & 0 & 0 & 0 & 0 & 0 & 0 & 0 & 0 & 0 \\
&&&&&&&&&&&&&&&&&&&&&&&&&&&&&& \\
0 & 0 & 0 & 0 & 0 & 0 & 0 & 0 & 0 & 0 & 0 & 0 & 0 & 0 & 0 & 1 & 0 & 0 &
0 & 0 & 0 & 0 & 0 & 0 & 0 & 0 & 0 & 0 & 0 & 0 & 0
\end{array}\right)
\nn
\ee}

\bigskip

One can also represent in the same form the simplest thick knot $9_{42}$ :
\be
H_R^{9_{42}} = \overline{\sum_{X,Y\in R\otimes\bar R}}\ \frac{d_R}{\bar S_{_{1X}}\bar S_{_{1Y}}}
(S T^{3}S^\dagger  )_{_{1X}}(S T^{-2}S^\dagger  )_{_{1X}}
 \Big(\bar S\bar T^2\bar S\Big)_{_{XY}}
(S TS^\dagger  )_{_{1Y}}(S TS^\dagger  )_{_{1Y}} = (\ref{trefinger942})
\ee

\bigskip

Note that in formulas of this section we do not make any difference between the
products $T_+^mT_-^n$ and $T_-^{m+n}$, see (\ref{Amatr}).
This is because such difference gets observable only for mutants,
which appear only at $11$ intersections (alternatively we can claim
that all these knots can be represented by diagrams without combinations
$T_+^mT_-^n$ with both $m,n\neq 0$).

\subsection{Mutant pairs}

\be
H_{[21]}^{11n73} =
\frac{1}{q^{34}A^{12}}\cdot
\ee
{\tiny
\be
\setlength{\arraycolsep}{1pt}
\nn
\left(\begin{array}{rrrrrrrrrrrrrrrrrrrrrrrrrrrrrrrrrrr}
0 & 0 & 0 & 0 & 0 & 0 & 0 & -1 & 0 & -4 & 2 & -8 & 4 & -13 & 6 & -16 & 8 & -20 & 8 & -16 & 6 & -13 & 4 & -8 & 2 & -4 & 0 & -1 & 0 & 0 & 0 & 0 & 0 & 0 & 0 \\
&&&&&&&&&&&&&&&&&&&&&&&&&&&&&&&&&& \\
0 & 0 & 0 & 0 & 1 & 2 & 4 & 7 & 5 & 20 & 2 & 46 & -8 & 75 & -19 & 103 & -28 & 108 & -28 & 103 & -19 & 75 & -8 & 46 & 2 & 20 & 5 & 7 & 4 & 2 & 1 & 0 & 0 & 0 & 0 \\
&&&&&&&&&&&&&&&&&&&&&&&&&&&&&&&&&& \\
0 & 0 & -2 & -1 & -8 & -6 & -20 & -14 & -44 & -31 & -80 & -48 & -120 & -81 & -139 & -109 & -148 & -134 & -148 & -109 & -139 & -81 & -120 & -48 & -80 & -31 & -44 & -14 & -20 & -6 & -8 & -1 & -2 & 0 & 0 \\
&&&&&&&&&&&&&&&&&&&&&&&&&&&&&&&&&& \\
1 & 0 & 7 & 3 & 19 & 8 & 52 & 17 & 115 & 21 & 203 & 44 & 279 & 91 & 326 & 161 & 329 & 187 & 329 & 161 & 326 & 91 & 279 & 44 & 203 & 21 & 115 & 17 & 52 & 8 & 19 & 3 & 7 & 0 & 1 \\
&&&&&&&&&&&&&&&&&&&&&&&&&&&&&&&&&& \\
-3 & 0 & -9 & -6 & -20 & -27 & -38 & -68 & -67 & -131 & -103 & -221 & -138 & -320 & -162 & -396 & -177 & -428 & -177 & -396 & -162 & -320 & -138 & -221 & -103 & -131 & -67 & -68 & -38 & -27 & -20 & -6 & -9 & 0 & -3 \\
&&&&&&&&&&&&&&&&&&&&&&&&&&&&&&&&&& \\
3 & 0 & 7 & 8 & 10 & 39 & 1 & 98 & -10 & 183 & -10 & 262 & 18 & 322 & 52 & 359 & 82 & 374 & 82 & 359 & 52 & 322 & 18 & 262 & -10 & 183 & -10 & 98 & 1 & 39 & 10 & 8 & 7 & 0 & 3 \\
&&&&&&&&&&&&&&&&&&&&&&&&&&&&&&&&&& \\
-1 & 0 & -5 & -3 & -8 & -12 & -14 & -33 & -24 & -51 & -50 & -61 & -100 & -39 & -176 & -14 & -223 & 12 & -223 & -14 & -176 & -39 & -100 & -61 & -50 & -51 & -24 & -33 & -14 & -12 & -8 & -3 & -5 & 0 & -1 \\
&&&&&&&&&&&&&&&&&&&&&&&&&&&&&&&&&& \\
0 & 0 & 2 & -1 & 7 & -4 & 18 & -9 & 32 & -10 & 45 & -11 & 62 & -12 & 83 & -35 & 107 & -32 & 107 & -35 & 83 & -12 & 62 & -11 & 45 & -10 & 32 & -9 & 18 & -4 & 7 & -1 & 2 & 0 & 0 \\
&&&&&&&&&&&&&&&&&&&&&&&&&&&&&&&&&& \\
0 & 0 & 0 & 0 & -1 & 0 & -3 & 2 & -5 & -2 & 0 & -17 & 23 & -50 & 59 & -91 & 90 & -106 & 90 & -91 & 59 & -50 & 23 & -17 & 0 & -2 & -5 & 2 & -3 & 0 & -1 & 0 & 0 & 0 & 0 \\
&&&&&&&&&&&&&&&&&&&&&&&&&&&&&&&&&& \\
0 & 0 & 0 & 0 & 0 & 0 & 0 & 1 & -2 & 5 & -9 & 15 & -20 & 27 & -32 & 38 & -40 & 42 & -40 & 38 & -32 & 27 & -20 & 15 & -9 & 5 & -2 & 1 & 0 & 0 & 0 & 0 & 0 & 0 & 0
\end{array}\right)\nn
\ee}

\begin{landscape}
\be
H_{[21]}^{11n74} =
\frac{1}{q^{36}A^{12}}
\cdot
\ee
{\tiny
\be
\hspace{-1.5cm}
\setlength{\arraycolsep}{1pt}
\nn
\left(\begin{array}{rrrrrrrrrrrrrrrrrrrrrrrrrrrrrrrrrrrrr}
0 & 0 & 0 & 0 & 0 & 0 & 0 & 0 & -1 & 0 & -4 & 2 & -8 & 4 & -13 & 6 & -16 & 8 & -20 & 8 & -16 & 6 & -13 & 4 & -8 & 2 & -4 & 0 & -1 & 0 & 0 & 0 & 0 & 0 & 0 & 0 & 0 \\
&&&&&&&&&&&&&&&&&&&&&&&&&&&&&&&&&&&& \\
0 & 0 & 0 & 0 & 0 & 1 & 2 & 4 & 8 & 2 & 25 & -7 & 61 & -29 & 103 & -54 & 143 & -72 & 154 & -72 & 143 & -54 & 103 & -29 & 61 & -7 & 25 & 2 & 8 & 4 & 2 & 1 & 0 & 0 & 0 & 0 & 0 \\
&&&&&&&&&&&&&&&&&&&&&&&&&&&&&&&&&&&& \\
0 & 0 & 0 & -2 & -1 & -10 & -1 & -27 & -2 & -62 & -10 & -105 & -20 & -145 & -58 & -161 & -93 & -161 & -120 & -161 & -93 & -161 & -58 & -145 & -20 & -105 & -10 & -62 & -2 & -27 & -1 & -10 & -1 & -2 & 0 & 0 & 0 \\
&&&&&&&&&&&&&&&&&&&&&&&&&&&&&&&&&&&& \\
0 & 1 & 1 & 6 & 2 & 22 & 0 & 70 & -11 & 154 & -34 & 268 & -26 & 356 & 15 & 396 & 92 & 394 & 127 & 394 & 92 & 396 & 15 & 356 & -26 & 268 & -34 & 154 & -11 & 70 & 0 & 22 & 2 & 6 & 1 & 1 & 0 \\
&&&&&&&&&&&&&&&&&&&&&&&&&&&&&&&&&&&& \\
-1 & 0 & -6 & 1 & -21 & -1 & -49 & -13 & -92 & -44 & -156 & -75 & -255 & -91 & -382 & -85 & -490 & -73 & -534 & -73 & -490 & -85 & -382 & -91 & -255 & -75 & -156 & -44 & -92 & -13 & -49 & -1 & -21 & 1 & -6 & 0 & -1 \\
&&&&&&&&&&&&&&&&&&&&&&&&&&&&&&&&&&&& \\
1 & 0 & 6 & -3 & 23 & -9 & 61 & -24 & 122 & -33 & 208 & -38 & 296 & -29 & 384 & -25 & 453 & -22 & 480 & -22 & 453 & -25 & 384 & -29 & 296 & -38 & 208 & -33 & 122 & -24 & 61 & -9 & 23 & -3 & 6 & 0 & 1 \\
&&&&&&&&&&&&&&&&&&&&&&&&&&&&&&&&&&&& \\
0 & -1 & -1 & -4 & -2 & -11 & -4 & -32 & -5 & -63 & 4 & -115 & 9 & -177 & 37 & -246 & 55 & -288 & 72 & -288 & 55 & -246 & 37 & -177 & 9 & -115 & 4 & -63 & -5 & -32 & -4 & -11 & -2 & -4 & -1 & -1 & 0 \\
&&&&&&&&&&&&&&&&&&&&&&&&&&&&&&&&&&&& \\
0 & 0 & 0 & 2 & -1 & 9 & -9 & 25 & -21 & 50 & -31 & 70 & -39 & 87 & -35 & 105 & -51 & 120 & -46 & 120 & -51 & 105 & -35 & 87 & -39 & 70 & -31 & 50 & -21 & 25 & -9 & 9 & -1 & 2 & 0 & 0 & 0 \\
&&&&&&&&&&&&&&&&&&&&&&&&&&&&&&&&&&&& \\
0 & 0 & 0 & 0 & 0 & -1 & 0 & -3 & 1 & -2 & -7 & 9 & -32 & 44 & -78 & 94 & -131 & 134 & -152 & 134 & -131 & 94 & -78 & 44 & -32 & 9 & -7 & -2 & 1 & -3 & 0 & -1 & 0 & 0 & 0 & 0 & 0 \\
&&&&&&&&&&&&&&&&&&&&&&&&&&&&&&&&&&&& \\
0 & 0 & 0 & 0 & 0 & 0 & 0 & 0 & 1 & -2 & 5 & -9 & 15 & -20 & 27 & -32 & 38 & -40 & 42 & -40 & 38 & -32 & 27 & -20 & 15 & -9 & 5 & -2 & 1 & 0 & 0 & 0 & 0 & 0 & 0 & 0 & 0
\end{array}\right)\nn
\ee}

\be
H_{[21]}^{11n76} =
\frac{1}{q^{40}A^{24}}
\cdot
\ee
{\tiny
\be
\hspace{-1cm}
\setlength{\arraycolsep}{0.7pt}
\nn
\left(\begin{array}{rrrrrrrrrrrrrrrrrrrrrrrrrrrrrrrrrrrrrrrrr}
0 & 0 & 0 & 0 & 0 & -1 & 2 & -7 & 14 & -29 & 47 & -78 & 110 & -156 & 198 & -254 & 297 & -347 & 374 & -404 & 404 & -404 & 374 & -347 & 297 & -254 & 198 & -156 & 110 & -78 & 47 & -29 & 14 & -7 & 2 & -1 & 0 & 0 & 0 & 0 & 0 \\
&&&&&&&&&&&&&&&&&&&&&&&&&&&&&&&&&&&&&&&& \\
0 & 0 & 1 & -1 & 6 & -8 & 23 & -32 & 73 & -101 & 186 & -244 & 382 & -456 & 624 & -689 & 858 & -880 & 1018 & -986 & 1076 & -986 & 1018 & -880 & 858 & -689 & 624 & -456 & 382 & -244 & 186 & -101 & 73 & -32 & 23 & -8 & 6 & -1 & 1 & 0 & 0 \\
&&&&&&&&&&&&&&&&&&&&&&&&&&&&&&&&&&&&&&&& \\
-1 & 2 & -10 & 18 & -52 & 83 & -176 & 251 & -444 & 567 & -875 & 1012 & -1396 & 1481 & -1904 & 1883 & -2295 & 2160 & -2547 & 2304 & -2630 & 2304 & -2547 & 2160 & -2295 & 1883 & -1904 & 1481 & -1396 & 1012 & -875 & 567 & -444 & 251 & -176 & 83 & -52 & 18 & -10 & 2 & -1 \\
&&&&&&&&&&&&&&&&&&&&&&&&&&&&&&&&&&&&&&&& \\
2 & -3 & 16 & -22 & 63 & -75 & 169 & -164 & 321 & -249 & 449 & -219 & 466 & -39 & 335 & 271 & 139 & 573 & -38 & 756 & -89 & 756 & -38 & 573 & 139 & 271 & 335 & -39 & 466 & -219 & 449 & -249 & 321 & -164 & 169 & -75 & 63 & -22 & 16 & -3 & 2 \\
&&&&&&&&&&&&&&&&&&&&&&&&&&&&&&&&&&&&&&&& \\
-1 & 0 & -7 & 2 & -21 & -8 & -28 & -69 & 22 & -263 & 193 & -645 & 517 & -1196 & 935 & -1801 & 1335 & -2296 & 1597 & -2563 & 1700 & -2563 & 1597 & -2296 & 1335 & -1801 & 935 & -1196 & 517 & -645 & 193 & -263 & 22 & -69 & -28 & -8 & -21 & 2 & -7 & 0 & -1 \\
&&&&&&&&&&&&&&&&&&&&&&&&&&&&&&&&&&&&&&&& \\
0 & 1 & 0 & 4 & 4 & 13 & 11 & 35 & 23 & 92 & 37 & 170 & 67 & 266 & 156 & 302 & 291 & 279 & 451 & 237 & 516 & 237 & 451 & 279 & 291 & 302 & 156 & 266 & 67 & 170 & 37 & 92 & 23 & 35 & 11 & 13 & 4 & 4 & 0 & 1 & 0 \\
&&&&&&&&&&&&&&&&&&&&&&&&&&&&&&&&&&&&&&&& \\
0 & 0 & 0 & -1 & 0 & -4 & -2 & -14 & -12 & -19 & -47 & -11 & -149 & 56 & -314 & 168 & -532 & 284 & -687 & 361 & -762 & 361 & -687 & 284 & -532 & 168 & -314 & 56 & -149 & -11 & -47 & -19 & -12 & -14 & -2 & -4 & 0 & -1 & 0 & 0 & 0 \\
&&&&&&&&&&&&&&&&&&&&&&&&&&&&&&&&&&&&&&&& \\
0 & 0 & 0 & 0 & 0 & 0 & 1 & 0 & 4 & 2 & 11 & 12 & 7 & 45 & -16 & 121 & -82 & 221 & -152 & 299 & -190 & 299 & -152 & 221 & -82 & 121 & -16 & 45 & 7 & 12 & 11 & 2 & 4 & 0 & 1 & 0 & 0 & 0 & 0 & 0 & 0 \\
&&&&&&&&&&&&&&&&&&&&&&&&&&&&&&&&&&&&&&&& \\
0 & 0 & 0 & 0 & 0 & 0 & 0 & 0 & 0 & 0 & -1 & 0 & -4 & -1 & -8 & -2 & -11 & -4 & -15 & -6 & -16 & -6 & -15 & -4 & -11 & -2 & -8 & -1 & -4 & 0 & -1 & 0 & 0 & 0 & 0 & 0 & 0 & 0 & 0 & 0 & 0 \\
&&&&&&&&&&&&&&&&&&&&&&&&&&&&&&&&&&&&&&&& \\
0 & 0 & 0 & 0 & 0 & 0 & 0 & 0 & 0 & 0 & 0 & 0 & 0 & 0 & 0 & 1 & 0 & 2 & -1 & 2 & 0 & 2 & -1 & 2 & 0 & 1 & 0 & 0 & 0 & 0 & 0 & 0 & 0 & 0 & 0 & 0 & 0 & 0 & 0 & 0 & 0
\end{array}\right)\nn
\ee}

\newpage

\be
H_{[21]}^{11n78} =
\frac{1}{q^{40}A^{24}}
\cdot
\ee
{\tiny
\be
\hspace{-1cm}
\setlength{\arraycolsep}{1pt}
\nn
\left(\begin{array}{rrrrrrrrrrrrrrrrrrrrrrrrrrrrrrrrrrrrrrrrr}
0 & 0 & 0 & 0 & 0 & -1 & 2 & -7 & 14 & -29 & 47 & -78 & 110 & -156 & 198 & -254 & 297 & -347 & 374 & -404 & 404 & -404 & 374 & -347 & 297 & -254 & 198 & -156 & 110 & -78 & 47 & -29 & 14 & -7 & 2 & -1 & 0 & 0 & 0 & 0 & 0 \\
&&&&&&&&&&&&&&&&&&&&&&&&&&&&&&&&&&&&&&&& \\
0 & 0 & 1 & -1 & 6 & -8 & 23 & -32 & 73 & -101 & 186 & -244 & 382 & -457 & 626 & -691 & 862 & -886 & 1024 & -993 & 1084 & -993 & 1024 & -886 & 862 & -691 & 626 & -457 & 382 & -244 & 186 & -101 & 73 & -32 & 23 & -8 & 6 & -1 & 1 & 0 & 0 \\
&&&&&&&&&&&&&&&&&&&&&&&&&&&&&&&&&&&&&&&& \\
-1 & 2 & -10 & 18 & -52 & 83 & -176 & 251 & -444 & 567 & -873 & 1009 & -1394 & 1476 & -1898 & 1880 & -2291 & 2155 & -2547 & 2304 & -2626 & 2304 & -2547 & 2155 & -2291 & 1880 & -1898 & 1476 & -1394 & 1009 & -873 & 567 & -444 & 251 & -176 & 83 & -52 & 18 & -10 & 2 & -1 \\
&&&&&&&&&&&&&&&&&&&&&&&&&&&&&&&&&&&&&&&& \\
2 & -3 & 16 & -22 & 63 & -75 & 169 & -165 & 321 & -247 & 447 & -214 & 456 & -29 & 325 & 287 & 127 & 580 & -50 & 765 & -93 & 765 & -50 & 580 & 127 & 287 & 325 & -29 & 456 & -214 & 447 & -247 & 321 & -165 & 169 & -75 & 63 & -22 & 16 & -3 & 2 \\
&&&&&&&&&&&&&&&&&&&&&&&&&&&&&&&&&&&&&&&& \\
-1 & 0 & -7 & 2 & -21 & -7 & -30 & -66 & 18 & -258 & 189 & -642 & 513 & -1195 & 933 & -1795 & 1327 & -2286 & 1581 & -2544 & 1684 & -2544 & 1581 & -2286 & 1327 & -1795 & 933 & -1195 & 513 & -642 & 189 & -258 & 18 & -66 & -30 & -7 & -21 & 2 & -7 & 0 & -1 \\
&&&&&&&&&&&&&&&&&&&&&&&&&&&&&&&&&&&&&&&& \\
0 & 1 & 0 & 4 & 4 & 12 & 13 & 32 & 27 & 87 & 41 & 167 & 71 & 265 & 158 & 296 & 299 & 269 & 467 & 218 & 532 & 218 & 467 & 269 & 299 & 296 & 158 & 265 & 71 & 167 & 41 & 87 & 27 & 32 & 13 & 12 & 4 & 4 & 0 & 1 & 0 \\
&&&&&&&&&&&&&&&&&&&&&&&&&&&&&&&&&&&&&&&& \\
0 & 0 & 0 & -1 & 0 & -4 & -2 & -13 & -12 & -21 & -45 & -16 & -139 & 46 & -304 & 152 & -520 & 277 & -675 & 352 & -758 & 352 & -675 & 277 & -520 & 152 & -304 & 46 & -139 & -16 & -45 & -21 & -12 & -13 & -2 & -4 & 0 & -1 & 0 & 0 & 0 \\
&&&&&&&&&&&&&&&&&&&&&&&&&&&&&&&&&&&&&&&& \\
0 & 0 & 0 & 0 & 0 & 0 & 1 & 0 & 4 & 2 & 9 & 15 & 5 & 50 & -22 & 124 & -86 & 226 & -152 & 299 & -194 & 299 & -152 & 226 & -86 & 124 & -22 & 50 & 5 & 15 & 9 & 2 & 4 & 0 & 1 & 0 & 0 & 0 & 0 & 0 & 0 \\
&&&&&&&&&&&&&&&&&&&&&&&&&&&&&&&&&&&&&&&& \\
0 & 0 & 0 & 0 & 0 & 0 & 0 & 0 & 0 & 0 & -1 & 0 & -4 & 0 & -10 & 0 & -15 & 2 & -21 & 1 & -24 & 1 & -21 & 2 & -15 & 0 & -10 & 0 & -4 & 0 & -1 & 0 & 0 & 0 & 0 & 0 & 0 & 0 & 0 & 0 & 0 \\
&&&&&&&&&&&&&&&&&&&&&&&&&&&&&&&&&&&&&&&& \\
0 & 0 & 0 & 0 & 0 & 0 & 0 & 0 & 0 & 0 & 0 & 0 & 0 & 0 & 0 & 1 & 0 & 2 & -1 & 2 & 0 & 2 & -1 & 2 & 0 & 1 & 0 & 0 & 0 & 0 & 0 & 0 & 0 & 0 & 0 & 0 & 0 & 0 & 0 & 0 & 0
\end{array}\right)\nn
\ee}

\be
H_{[21]}^{11a57} =
\frac{1}{q^{40}A^{12}}
\cdot
\ee
{\tiny
\be
\hspace{-1.3cm}
\setlength{\arraycolsep}{0.3pt}
\nn
\left(\begin{array}{rrrrrrrrrrrrrrrrrrrrrrrrrrrrrrrrrrrrrrrrr}
0 & 0 & 0 & 0 & 0 & 0 & 0 & 0 & 0 & 0 & -1 & 0 & -4 & 2 & -8 & 4 & -13 & 6 & -16 & 8 & -20 & 8 & -16 & 6 & -13 & 4 & -8 & 2 & -4 & 0 & -1 & 0 & 0 & 0 & 0 & 0 & 0 & 0 & 0 & 0 & 0 \\
&&&&&&&&&&&&&&&&&&&&&&&&&&&&&&&&&&&&&&&& \\
0 & 0 & 0 & 0 & 0 & 0 & 0 & 2 & 1 & 9 & 0 & 19 & -1 & 38 & 1 & 57 & -2 & 77 & 3 & 86 & -4 & 86 & 3 & 77 & -2 & 57 & 1 & 38 & -1 & 19 & 0 & 9 & 1 & 2 & 0 & 0 & 0 & 0 & 0 & 0 & 0 \\
&&&&&&&&&&&&&&&&&&&&&&&&&&&&&&&&&&&&&&&& \\
0 & 0 & 0 & 0 & -1 & -3 & -3 & -16 & 6 & -56 & 35 & -148 & 99 & -292 & 203 & -479 & 320 & -651 & 419 & -762 & 450 & -762 & 419 & -651 & 320 & -479 & 203 & -292 & 99 & -148 & 35 & -56 & 6 & -16 & -3 & -3 & -1 & 0 & 0 & 0 & 0 \\
&&&&&&&&&&&&&&&&&&&&&&&&&&&&&&&&&&&&&&&& \\
0 & 0 & 2 & -1 & 13 & -8 & 40 & -13 & 75 & 14 & 97 & 97 & 97 & 220 & 99 & 346 & 131 & 439 & 161 & 475 & 184 & 475 & 161 & 439 & 131 & 346 & 99 & 220 & 97 & 97 & 97 & 14 & 75 & -13 & 40 & -8 & 13 & -1 & 2 & 0 & 0 \\
&&&&&&&&&&&&&&&&&&&&&&&&&&&&&&&&&&&&&&&& \\
-1 & 2 & -13 & 24 & -78 & 129 & -293 & 437 & -836 & 1141 & -1923 & 2472 & -3750 & 4492 & -6241 & 6974 & -8862 & 9247 & -10899 & 10608 & -11644 & 10608 & -10899 & 9247 & -8862 & 6974 & -6241 & 4492 & -3750 & 2472 & -1923 & 1141 & -836 & 437 & -293 & 129 & -78 & 24 & -13 & 2 & -1 \\
&&&&&&&&&&&&&&&&&&&&&&&&&&&&&&&&&&&&&&&& \\
3 & -9 & 39 & -92 & 233 & -441 & 872 & -1412 & 2376 & -3452 & 5159 & -6816 & 9305 & -11367 & 14282 & -16253 & 19100 & -20389 & 22510 & -22725 & 23770 & -22725 & 22510 & -20389 & 19100 & -16253 & 14282 & -11367 & 9305 & -6816 & 5159 & -3452 & 2376 & -1412 & 872 & -441 & 233 & -92 & 39 & -9 & 3 \\
&&&&&&&&&&&&&&&&&&&&&&&&&&&&&&&&&&&&&&&& \\
-3 & 11 & -42 & 100 & -235 & 437 & -810 & 1277 & -2012 & 2798 & -3938 & 4966 & -6373 & 7424 & -8878 & 9657 & -10881 & 11240 & -12129 & 12016 & -12518 & 12016 & -12129 & 11240 & -10881 & 9657 & -8878 & 7424 & -6373 & 4966 & -3938 & 2798 & -2012 & 1277 & -810 & 437 & -235 & 100 & -42 & 11 & -3 \\
&&&&&&&&&&&&&&&&&&&&&&&&&&&&&&&&&&&&&&&& \\
1 & -4 & 15 & -34 & 76 & -126 & 209 & -274 & 354 & -328 & 290 & -40 & -256 & 855 & -1407 & 2278 & -2980 & 3878 & -4357 & 4932 & -4940 & 4932 & -4357 & 3878 & -2980 & 2278 & -1407 & 855 & -256 & -40 & 290 & -328 & 354 & -274 & 209 & -126 & 76 & -34 & 15 & -4 & 1 \\
&&&&&&&&&&&&&&&&&&&&&&&&&&&&&&&&&&&&&&&& \\
0 & 0 & -1 & 3 & -8 & 11 & -11 & -13 & 61 & -174 & 354 & -646 & 1001 & -1486 & 2012 & -2596 & 3101 & -3625 & 3991 & -4252 & 4286 & -4252 & 3991 & -3625 & 3101 & -2596 & 2012 & -1486 & 1001 & -646 & 354 & -174 & 61 & -13 & -11 & 11 & -8 & 3 & -1 & 0 & 0 \\
&&&&&&&&&&&&&&&&&&&&&&&&&&&&&&&&&&&&&&&& \\
0 & 0 & 0 & 0 & 0 & 1 & -4 & 12 & -26 & 48 & -73 & 101 & -118 & 114 & -75 & 12 & 86 & -202 & 317 & -386 & 413 & -386 & 317 & -202 & 86 & 12 & -75 & 114 & -118 & 101 & -73 & 48 & -26 & 12 & -4 & 1 & 0 & 0 & 0 & 0 & 0
\end{array}\right)\nn
\ee}

\newpage

\be
H_{[21]}^{11a231} =
\frac{1}{q^{40}A^{12}}
\cdot
\ee
{\tiny
\be
\hspace{-1.3cm}
\setlength{\arraycolsep}{0.3pt}
\nn
\left(\begin{array}{rrrrrrrrrrrrrrrrrrrrrrrrrrrrrrrrrrrrrrrrr}
0 & 0 & 0 & 0 & 0 & 0 & 0 & 0 & 0 & 0 & -1 & 0 & -4 & 2 & -8 & 4 & -13 & 6 & -16 & 8 & -20 & 8 & -16 & 6 & -13 & 4 & -8 & 2 & -4 & 0 & -1 & 0 & 0 & 0 & 0 & 0 & 0 & 0 & 0 & 0 & 0 \\
&&&&&&&&&&&&&&&&&&&&&&&&&&&&&&&&&&&&&&&& \\
0 & 0 & 0 & 0 & 0 & 0 & 0 & 2 & 1 & 9 & 0 & 19 & -1 & 38 & 1 & 57 & -2 & 77 & 3 & 86 & -4 & 86 & 3 & 77 & -2 & 57 & 1 & 38 & -1 & 19 & 0 & 9 & 1 & 2 & 0 & 0 & 0 & 0 & 0 & 0 & 0 \\
&&&&&&&&&&&&&&&&&&&&&&&&&&&&&&&&&&&&&&&& \\
0 & 0 & 0 & 0 & -1 & -3 & -3 & -16 & 6 & -56 & 36 & -151 & 104 & -301 & 218 & -500 & 348 & -686 & 459 & -806 & 496 & -806 & 459 & -686 & 348 & -500 & 218 & -301 & 104 & -151 & 36 & -56 & 6 & -16 & -3 & -3 & -1 & 0 & 0 & 0 & 0 \\
&&&&&&&&&&&&&&&&&&&&&&&&&&&&&&&&&&&&&&&& \\
0 & 0 & 2 & -1 & 13 & -8 & 40 & -15 & 80 & 7 & 109 & 79 & 118 & 195 & 127 & 321 & 154 & 417 & 177 & 462 & 198 & 462 & 177 & 417 & 154 & 321 & 127 & 195 & 118 & 79 & 109 & 7 & 80 & -15 & 40 & -8 & 13 & -1 & 2 & 0 & 0 \\
&&&&&&&&&&&&&&&&&&&&&&&&&&&&&&&&&&&&&&&& \\
-1 & 2 & -13 & 24 & -77 & 128 & -294 & 440 & -844 & 1159 & -1951 & 2511 & -3805 & 4557 & -6311 & 7051 & -8938 & 9317 & -10968 & 10673 & -11704 & 10673 & -10968 & 9317 & -8938 & 7051 & -6311 & 4557 & -3805 & 2511 & -1951 & 1159 & -844 & 440 & -294 & 128 & -77 & 24 & -13 & 2 & -1 \\
&&&&&&&&&&&&&&&&&&&&&&&&&&&&&&&&&&&&&&&& \\
3 & -9 & 38 & -89 & 227 & -431 & 857 & -1393 & 2354 & -3427 & 5135 & -6793 & 9280 & -11339 & 14248 & -16206 & 19038 & -20312 & 22416 & -22621 & 23664 & -22621 & 22416 & -20312 & 19038 & -16206 & 14248 & -11339 & 9280 & -6793 & 5135 & -3427 & 2354 & -1393 & 857 & -431 & 227 & -89 & 38 & -9 & 3 \\
&&&&&&&&&&&&&&&&&&&&&&&&&&&&&&&&&&&&&&&& \\
-3 & 11 & -41 & 97 & -229 & 427 & -795 & 1258 & -1990 & 2773 & -3914 & 4943 & -6348 & 7396 & -8844 & 9610 & -10819 & 11163 & -12035 & 11912 & -12412 & 11912 & -12035 & 11163 & -10819 & 9610 & -8844 & 7396 & -6348 & 4943 & -3914 & 2773 & -1990 & 1258 & -795 & 427 & -229 & 97 & -41 & 11 & -3 \\
&&&&&&&&&&&&&&&&&&&&&&&&&&&&&&&&&&&&&&&& \\
1 & -4 & 15 & -34 & 75 & -125 & 210 & -277 & 362 & -346 & 318 & -79 & -201 & 790 & -1337 & 2201 & -2904 & 3808 & -4288 & 4867 & -4880 & 4867 & -4288 & 3808 & -2904 & 2201 & -1337 & 790 & -201 & -79 & 318 & -346 & 362 & -277 & 210 & -125 & 75 & -34 & 15 & -4 & 1 \\
&&&&&&&&&&&&&&&&&&&&&&&&&&&&&&&&&&&&&&&& \\
0 & 0 & -1 & 3 & -8 & 11 & -11 & -11 & 56 & -167 & 342 & -628 & 980 & -1461 & 1984 & -2571 & 3078 & -3603 & 3975 & -4239 & 4272 & -4239 & 3975 & -3603 & 3078 & -2571 & 1984 & -1461 & 980 & -628 & 342 & -167 & 56 & -11 & -11 & 11 & -8 & 3 & -1 & 0 & 0 \\
&&&&&&&&&&&&&&&&&&&&&&&&&&&&&&&&&&&&&&&& \\
0 & 0 & 0 & 0 & 0 & 1 & -4 & 12 & -26 & 48 & -74 & 104 & -123 & 123 & -90 & 33 & 58 & -167 & 277 & -342 & 367 & -342 & 277 & -167 & 58 & 33 & -90 & 123 & -123 & 104 & -74 & 48 & -26 & 12 & -4 & 1 & 0 & 0 & 0 & 0 & 0
\end{array}\right)\nn
\ee}

\be
H_{[21]}^{11n71} =
\frac{A^{6}}{q^{34}}
\cdot
\ee
{\tiny
\be
\hspace{-1cm}
\setlength{\arraycolsep}{1pt}
\nn
\left(\begin{array}{rrrrrrrrrrrrrrrrrrrrrrrrrrrrrrrrrrr}
0 & 0 & 0 & 0 & 0 & 0 & 0 & 0 & 0 & 0 & 0 & 0 & 1 & 0 & 2 & -1 & 2 & 0 & 2 & -1 & 2 & 0 & 1 & 0 & 0 & 0 & 0 & 0 & 0 & 0 & 0 & 0 & 0 & 0 & 0 \\
&&&&&&&&&&&&&&&&&&&&&&&&&&&&&&&&&& \\
0 & 0 & 0 & 0 & 0 & 0 & 0 & 0 & 0 & -3 & -2 & -8 & 0 & -10 & -2 & -19 & -2 & -16 & -2 & -19 & -2 & -10 & 0 & -8 & -2 & -3 & 0 & 0 & 0 & 0 & 0 & 0 & 0 & 0 & 0 \\
&&&&&&&&&&&&&&&&&&&&&&&&&&&&&&&&&& \\
0 & 0 & 0 & 0 & 0 & 0 & 3 & 6 & 9 & 16 & 6 & 36 & 13 & 69 & 8 & 85 & 8 & 100 & 8 & 85 & 8 & 69 & 13 & 36 & 6 & 16 & 9 & 6 & 3 & 0 & 0 & 0 & 0 & 0 & 0 \\
&&&&&&&&&&&&&&&&&&&&&&&&&&&&&&&&&& \\
0 & 0 & 0 & -1 & -5 & -5 & -25 & 2 & -67 & 20 & -164 & 59 & -287 & 141 & -437 & 209 & -538 & 243 & -538 & 209 & -437 & 141 & -287 & 59 & -164 & 20 & -67 & 2 & -25 & -5 & -5 & -1 & 0 & 0 & 0 \\
&&&&&&&&&&&&&&&&&&&&&&&&&&&&&&&&&& \\
0 & 1 & 2 & 10 & 8 & 26 & 23 & 51 & 77 & 78 & 153 & 96 & 239 & 161 & 293 & 218 & 299 & 253 & 299 & 218 & 293 & 161 & 239 & 96 & 153 & 78 & 77 & 51 & 23 & 26 & 8 & 10 & 2 & 1 & 0 \\
&&&&&&&&&&&&&&&&&&&&&&&&&&&&&&&&&& \\
0 & -8 & 8 & -46 & 59 & -172 & 205 & -475 & 542 & -1026 & 1123 & -1866 & 1926 & -2863 & 2796 & -3722 & 3363 & -4080 & 3363 & -3722 & 2796 & -2863 & 1926 & -1866 & 1123 & -1026 & 542 & -475 & 205 & -172 & 59 & -46 & 8 & -8 & 0 \\
&&&&&&&&&&&&&&&&&&&&&&&&&&&&&&&&&& \\
1 & 10 & -22 & 79 & -142 & 334 & -506 & 921 & -1257 & 1966 & -2431 & 3408 & -3892 & 4954 & -5259 & 6203 & -6111 & 6655 & -6111 & 6203 & -5259 & 4954 & -3892 & 3408 & -2431 & 1966 & -1257 & 921 & -506 & 334 & -142 & 79 & -22 & 10 & 1 \\
&&&&&&&&&&&&&&&&&&&&&&&&&&&&&&&&&& \\
-1 & -3 & 11 & -46 & 89 & -204 & 333 & -573 & 773 & -1134 & 1368 & -1773 & 1952 & -2357 & 2418 & -2747 & 2675 & -2894 & 2675 & -2747 & 2418 & -2357 & 1952 & -1773 & 1368 & -1134 & 773 & -573 & 333 & -204 & 89 & -46 & 11 & -3 & -1 \\
&&&&&&&&&&&&&&&&&&&&&&&&&&&&&&&&&& \\
0 & 0 & 1 & 4 & -9 & 22 & -30 & 54 & -38 & 4 & 108 & -237 & 479 & -700 & 987 & -1173 & 1362 & -1371 & 1362 & -1173 & 987 & -700 & 479 & -237 & 108 & 4 & -38 & 54 & -30 & 22 & -9 & 4 & 1 & 0 & 0 \\
&&&&&&&&&&&&&&&&&&&&&&&&&&&&&&&&&& \\
0 & 0 & 0 & 0 & 0 & -1 & -3 & 14 & -37 & 79 & -161 & 278 & -431 & 605 & -792 & 947 & -1058 & 1093 & -1058 & 947 & -792 & 605 & -431 & 278 & -161 & 79 & -37 & 14 & -3 & -1 & 0 & 0 & 0 & 0 & 0
\end{array}\right)\nn
\ee}

\newpage

\be
H_{[21]}^{11n75} =
\frac{A^{6}}{q^{34}}
\cdot
\ee
{\tiny
\be
\hspace{-1cm}
\setlength{\arraycolsep}{0.6pt}
\nn
\left(\begin{array}{rrrrrrrrrrrrrrrrrrrrrrrrrrrrrrrrrrr}
0 & 0 & 0 & 0 & 0 & 0 & 0 & 0 & 0 & 0 & 0 & 0 & 1 & 0 & 2 & -1 & 2 & 0 & 2 & -1 & 2 & 0 & 1 & 0 & 0 & 0 & 0 & 0 & 0 & 0 & 0 & 0 & 0 & 0 & 0 \\
&&&&&&&&&&&&&&&&&&&&&&&&&&&&&&&&&& \\
0 & 0 & 0 & 0 & 0 & 0 & 0 & 0 & 0 & -3 & -2 & -8 & 0 & -10 & -2 & -19 & -2 & -16 & -2 & -19 & -2 & -10 & 0 & -8 & -2 & -3 & 0 & 0 & 0 & 0 & 0 & 0 & 0 & 0 & 0 \\
&&&&&&&&&&&&&&&&&&&&&&&&&&&&&&&&&& \\
0 & 0 & 0 & 0 & 0 & 0 & 3 & 6 & 9 & 16 & 5 & 38 & 11 & 73 & 2 & 91 & 1 & 108 & 1 & 91 & 2 & 73 & 11 & 38 & 5 & 16 & 9 & 6 & 3 & 0 & 0 & 0 & 0 & 0 & 0 \\
&&&&&&&&&&&&&&&&&&&&&&&&&&&&&&&&&& \\
0 & 0 & 0 & -1 & -5 & -5 & -25 & 4 & -70 & 22 & -169 & 65 & -290 & 145 & -442 & 209 & -538 & 247 & -538 & 209 & -442 & 145 & -290 & 65 & -169 & 22 & -70 & 4 & -25 & -5 & -5 & -1 & 0 & 0 & 0 \\
&&&&&&&&&&&&&&&&&&&&&&&&&&&&&&&&&& \\
0 & 1 & 2 & 10 & 7 & 26 & 25 & 49 & 82 & 68 & 163 & 86 & 255 & 149 & 300 & 206 & 308 & 249 & 308 & 206 & 300 & 149 & 255 & 86 & 163 & 68 & 82 & 49 & 25 & 26 & 7 & 10 & 2 & 1 & 0 \\
&&&&&&&&&&&&&&&&&&&&&&&&&&&&&&&&&& \\
0 & -8 & 9 & -48 & 62 & -176 & 210 & -479 & 545 & -1030 & 1124 & -1868 & 1932 & -2871 & 2806 & -3738 & 3382 & -4096 & 3382 & -3738 & 2806 & -2871 & 1932 & -1868 & 1124 & -1030 & 545 & -479 & 210 & -176 & 62 & -48 & 9 & -8 & 0 \\
&&&&&&&&&&&&&&&&&&&&&&&&&&&&&&&&&& \\
1 & 10 & -23 & 81 & -145 & 338 & -511 & 925 & -1260 & 1970 & -2432 & 3410 & -3898 & 4962 & -5269 & 6219 & -6130 & 6671 & -6130 & 6219 & -5269 & 4962 & -3898 & 3410 & -2432 & 1970 & -1260 & 925 & -511 & 338 & -145 & 81 & -23 & 10 & 1 \\
&&&&&&&&&&&&&&&&&&&&&&&&&&&&&&&&&& \\
-1 & -3 & 11 & -46 & 90 & -204 & 331 & -571 & 768 & -1124 & 1358 & -1763 & 1936 & -2345 & 2411 & -2735 & 2666 & -2890 & 2666 & -2735 & 2411 & -2345 & 1936 & -1763 & 1358 & -1124 & 768 & -571 & 331 & -204 & 90 & -46 & 11 & -3 & -1 \\
&&&&&&&&&&&&&&&&&&&&&&&&&&&&&&&&&& \\
0 & 0 & 1 & 4 & -9 & 22 & -30 & 52 & -35 & 2 & 113 & -243 & 482 & -704 & 992 & -1173 & 1362 & -1375 & 1362 & -1173 & 992 & -704 & 482 & -243 & 113 & 2 & -35 & 52 & -30 & 22 & -9 & 4 & 1 & 0 & 0 \\
&&&&&&&&&&&&&&&&&&&&&&&&&&&&&&&&&& \\
0 & 0 & 0 & 0 & 0 & -1 & -3 & 14 & -37 & 79 & -160 & 276 & -429 & 601 & -786 & 941 & -1051 & 1085 & -1051 & 941 & -786 & 601 & -429 & 276 & -160 & 79 & -37 & 14 & -3 & -1 & 0 & 0 & 0 & 0 & 0
\end{array}\right) \nn
\ee}
\end{landscape}
\end{document}